\documentclass[english,notitlepage,nofootinbib]{revtex4-1}
\usepackage[T1]{fontenc}
\usepackage[latin9]{inputenc}
\usepackage{xcolor}
\usepackage{amsmath}
\usepackage{amssymb}
\usepackage{graphicx}
\usepackage{geometry}
\makeatletter

\providecommand{\tabularnewline}{\\}

\pdfoutput=1
\usepackage[colorlinks=true,citecolor=blue,linkcolor=blue]{hyperref}
\usepackage{slashed}

\allowdisplaybreaks

\makeatother

\usepackage{babel}
\begin{document}
\title{Neutrino bound states and bound systems }
\author{Alexei Yu. Smirnov$^a$ and Xun-Jie Xu$^b$}
\affiliation{
$^a$ Max-Planck-Institut f\"ur Kernphysik, Postfach 103980, D-69029 Heidelberg, Germany\\
$^b$ Institute of High Energy Physics, Chinese Academy of Sciences, Beijing 100049, China}
\date{\today}

\begin{abstract}
%%%%%%%%%%%%%%%%%%%%%%%%%%%%%%%%%%%%%%%%%%%%%%%%%%%%%%%%%%%%%
\noindent 

Yukawa interactions of neutrinos with a new light scalar boson $\phi$ 
can lead to formation of  stable bound states
and bound systems of many neutrinos ($\nu$-clusters). 
For allowed values
of the coupling $y$ and the scalar mass $m_\phi$,
the bound state of two neutrinos would have the size larger than $10^{12}$ cm.
Bound states with sub-cm sizes are possible for keV scale 
sterile neutrinos with coupling $y > 10^{-4}$.
For $\nu$-clusters we study in detail the properties of final stable configurations.
If there is an efficient cooling mechanism, these configurations
are in the state of degenerate Fermi gas. We formulate
and solve equations of the density distributions
in $\nu$-clusters.  
In the non-relativistic case, they are reduced to the Lane-Emden equation. 
We find that (i) stable configurations exist for
any number of neutrinos, $N$; (ii) there is a maximal central density
$\sim 10^9$ cm$^{-3}$ determined by the neutrino mass;
(iii) for a given $m_\phi$ there is a minimal value of $Ny^3$
for which stable configurations can be formed;
(iv) for a given strength of interaction, $S_\phi = (ym_\nu/m_\phi)^2$, 
the minimal radius of $\nu$-clusters exists. 
We discuss the formation of $\nu$-clusters 
from  relic neutrino background in the process of expansion and 
cooling of the Universe. 
One possibility realized for $S_\phi > 700$ is the development of instabilities 
in the $\nu$-background at $T < m_\nu$
which leads to its fragmentation.  
For $S_\phi \in [70, \, 700]$ they might be formed via the growth of initial density perturbations in the 
$\nu$-background and virialiazation, in analogy with the formation of Dark Matter halos. 
For allowed values of $y$,  cooling of $\nu$-clusters due to $\phi$-bremsstrahlung 
and neutrino annihilation is negligible.  The sizes of $\nu$-clusters may range from  $\sim$ km to $\sim 5$ Mpc.
Formation of clusters affects perspectives of detection of relic neutrinos.

\end{abstract}
\maketitle
\newpage
\tableofcontents

\newpage

\section{Introduction}
%%%%%%%%%%%%%%%%%%%%%%%%%%%%%%%%%%%%%%%%%%%%%%%%%%%%%%%%%%

In recent years 
neutrinos and new  physics at low energy scales
became one of the most active areas of research.
% \cite{Samanta:2010zh,Heeck:2010pg,Boehm:2012gr,Heeck:2014zfa,Bilmis:2015lja,Huang:2017egl,
% Farzan:2018gtr,Lindner:2018kjo,Abdullah:2018ykz,Bustamante:2018mzu,Wise:2018rnb,
% Smirnov:2019cae,Babu:2019iml,Luo:2020fdt,Esteban:2021ozz}. 
In particular,  neutrino interactions with new and very light bosons have 
been actively explored, with considerations covering 
e.g.~oscillation 
phenomena~\cite{Samanta:2010zh,Heeck:2010pg,Wise:2018rnb,Bustamante:2018mzu,Smirnov:2019cae,Babu:2019iml} 
and various astrophysical and  cosmological 
consequences~\cite{Boehm:2012gr,Heeck:2014zfa,Huang:2017egl,Luo:2020fdt,Esteban:2021ozz,Green:2021gdc}.
Yet another possible consequence of such interactions (the subject of this paper) is the 
formation of bound states and bound systems of neutrinos.\\

The exploration of possible existence
of  $N$-body bound states of neutrinos (neutrino  clouds, balls,  
stars, or halos) has  a long story.
In 1964 Markov proposed the existence of ``neutrino superstars'' formed by
massive neutrinos due to the usual gravity~\cite{Markov:1964fjm}.
The system is similar to neutron stars and
can be described as the degenerate Fermi gas in thermodynamical
and mechanical equilibrium. Substituting the
mass of neutron, $m_n$, by the mass of neutrino, $m_n \rightarrow m_\nu$ in  
results for neutron stars, Markov obtained the radius
of a neutrino star:
\begin{equation}
R \simeq  \sqrt{\frac{8\pi}{G_N}} \frac{1}{m_\nu^2}, 
\label{eq:nsrad}
\end{equation}
which corresponds to the Oppenheimer-Volkoff limit.  
In comparison to the radius of neutron star,  
$R$ increases by a factor of $\beta^2$  
where
$$
\beta \equiv \frac{m_n}{m_\nu} = 1.9 \cdot 10^{10}, 
$$
and for numerical estimation we take $m_\nu = 0.05$ eV.
Correspondingly, the mass of neutrino star increases
and the central density decreases as
\begin{equation}
M_\nu \simeq  \beta^2 M_{\rm NS}, \, \, \, \, \, 
n_\nu (0) \simeq \beta^{-3} n_{\rm NS}(0).
\label{eq:nsmassden}
\end{equation}
For $m_\nu \approx m_e $  used by Markov, the size, the mass and the number density of the superstars 
would be  $R = 10^{12}$ cm,  $M = 10^6 M_\odot$ and $n_\nu  \simeq  10^{29} \, {\rm cm}^{-3}$ correspondingly. 
For the present values of neutrino mass (0.1 eV), the relations  
(\ref{eq:nsrad}) and (\ref{eq:nsmassden}) give
%\begin{equation}
$$
R \simeq 5 \cdot 10^{26} \, {\rm cm}, \, \, \, \, \,
M_\nu = 4 \cdot 10^{20} M_{\odot}, \, \, \, \, \, 
n_\nu (0) \simeq  10^{8} \, {\rm cm}^{-3}.
%\label{eq:nsrad1}
%\end{equation}
$$
The radius is only  20 times smaller
than the size of observable Universe (the present Hubble scale), and
the total number of neutrinos is 10 times larger than the
number of relic neutrinos in whole the Universe with the
standard density. The  bound states of much smaller sizes and 
higher densities with substantial implications 
for cosmology and structure formation 
as well as for laboratory experiments require new physics 
beyond the Standard Model. \\

One can also consider heavy (sterile) neutrinos and gravity only. 
Masses of $(10 - 100)$ keV correspond to $\beta = (10^{5} - 10^4)$,  
and consequently, 
 stars have radii of $(10^{16} - 10^{14})$ cm and masses of 
$(10^{10} - 10^8) M_{\odot}$,
{\it i.e.} much below the galactic scales. 
The central density would be
$(10^{24} - 10^{27}) \, {\rm cm}^{-3}$. 
Such a possibility was explored 
in a series of papers~\cite{Viollier:1993mm,Bilic:1998kn,Bilic:2001iv}.
This is essentially the warm dark matter for which final 
configurations of degenerate gas cannot be achieved.  \\

Another way to reduce the size and mass of a neutrino star  
is to consider new long-range interactions between neutrinos 
which are much stronger than gravity. 
%%Much smaller sizes and masses for neutrino bound states
%%are possible  if new long range forces acting on neutrinos
%%exist. 
In the case of Yukawa forces due to a new light scalar boson  $\phi$
with the  coupling constant to neutrinos, $y$, the  Newton coupling $G_N$  
should be substituted in (\ref{eq:nsrad}) by
\begin{equation}
G_\nu = \frac{y^2}{4\pi} \frac{1}{m_\nu^2}.
\label{eq:gphi}
\end{equation}
As a result, we obtain
\begin{equation}
R =  4 \pi \sqrt{2} \frac{1}{y m_\nu}  =
\frac{17.8}{y m_\nu}.
\label{eq:nradnew}
\end{equation}
Notice that this expression does not depend
on the mass of mediator $m_\phi$
as long as $m_\phi \ll 1/R$. 
For $y = 10^{-7}$ and $m_\nu = 0.05$ eV, one finds
from (\ref{eq:gphi}) $\sqrt{G_\nu} = 0.56$ MeV$^{-1}$
(that is, 22 orders of magnitude larger than $\sqrt{G_N}$), 
and correspondingly Eq. (\ref{eq:nradnew}) gives $R = 0.7$ km. 

Properties and formation of the $\nu$-clusters due to new 
scalar interactions
were studied in \cite{Stephenson:1996qj,Stephenson:1993rx} 
using equations of Quantum 
Hadrodynamics. 
%Hydrodynamics.
The central notion is the effective neutrino mass
$\tilde{m}_{\nu}$ in the background of degenerate neutrino 
gas and scalar field.
%%From the equation of motion for $\nu$ and $\phi$
The equation for $\tilde{m}_{\nu}$ was obtained in the limit
of static and uniform medium. This non-linear equation depends
on the Fermi momentum $k_F$ and the strength 
of interactions which is defined as 
$$
%\begin{equation}
G_\phi \equiv  \frac{y^2 \omega}{2\pi^2 m_\phi^2}. 
%\label{eq:strength}
%\end{equation}
$$
Here $\omega$ is the number of degrees of freedom. 
For large densities before neutrino clustering,  
the effective mass   
is close to zero \cite{Stephenson:1996qj}: 
$$
%\begin{equation}
\tilde{m}_{\nu} \approx \frac{2m_\nu}{G_\phi k_F^2}. 
%\label{eq:meff}
%\end{equation}
$$
At small densities: $\tilde{m}_{\nu} \rightarrow m_\nu$. 
Using $\tilde{m}_{\nu}$,  the energy density
of neutrinos, $\rho_\nu$,  and   the energy density of the scalar field, $\rho_\phi$,  
were computed in \cite{Stephenson:1996qj} which gives the total 
energy per neutrino
$$
%\begin{equation}
 \epsilon^{\rm tot} = \epsilon_\nu +\epsilon_\phi =
\frac{1}{n_F}(\rho_\nu + \rho_\phi),
%\label{eq:toten}
%\end{equation}
$$
here $n_F = k_F^3/6 \pi^2$. For large enough strength 
$G_\phi$ the energy
$\epsilon^{\rm tot}$ as a function of $k_F$ acquires a minimum  
with $\epsilon^{\rm tot} < m_\nu$ around $k_F^{\min} \approx 0.8 m_\nu$.  
Clustering of neutrinos starts when $k_F$ decreases due to the expansion of the Universe 
 down to 
$$
%\begin{equation}
k_F \sim k_F^{\min}. 
%\end{equation}
$$
%%where $k_F^{\min}$ corresponds to minimum of total energy
%%$\epsilon^{\rm tot}$.
It is also assumed that with the expansion (decrease of $n$),   
the kinetic energy could be  converted into the increasing effective 
mass $\tilde{m}_{\nu}$ 
and the gas became strongly degenerate.

To describe finite size objects, 
the static equations of motion  were used 
with non-zero spatial derivatives 
of the fields,  and consequently,   $\tilde{m}_{\nu}$.
%\footnote{ 
%    \color{blue} 
%The same set of equations was also applied to dark matter 
%%in Ref.~\cite{Wise:2014ola} which however did not exactly solve the equations. 
%%Only approximate solutions were obtained in Ref.~\cite{Wise:2014ola} by taking 
%%two ansatzs on the distribution. Solving the equations in this approach would 
%%miss the step of  matching solutions to the boundary conditions which provide additional constraints. 
%%In our work, we find that the matching can affect the existence of solutions.
%    } 
It was assumed that the distribution of clusters on the number of neutrinos $N$ 
follows the density fluctuations and has the Harrison-Zeldovich spectrum.
In \cite{Stephenson:1996qj}, the neutrino mass  $m_\nu = 13$ eV
motivated by the Tritium experiment anomaly was used.  
Very small  mediator masses $ m_\phi \sim 10^{-17}$ eV 
and couplings $g \sim  10^{-14}$ 
were taken. Consequently, the sizes of clusters equal 
$\sim 10^{13}$ cm, which is roughly the size of the Earth orbit, and the central densities 
could reach $10^{15}$ cm$^{-3}$.

%%The idea that neutrinos could form bound states has been considered
%%in a few early studies. In the 1960s, M. Markov~\cite{Markov:1964fjm}
%%studied the possibility that neutrino stars could be formed due to
%%gravitational attraction, assuming upper bounds of $\nu_{e}$ and
%%$\nu_{\mu}$ masses are $m_{\nu_e} < 4 \times 10^{-4} m_{e}$ 
%%and $m_{\nu_{\mu}}<8m_{e}$,
%%respectively. Although today's experimental data have constrained
%%neutrino masses to be below ${\cal O}(0.1)$ eV~\cite{Aghanim:2018eyx,Aker:2019uuj},
%%similar ideas have been pursued for ${\cal O}(10)$ keV sterile 
%%neutrinos~\cite{Viollier:1993mm,Bilic:1998kn,Bilic:2001iv}.
%Considering new forces, the earliest discussion appeared in the seminal
%%work on majorons~\cite{Chikashige:1980ui} 
%%[[???]]
%%and more recently there
%%were studies on neutrino clouds~\cite{Stephenson:1996qj,Stephenson:1993rx}.

%\noindent
%{\bf insertion, p. 4, before paragraph: ``In the paper..."}
%%%%%%%%%%%%%%%%%%%%%%%%%%%%%%%%%%%%%%%%%%%%%%%%%%%%%%%%%%%
%{\color{red}
Later the bound systems of heavy Dirac fermions (``nuggets'') due to 
Yukawa couplings with light or massless scalar were explored
\cite{Wise:2014jva,Wise:2014ola,Gresham:2017zqi,Gresham:2018anj}.  With fermion masses 
$m_D \sim 100$ GeV and coupling $\alpha_\phi \sim (0.01 - 0.1)$ such 
systems  were applied to the Asymmetric Dark Matter (ADM).
The description of nuggets in \cite{Wise:2014ola},
\cite{Gresham:2017zqi} is similar to that of neutrino
stars \cite{Stephenson:1996qj}\footnote{
	The authors of those ADM papers  overlooked (and do not refer to) similar 
	papers on neutrinos
	\cite{Stephenson:1996qj} published 20 years before.
	In the first version of our paper we, being focussed on neutrinos, overlooked the papers on 
	ADM.}.
The effective mass of fermion was introduced
as in \cite{Stephenson:1996qj} and equations for its dependence on $r$ 
was derived.
The  system of equations for the scalar field and Fermi
momentum of ADM particles  $p_F(r)$ (and therefore their density) was presented.
In \cite{Wise:2014ola} the equations were solved by introducing
an ansatz for $p_F(r)$, thus reducing the system to a single equation. In 
\cite{Gresham:2017zqi} the
complete system of equations was solved numerically.
Dependence of properties of nuggets on $N$,
in particular $R(N)$, were studied.
It was established  in \cite{Wise:2014ola}
that with the increase of  $N$ the radius first decreases,
reaches minimum and then increases. The increase
is associated to transition from non-relativistic to relativistic regime.
While in \cite{Wise:2014ola} the mediator was assumed to be massless, in 
\cite{Gresham:2017zqi} dependence of
characteristics of nuggets on $m_\phi$ was studied and, in particular, 
the case of  ``saturation'', which corresponds to $R > 1/m_\phi$, 
was noticed.  

Due to strongly different values of masses and coupling constant 
the formation and cosmological consequences of nuggets 
are completely different from those of neutrino stars. Also observational 
methods and perspectives of detection are different.  
Our paper is continuation of the work in 
\cite{Stephenson:1996qj} for neutrinos. 
Where possible,  we compare our results with those for ADM.

We perform a systematic
study of various aspects of physics of neutrino clusters. The paper 
contains a number of derivations and auxiliary material which will be 
useful for further studies and applications.

%}

In this paper, we revisit the possibility of formation of the neutrino 
bound states and bound systems due to neutrino interactions with 
very light scalar bosons. 
First the bound states of two neutrinos are considered. 
We formulate conditions for their existence and find their parameters 
such as eigenstates and eigenvalues. For usual neutrinos and allowed values  
of $y$ and $m_\phi$, the sizes of these bound states are of the astronomical  
distances  which has no sense. 
For sterile neutrinos with mass $m_\nu \geq 10$ keV and 
$y > 10^{-4}$ the size can be in sub-cm range and formation of such  states 
may have implications for dark matter. 

Then we study properties and formation of the $N$-neutrino  
bound systems \textemdash\,the $\nu$-clusters with a sufficiently large $N$, 
so that the system can be quantitatively described  using Fermi-Dirac
statistics.  
We compute characteristics of possible stable final configurations 
of the $\nu$-clusters as function of $N$ \textemdash\,mostly in the state 
of degenerate Fermi gas (ground state) 
but also  with thermal distributions. 
Energy loss of the $\nu$-clusters is estimated. 
Depending on $N$ (we explore the entire possible range) the 
clusters can be in non-relativistic or relativistic regimes and we 
study the transition between these two regimes. 
In the non-relativistic case, using equations of
motions  for $\nu$ and $\phi$,  
we rederive the Lane-Emden equation for the neutrino density.  
%%which is essentially the equation 
%%for Hydrostatic equilibrium for degenerate fermionic gas. 
We obtain qualitative analytic results for parameters 
of the neutrino cluster. 
The equation for the relativistic case is derived 
which is similar to the one used in 
\cite{Stephenson:1996qj}. 
We use presently known values of neutrino masses which 
are two orders of magnitude smaller than in 
\cite{Stephenson:1996qj}, and consequently,  parameters 
of the clouds are different. 

Understanding the formation of neutrino clusters from 
homogeneous relativistic neutrino gas requires numerical simulations. 
In this connection we present some relevant analytic results. 
One possible mechanism of formation is the development of instabilities 
in the uniform neutrino background, which leads to fragmentation and 
further redistribution of neutrino density, thus 
approaching final degenerate configurations.  
In some parameter space,  one can use an analogy between Yukawa forces and gravity 
and therefore compare it with  the formation of dark matter 
halos due to gravity.  \\

The paper is organized as follows. 
We start with re-derivation of the Yukawa potential 
in Sec.~\ref{sec:Yukawa-potential}, 
where we adopt the approach  developed 
in Ref.~\cite{Smirnov:2019cae}
and generalize it to  the relativistic case.  
In Sec.~\ref{sec:2-body}, we study the two-body bound states 
in the framework of quantum mechanics. 
%%since neutrinos in two-body bound states cannot be relativistic due
%%to various known experiments bounds.  
In Sec.~\ref{sec:N-body},
the $N$-body bound systems are considered.  
The results of Sec.~\ref{sec:2-body} and Sec.~\ref{sec:N-body}
apply to the non-relativistic neutrinos, while a study
of the relativistic case is presented in  Sec.~\ref{sec:Relativistic}.
In Sec.~\ref{sec:Formation-and-lifetime}, 
we speculate on possible mechanisms of formation of the $\nu$-clusters
which include fragmentation of the cosmic neutrino background 
induced by the expansion of the Universe. 
Further evolution and formation of final configurations are 
outlined. Energy loss of the clusters is estimated. 
Analogy with  formation of DM halos is outlined.  
%we estimate the time it
%takes for neutrinos in the early universe to collapse and form the
%%degenerate state and compute the neutrino annihilation rates.]] 
%%Combining the formation and annihilation time with current 
%%known bounds on neutrinophilic
%%scalar interactions, we find in Sec.~\ref{sec:Known-constraints}
%%the viable parameter space. 
Finally, we conclude in Sec.~\ref{sec:Conclusion}. 
Various details and explanations are presented in the appendices.

\section{Yukawa potential for neutrinos \label{sec:Yukawa-potential}}
%%%%%%%%%%%%%%%%%%%%%%%%%%%%%%%%%%%%%%%%%%%%%%%%%%%%%%%%%%%%%%%%

In general, to form a bound state, the range of the force, $R_{{\rm force}}$, 
 needs to be longer than the de Broglie wavelength of the neutrino, 
$\lambda_{\nu}$:
\begin{equation}
R_{{\rm force}}>\lambda_{\nu}.
\label{eq:m}
\end{equation}
For a non-relativistic neutrino  $\lambda_{\nu} = (m_{\nu}v)^{-1}$, 
where $v$ is the neutrino velocity,  and  $R_{{\rm force}} \sim1/n_\phi$ 
is given by the inverse of the mediator mass. Therefore Eq.~\eqref{eq:m} 
can be written as $m_\phi < m_\nu v$, which implies that 
mediator should be  lighter than the neutrino 
to form a bound state\footnote{In the
standard model (SM), the $Z$ boson does mediate an attractive force
between a neutrino and an antineutrino but due to its large mass,
the range of this force is too short to bind neutrinos together.}.

Let us consider a light real scalar boson $\phi$ interacting
with neutrinos 
\begin{equation}
{\cal L}=\frac{1}{2}\partial^{\mu}\phi\partial_{\mu}\phi-
\frac{1}{2}m_{\phi}^{2}\phi^{2} + 
\overline{\nu}i\slashed{\partial}\nu - m_{\nu}\overline{\nu}\nu - 
y\overline{\nu}\phi\nu.
\label{eq:L}
\end{equation}
We assume $m_{\phi}\ll m_{\nu}$,  so that 
$\nu$ can be treated  as point-like particles while  
$\phi$ as a classical field.
This is valid as long as we are not
concerned with hard scattering processes. 
In this regime, the effect of the $\phi$ field on $\nu$ particles  can be described using Yukawa
potential. 

From  Eq.~\eqref{eq:L}, we obtain the Euler-Lagrange equations 
of motion (EOM) for $\nu$ and $\phi$:
\begin{align}
i\slashed{\partial}\nu - (m_{\nu}+y\phi)\nu & =0\thinspace,
\label{eq:-39}\\
(\partial^{2}+m_{\phi}^{2})\phi + y\overline{\nu}\nu & = 0\thinspace.
\label{eq:-40}
\end{align}
According to Eq.~\eqref{eq:-39} the effect of  the $\phi$ field
on the motion of $\nu$ can be accounted by shifting its mass:
\begin{equation}
\tilde{m}_{\nu} \equiv   m_{\nu} + V,\ ~~~~~ V\equiv y\phi\thinspace.
\label{eq:-59}
\end{equation}
Here $V$ is the potential, and $\tilde{m}_{\nu}$ is treated as  
the effective mass of neutrino in background.

At quantum level 
%%When $\psi$ is quantized and the quanta are interpreted as particles,
$\overline{\nu}\nu$ is replaced by the expectation value computed as (see
Appendix~\ref{sec:QFT}):
\begin{equation}
\overline{\nu}\nu \rightarrow \langle\overline{\nu}\nu\rangle = 
\int\frac{d^3 \mathbf{p}}{(2\pi)^{3}}\frac{\tilde{m}_{\nu}}{E_{\mathbf{p}}}
f(t,\ \mathbf{x},\ \mathbf{p})\thinspace,
\label{eq:-57}
\end{equation}
where $E_{\mathbf{p}}=
\sqrt{\tilde{m}_{\nu}^{2}+|\mathbf{p}|^{2}}$ 
and  $f(t,\ \mathbf{x},\ \mathbf{p})$
is the neutrino distribution function  
normalized in such a way that  the number density, 
$n$, and the total number of $\nu$, $N$,  are given by 
$$
%\begin{equation}
n =\int f(t,\ \mathbf{x},\ 
\mathbf{p})\frac{d^{3}\mathbf{p}}{(2\pi)^{3}}\thinspace,\ \ N = 
\int n(t,\ \mathbf{x})d^{3}\mathbf{x}\thinspace.
%\label{eq:-58}
%\end{equation}
$$

{\it In the non-relativistic case},  $\tilde{m}_{\nu}/E_{\mathbf{p}}\approx 1$, 
Eq.~\eqref{eq:-57} gives  
\begin{equation}
\langle\overline{\nu}\nu\rangle\approx n.
\end{equation}
We assume a static distribution, so that $\partial_{t}\phi=0$.
Replacing  $\overline{\psi}\psi$ by $\langle\overline{\psi}\psi\rangle \approx n$ 
in Eq.~\eqref{eq:-40} and using relation    $\phi = V/y$, 
(see Eq.~\eqref{eq:-59}) we obtain equation for the potential 
\begin{equation}
\left[ \nabla^{2} - m_{\phi}^{2} \right] V=y^2 n.
\label{eq:-60}
\end{equation}
For spherically symmetric distribution of neutrinos, $n=n(r)$,  
where $r$ is the distance from the center, Eq.~\eqref{eq:-60} can be solved 
analytically~\cite{Smirnov:2019cae}, giving 
\begin{equation}
V=-\frac{y^{2}}{m_{\phi}r}\left[e^{-m_{\phi}r}\int_{0}^{r}dr'r'n(r')
\sinh(m_{\phi}r')  + 
\sinh(m_{\phi}r')\int_{r}^{\infty}dr'e^{-m_{\phi}r'}r'n(r') \right].
\label{eq:-169}
\end{equation}
For $n=N\delta^{3}(\mathbf{x})$ (all neutrinos are in origin) 
the solution reproduces 
the Yukawa potential  
\begin{equation}
V(r)=-N\frac{y^{2}}{4\pi}\frac{1}{r}e^{-rm_{\phi}}.  
\label{eq:V_NR}
\end{equation}
Note that $V$ is the potential experienced by a given neutrino, and  
for two neutrinos separated by a distance $r$,  
%%Since each of them causes a potential for the other,
%%both particles have $m_{\psi}\rightarrow m_{\psi}+V$. 
%%[[edit]] So, one might
%%be confused about a factor of two in the total binding energy. In fact, 
the total binding energy is given by Eq.~\eqref{eq:V_NR} with 
$N=1$ (see Appendix~\ref{sec:energy-issues} for details).

%%Eq.~\eqref{eq:-60} only applies to slow moving $\psi$ particles.
{\it For relativistic neutrinos}, if  $\partial_{t}\phi$ can be
neglected\footnote{The time derivative $\partial_{t}\phi$ can only be neglected for
stationary (e.g. the ground state of a binary system) or approximately
stationary distributions (e.g., a many-body system with a low temperature).
 Otherwise, $\partial_{t}\phi$ plays an important role in quantum
state transitions or in energy loss of a many-body system due to emission
of $\phi$ waves. This is similar to the familiar example of an
excited atom that will undergo transition to the ground state in finite 
time via spontaneous emission of electromagnetic waves.}, 
one should use general expression in (\ref{eq:-57}) 
and the equation for the potential becomes 
\begin{equation}
\left[ \nabla^{2} - m_{\phi}^2 \right] V = 
y^2 \int\frac{d^{3}\mathbf{p}}{(2\pi)^{3}}
\frac{\tilde{m}_{\nu}}{E_{\mathbf{p}}}f(\mathbf{x},\ \mathbf{p}) .
\label{eq:-61}
\end{equation}
The resulting $V$ is suppressed  by a factor of $\tilde{m}_{\nu}/E_{\mathbf{p}}$
in comparison  to the non-relativistic case. 
This equation can be written in the form Eq.~\eqref{eq:-60}
with $n$ substituted by the effective number density, $\tilde{n}$, 
defined as 
$$
%\begin{equation} 
\tilde{n} \equiv \int\frac{d^{3}\mathbf{p}}{(2\pi)^{3}}
\frac{\tilde{m}_{\nu}}{E_{\mathbf{p}}}f(\mathbf{x},\ \mathbf{p}). 
\label{eq:-eff}
%\end{equation}
$$
In the non-relativistic case $\tilde{n} \rightarrow n$.

In this paper we focus on the real scalar field. Pseudo-scalar mediators 
(e.g.~the majoron) cause spin-dependent forces, and the corresponding potential can be found in \cite{Chikashige:1980ui}.
For a two-body system, the
potential might be able to bind the pair if the mediator mass is
sufficiently light. In an $N$-body system with large $N$, the effect of spin-dependent
forces is suppressed by spin averaging. 
%The conditions for the formation of bound 
%systems are more complicated and will be addressed elsewhere. 

In the case of  a complex scalar, our analysis 
can be applied to its real part, while the effect of its imaginary
part, which is a pseudo-scalar, can be neglected if the two parts have
the same mass. In models with spontaneous symmetry
breaking such as the one in \cite{Chikashige:1980ui}, only the
pseudo-scalar component is massless and therefore leads to long-range effects. 
%%We leave explorations on this possibility to future work.

%%%%%%%%%%%%%%%%%%%%%%%%%%%%%%%%%%%%%%%%%%%%%%%%%%%%%%%%%%%%%%%%%
\section{Two-neutrino bound states\label{sec:2-body}}
%%%%%%%%%%%%%%%%%%%%%%%%%%%%%%%%%%%%%%%%%%%%%%%%%%%%%%%%%%%%%%%%%%%%%%%%%

\noindent 
%%Now we apply the Yukawa potential to neutrinos and consider
%%the possibility that two neutrinos are bound by the $\phi$ mediator
%%and form a two-body bound state.   

In this section, we consider bound states of two  
non-relativistic neutrinos due to the attractive $\phi$ potential. 
%{\color{red}
	For DM particles this was considered in a number of papers before 
	(see \cite{Wise:2014jva} and references therein). 
%} 

%%The relativistic
%%case will be studied in Sec.~\ref{sec:Relativistic}.   

% \subsection{Qualitative consideration}
%%%%%%%%%%%%%%%%%%%%%%%%%%%%%%%%%%%%%%%%%%%%%%%%%%%%%%%%%%%%
% [[without subsection]]
%For two non-relativistic neutrinos bound by an attractive
%potential, it is a typical two-body problem in quantum mechanics,
%which can be studied using the Schr\"odinger equation. Without quantitatively
%solving the Schr\"odinger equation, 
The strength of the interaction required to form bound states 
can be estimated in the following way. 
If the  wave function of the bound state is localized in 
the region of radius $R$, then 
according to the Heisenberg uncertainty principle the momentum of neutrino is
$p \sim 1/R$, and consequently, the  kinetic energy, 
$E_{kin} \approx p^2/2 m_\nu$, equals  
%%\footnote{The uncertainty principle $\Delta x\Delta p\sim1$ 
%%implies that the momentum $p=\sqrt{2E_{{\rm kin}}m_{\nu}}$ 
%%then the kinetic energy of this neutrino $E_{{\rm kin}}$
%%can be estimated should be comparable to
%%$\Delta x^{-1}=R^{-1}$. Hence $E_{{\rm kin}}$ is determined by Eq.~\eqref{eq:-10}.}
\begin{equation}
E_{{\rm kin}}\sim\frac{1}{2m_{\nu}}\frac{1}{R^{2}}.
\label{eq:-10}
\end{equation}
The condition of neutrino trapping in the potential $V$ reads 
\begin{equation}
E_{{\rm kin}}\lesssim-V,\ \, \, (r < R).  
\label{eq:-11}
\end{equation}
Plugging 
$E_{{\rm kin}}$ from Eq.~\eqref{eq:-10} and $V$ from 
Eq.~\eqref{eq:V_NR} with  $N = 1$ into this equation we obtain 
\begin{equation}
\frac{1}{2m_{\nu}}\lesssim\frac{y^{2}R}{4\pi}e^{-Rm_{\phi}}.
\label{eq:-13}
\end{equation}
%%Depending on the mass $m_{\phi}$ and the coupling $y$, Eq.~\eqref{eq:-13}
%%may or may not have a solution. 
Maximal value of the r.h.s. of this equation equals  
$y^2/(4 \pi e m_\phi)$, which  is achieved at 
$R=m_{\phi}^{-1}$. Then, according to \eqref{eq:-13}, 
the sufficient condition for existence of bound state can be written as 
%%Hence a function of $R$ has a maximum at 
%%to satisfy Eq.~\eqref{eq:-13}, at least the maximum of the r.h.s
%%should be larger than the l.h.s. Substituting $R=m_{\phi}^{-1}$ in
%%Eq.~\eqref{eq:-13}, we obtain
\begin{equation}
\lambda \equiv  \frac{y^{2}}{8\pi}\frac{m_{\nu}}{m_{\phi}} 
\gtrsim  0.7, 
%%{\cal O}(1).
\label{eq:-14}
\end{equation}
where   $\lambda$ can be interpreted as  strength of interaction. 
Quantitative study below gives  $\lambda > 0.84$. 
The condition can be rewritten as the upper bound on $m_\phi$:  
$$
%\begin{equation}
m_\phi \lesssim \frac{y^2}{4\pi} m_\nu = 
4 \cdot 10^{-17} \left(\frac{y}{10^{-7}}\right)^2 {\rm eV}.
%\label{eq:-nn}
%\end{equation} 
$$

\subsection{Quantitative study 
\label{subsec:Quantitative-solve}}
%%%%%%%%%%%%%%%%%%%%%%%%%%%%%%%%%%%%%%%%%%%%%%%%%%%%%%%%%%%%%%%%%%%

%%%%%%%%%%%%%%%%%%%%%%%%%%%%%%%%%%%%%%%%ffff1  %%%%%%%%%%%%%%%%%%%%%%%
\begin{figure}
\centering
\includegraphics[width=0.88\textwidth]{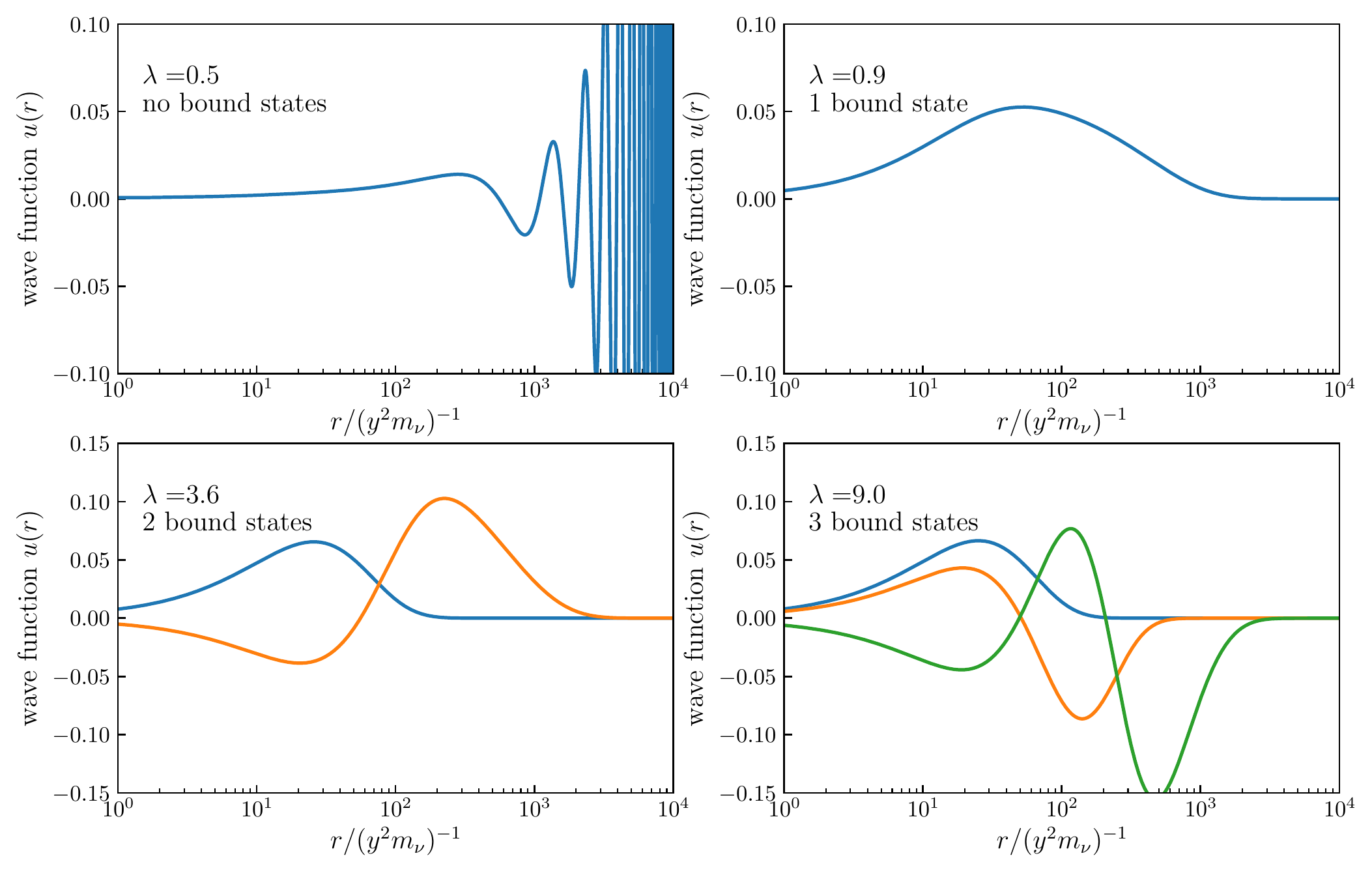}
\caption{\label{fig:Solutions} The radial dependences of 
the wave functions of neutrino bound states  
for different values of the interaction strength $\lambda$ defined in Eq.~\eqref{eq:-14}.
In all the panels we take the orbital momentum $l=0$.  
For $\lambda=0.5$, since no bound states can be formed, we plot the
wave function with the lowest energy.}
\end{figure}
%%%%%%%%%%%%%%%%%%%%%%%%%%%%%%%%%%%%%%%%%%%%%%%%%%%%%%%%%%%%%%%%%%%%%%%%%%%

Quantitative treatment of  neutrino bound states 
requires solution of the two-body Schr\"odinger equation. 
Let $\mathbf{r}_{1}$ and $\mathbf{r}_{2}$ be the coordinates
of two neutrinos. 
The potential depends  on the relative
distance between neutrinos $\mathbf{r} \equiv \mathbf{r}_{1}-\mathbf{r}_{2}$ 
only. Consequently,  the wave function of bound state depends on $\mathbf{r}$,  
$\nu = \nu(\mathbf{r})$,  and it  obeys the Schr\"odinger equation 
\begin{equation}
\left[-\frac{1}{m_\nu} \nabla^2 + V(r)\right]\nu(\mathbf{r})  =
E\nu(\mathbf{r}), 
\label{eq:-3}\\
%%V(r) & =-\frac{y^{2}}{4\pi}\frac{1}{r}e^{-rm_{\phi}}. 
%%\label{eq:V_NRx}
\end{equation}
with $V(r) = V(r, N = 1)$ given in  Eq.~\eqref{eq:V_NR},  
$r = |\mathbf{r}|$ and   $E$ being  the energy eigenvalue. 
The numeric coefficient of the kinetic term takes into account 
the reduced mass of neutrino 
(see Appendix~\ref{sec:two-body-sch}). 
Since the potential $V(r)$ is independent of the direction, 
the wave function can be factorized as 
\begin{equation}
\nu = \frac{u(r)}{r}Y_{l}^{m}(\theta,\ \varphi).
\label{eq:psi_sol}
\end{equation}
Here $u(r)$ is a radial component and $Y_{l}^{m}(\theta,\ \varphi)$
is a spherical harmonic function. Plugging Eq.~\eqref{eq:psi_sol}
into Eq.~\eqref{eq:-3}, we obtain the equation for  $u(r)$: 
$$
%\begin{equation}
-\frac{1}{m_{\nu}}\frac{d^{2}u(r)}{dr^{2}}+\left[V(r) + 
\frac{1}{m_{\nu}}\frac{l(l+1)}{r^{2}}\right]u(r)=Eu(r).
%\label{eq:-4}
%\end{equation}
$$
For zero  angular momentum, $l=0$, the equation  simplifies  further: 
\begin{equation}
-u''(r)+m_{\nu}V(r)u(r)=m_{\nu}Eu(r). 
\label{eq:-5}
\end{equation} 
We solve this equation numerically   
(see details in Appendix~\ref{sec:two-body-sch}).  
$u(r)$  for different values of $\lambda$ 
are shown in Fig.~\ref{fig:Solutions}.  
According to this figure
%% which is defined as
%%\begin{equation}
%%\lambda\equiv\frac{y^{2}}{8\pi}\frac{m_{\nu}}{m_{\phi}}.\label{eq:lambda}
%%\end{equation}
there is no bound state  for $\lambda=0.5$:  the solution
with the lowest energy level does not converge 
to zero  when $r\rightarrow\infty$.
For other specific values, 
$\lambda= 0.9$, $3.6$, and $9.0$, the figure shows 
one, two, and three bound states.  
With further increase of $\lambda$,  
the number of allowed energy levels of increases.  
%%for respectively.  Varying $\lambda$, we obtain  that 
Varying $\lambda$  we find that for
%\footnote{\color{blue}
%	In \cite{Wise:2014jva} the bound $\lambda \geq 0.84$
%	was obtained for the Yukawa interaction to form a bound state.}
\begin{itemize}
\item $\lambda<0.84$, no bound states can be formed;
\item $0.84<\lambda<3.2$, there is one bound state (the 1s state);
\item $3.2< \lambda < 7.2$, two bound states with $l=0$ (the 2s
state) can be formed;
\item $ \lambda > 7.2$, three or more bound states exist, including
states with $l\neq 0$. 

\end{itemize}

%%%%%%%%%%%%%%%%%%%%%%%%%%%%%%%%%%%%%%%%%%%%%%%%%%%%%%%%%%%%%%%%%%%%%%%%%%%%%
\subsection{Coulomb limit }
%%%%%%%%%%%%%%%%%%%%%%%%%%%%%%%%%%%%%%%%%%%%%%%%%%%%%%%%%%%%%%%%%%%%%%%%%%%%%%%

\noindent The limit of   $m_{\phi}\rightarrow0$ ($\lambda\rightarrow\infty$)
 corresponds to a Coulomb-like potential,  and  in this case,
the Schr\"odinger equation~\eqref{eq:-5} can be solved analytically. 
%%and as long as $y$ is finite, 
There is an infinite number of levels and 
the wave function of ground state equals 
\begin{equation}
u(r)  = 2 R_{0}^{3/2}r\exp\left(-\frac{r}{R_{0}}\right),
\label{eq:-143}
\end{equation}
where 
\begin{equation}
R_{0} \equiv\frac{8\pi}{m_{\nu}y^{2}}\thinspace
\label{eq:-145}
\end{equation}
is the radius of the bound state, and the corresponding eigenvalue equals 
\begin{equation}
E =-\frac{y^{4}m_{\nu}}{64\pi^{2}}\thinspace. 
\label{eq:-144}
\end{equation}
This $E$  can  be
interpreted as the binding energy of the state. 
%% and $R_{0}$ in  \eqref{eq:-145}  represents the radius of the bound state. 
For non-zero $m_{\phi}$ such that  $R_{0}\ll m_{\phi}^{-1}$
(equivalent to $\lambda\gg1$), the range of the force is much
larger than the radius of the bound state in the Coulomb limit. In
this case,  Eq.~\eqref{eq:-145} and \eqref{eq:-144} can be used as a good approximation.
The correction to the Coulomb limit result due to  non-zero
$m_{\phi}$ can be computed using the variation principle (see below).

%%%%%%%%%%%%%%%%%%%%%%%%%%%%%%%%%%%%%%%%%%%%%%%%%%%%%%%%%%%%%%%%%%%%%%%%%%%%%%%%%%
\subsection{Approximate solutions for non-zero mass of scalar 
from the variation principle}
%%%%%%%%%%%%%%%%%%%%%%%%%%%%%%%%%%%%%%%%%%%%%%%%%%%%%%%%%%%%%%%%%%%%%

According to the variation principle 
the expectation of the Hamiltonian
$$
%\begin{equation}
\langle H\rangle\equiv\frac{\int u^{*}(r)Hu(r)
\thinspace dr}{\int u^{*}(r)u(r)\thinspace dr}
%\label{eq:-146}
%\end{equation}
$$
reaches the minimal value when $u$ is the wave function of the ground state, 
while  $\langle H\rangle$ itself gives the 
corresponding eigenvalue. The Hamiltonian 
for the radial wave function and $l=0$  reads 
$$
%\begin{equation}
H=-\frac{1}{m_{\nu}}\frac{d^{2}}{dr^{2}}+V(r)\thinspace.
%\label{eq:-147}
%\end{equation}
$$
Assuming solution in the form  Eq.~\eqref{eq:-143}:  
$u(r)=2a^{3/2}re^{-r/a}$ with $a$ being a free parameter, we
find %Eq.~\eqref{eq:-146} 
\begin{equation}
\langle H\rangle = \frac{1}{a^2 m_\nu}\left[1 
- \frac{ay^2 m_\nu}{\pi(am_{\phi}+2)^2}\right]. 
\label{eq:-149}
\end{equation}
Minimization of $\langle H\rangle$ with respect to $a$,  
that is  $d\langle H \rangle / da = 0$,  gives 
\begin{equation}
% (16\pi\lambda + x)^3 -  32\pi\lambda^2 x (16\pi \lambda + 3x) = 0,
(16\pi\lambda + a m_{\nu}y^{2})^3 -  32\pi\lambda^2 a m_{\nu}y^{2} (16\pi \lambda + 3 am_{\nu}y^{2} ) = 0,
\label{eq:-150}
\end{equation}
% [[write more explicitly, do not introduce $x$]]
% where we introduced $x\equiv am_{\nu}y^{2}$.  
which is a cubic equation with respect to $a$.
The cubic equation has
one negative solution which can be ignored, and two  others which 
% can be real [[posivive??]], if the discriminant is positive: 
can be either both complex (in this case the two solutions are conjugate to each other) 
or both real and positive (in this case the two solutions are identical). 
Only the later case leads to a physical solution and requires that the discriminant is positive: 
$$
%\begin{equation}
\lambda^{2}+\frac{20}{9}\lambda-3\geq0\thinspace.
%\label{eq:-151}
%\end{equation}
$$
From this inequality we obtain 
$$
%\begin{equation}
\lambda\geq\frac{1}{9}\left(7\sqrt{7}-10\right)\approx0.95\thinspace, 
%\label{eq:-152}
%\end{equation}
$$
which  is about 13\% larger than  the exact value $0.84$ obtained in Sec.~\ref{subsec:Quantitative-solve}.  
In \cite{Wise:2014jva} the bound $\lambda \geq 0.84$ was also obtained. 

Using the variation
principle in this way may not give an accurate wave function, but
usually leads to a very accurate result for  
eigenvalues~\cite{griffiths2016introduction}.
Note that here we do not require that $\lambda^{-1}\propto m_{\phi}$
is perturbatively small. For very small $\lambda^{-1}$ the result
is expected to be more accurate. 
The series expansion of the positive physical solution 
of Eq.~\eqref{eq:-150} in $\lambda^{-1}$ gives  
$$
%\begin{equation}
x = 8\pi\left[1+ \frac{3}{4\lambda^2} + 
{\cal O} \left(\frac{1}{\lambda^3} \right)\right].
%\label{eq:-153}
%\end{equation}
$$
Therefore, the radius of the system $R = a = x/y^2 m_\nu$ equals 
$$
%\begin{equation} 
R  \approx\frac{8\pi}{m_{\nu}y^{2}}\left[1+\frac{3}{4\lambda^{2}}\right].
%\label{eq:-145-1}
%\end{equation}
$$
The eigenvalue is corrected according to 
\eqref{eq:-149} as 
\begin{equation}
E =   \langle H \rangle  \approx   - 
\frac{y^{4}m_{\nu}}{64\pi^{2}}\left[1-\frac{2}{\lambda}+\frac{3}{2\lambda^{2}}\right]. 
\label{eq:-144-1}
\end{equation}
%{\color{red}
	The lowest order expressions for $R$ and binding energy~\eqref{eq:-144-1} coincide with those in \cite{Wise:2014jva}.
%}

Numerically, the radius equals 
\begin{equation}
R  =   5 \cdot 10^{11} {\rm cm}
\left(\frac{0.1 {\rm eV}}{m_\nu}\right) \left(\frac{10^{-7}}{y}\right)^2.   
\label{eq:rnum}
\end{equation}
This is much larger than the distance between 
relic neutrinos (without clustering) $\sim 0.16$ cm,  
and therefore existence of bound states of such size 
has no practical sense. 
 
For two-body bound states, it is impossible to enter the relativistic
regime. Indeed, according to 
Eq.~\eqref{eq:-144} or  Eq.~\eqref{eq:-144-1} 
the kinetic energy  being of the order of  binding energy  
equals  $y^4 m_\nu/64\pi^2$. Therefore the 
relativistic case, $E_{\rm kin}/m_\nu \geq 1$ implies   
$y\gtrsim(64\pi^{2})^{1/4}$ which would not only violate the perturbativity 
bound but also be excluded by various experimental 
limits\,\textemdash\,see Appendix~\ref{sec:Known-constraints}.

%{\color{blue}
%	Formation of the two-body bound states
%	is an open question: due to the smallness 
%	of coupling the process of $\phi$ emission, 
%	$\nu + \nu  \rightarrow (\nu \nu)_{\rm bound} + \phi$, 
%	is negligible (see Sec.~\ref{sub:energy-loss}).
%}

The results of this section may have applications 
for sterile neutrinos with mass $m_\nu \geq 10$ keV. These neutrinos 
could compose dark matter of the Universe. 
For $m_\nu \geq 10$ keV and $y = 10^{-3}$ we find from (\ref{eq:rnum})
$R = 0.05$ cm.

%%%%%%%%%%%%%%%%%%%%%%%%%%%%%%%%%%%%%%%%%%%%%%%%%%%%%%%%%%%%%%%%%%%%%%%%%%%
\section{$N$-neutrino bound systems \label{sec:N-body}}
%%%%%%%%%%%%%%%%%%%%%%%%%%%%%%%%%%%%%%%%%%%%%%%%%%%%%%%%%%%%%%%

Let us consider
$N$  neutrinos trapped in the Yukawa  potential  
with $N$ being large enough,  so that neutrinos form a statistical 
system which is described by the Fermi-Dirac distribution 
function
\begin{equation}
f=\frac{1}{\exp\frac{E-\mu}{T}+1}. 
\label{eq:-16}
\end{equation}
Here $E$ is the total neutrino energy,  $\mu$ is the chemical potential 
and $T$ is the temperature. 
Recall that for this distribution the  number density, $n$, 
the energy density, $\rho$, and pressure
${\cal P}$ of the system equal respectively 
$$
%\begin{equation}
n=\int f\frac{d^{3}\mathbf{p}}{(2\pi)^{3}}\thinspace, \,  \, \, \, 
\rho = \int Ef\thinspace\frac{d^{3}\mathbf{p}}{(2\pi)^{3}}
\thinspace, \, \, \, \,  
{\cal P}=\frac{1}{3}
\int\frac{|\mathbf{p}|^{2}}{E}f
\thinspace\frac{d^{3}\mathbf{p}}{(2\pi)^{3}}\thinspace.
%\label{eq:-2}
%\end{equation}
$$
The ground state of this system ($T\rightarrow0$) is the degenerate 
Fermi gas with 
distribution $f(p) = 1$,  if $E< \mu$,  and  $f(p) = 0$ if $E > \mu$ 
according to  Eq.~\eqref{eq:-16}. 
(The case of finite $T$ will be discussed
in Sec.~\ref{sec:Formation-and-lifetime}). Consequently, 
in the degenerate state  
\begin{equation}
n=\int_{0}^{p_{F}}\frac{4\pi p^{2}dp}{(2\pi)^{3}}=\frac{p_{F}^{3}}{6\pi^{2}}\thinspace,
\label{eq:-12}
\end{equation}
and  the total number of neutrinos $N$ for uniform distribution 
in the sphere of radius $R$ equals 
%%according to Eq.~\eqref{eq:-12}  
\begin{equation}
N=\frac{2 p_{F}^{3}}{9\pi}  R^{3}\thinspace.
\label{eq:-72}
\end{equation}
Eq.~\eqref{eq:-12} applies to both relativistic and non-relativistic
cases. For $\rho$ and ${\cal P}$, the non-relativistic results are
\begin{equation}
{\cal P}_{\rm deg} = \frac{1}{6\pi^2 }  \int_{0}^{p_{F}}\frac{p^2}{E} p^2dp = 
\frac{p_{F}^{5}}{30\pi^{2}m_{\nu}}\thinspace,\, \, \,  \ \rho=\frac{3}{2}{\cal P},
\label{eq:-71}
\end{equation}
while the relativistic results can be obtained by computing the integrals
numerically.

%%%%%%%%%%%%%%%%%%%%%%%%%%%%%%%%%%%%%%%%%%%%%%%%%%%%%%%%%%%%%%%%%%%%
\subsection{Qualitative estimation}
\label{sub:N-qualitative}
%%%%%%%%%%%%%%%%%%%%%%%%%%%%%%%%%%%%%%%%%%%%%%%%%%%%%%%%%%%%%%%%

Let us consider $N$ neutrinos uniformly  distributed in
a sphere of radius $R$. The Yukawa attraction produces  a pressure,
${\cal P}_{{\rm Yuk}}$,  which for $m_{\phi}=0$ can be estimated 
as 
\begin{equation}
{\cal P}_{{\rm Yuk}}\simeq 
\frac{1}{{\cal V}}\left[-\frac{1}{2}\frac{y^{2}N^{2}}{4\pi R}\right]\thinspace,
\label{eq:-30-1}
\end{equation}
where ${\cal V}=\frac{4}{3}\pi R^{3}$ is the volume and the quantity in the 
brackets is the total potential energy. 

Static configuration corresponds to the equilibrium  between this Yukawa  
pressure and the pressure of degenerate fermion gas 
(Pauli repulsion):   
$$
%\begin{equation}
{\cal P}_{{\rm deg}}\simeq {\cal P}_{{\rm Yuk}}\thinspace. 
%\label{eq:-29}
%\end{equation}
$$
Using  Eq.~\eqref{eq:-71} and 
Eq.~\eqref{eq:-30-1} this equality can be rewritten as  
\begin{equation}
\frac{p_{F}^{5}}{30\pi^{2}m_{\nu}} \simeq \frac{3y^{2}N^{2}}{2\left(4\pi\right)^{2}R^{4}}\thinspace. 
\label{eq:-32}
\end{equation}
Inserting Eq.~\eqref{eq:-72} in Eq.~\eqref{eq:-32} we find 
\begin{equation}
R \simeq \frac{8}{y\sqrt{m_{\nu}p_{F}}}\thinspace.
\label{eq:-73}
\end{equation}
Alternatively, we can express $p_{F}$ in terms of $N$ and obtain
\begin{equation}
R \simeq \frac{30}{m_{\nu}y^{2}}\left(\frac{1}{N}\right)^{1/3}\thinspace.
\label{eq:-34}
\end{equation}
Plugging $R$ from Eq.~\eqref{eq:-34} back into Eq.~\eqref{eq:-72}, we obtain
a relation between $N$ and $p_{F}$:
\begin{equation}
N \simeq \frac{40}{y^3} \left(\frac{p_F}{m_\nu} \right)^{3/2}
\thinspace.
\label{eq:-154}
\end{equation}

Notice that for $N \rightarrow 2$, Eq.~\eqref{eq:-34}
recovers  approximately the result in Eq.~\eqref{eq:-145} 
for the two-body system. A noteworthy
feature  of Eq.~\eqref{eq:-34} is that the radius decreases with $N$: 
$R\propto N^{-1/3}$. 
This implies that increasing $N$  leads to a more compact cluster. 
This dependence is valid only in the non-relativistic regime. When $N$ is too large, the Fermi momentum  $p_{F}$ which 
increases with $N$ according to Eq.~\eqref{eq:-154}  will exceed
$m_\nu$ and  the system  enters the
relativistic regime. 
As we will show in Sec.~\ref{sec:Relativistic}, in the relativistic regime, 
the attractive force becomes suppressed, causing $R$ to increases with $N$.

%%%%%%%%%%%%%%%%%%%%%%%%%%%%%%%%%%%%%%%%%%%%%%%%%%%%%%%%%%%%%%%%%%%
\subsection{Quantitative study
\label{subsec:solveLE}}
%%%%%%%%%%%%%%%%%%%%%%%%%%%%%%%%%%%%%%%%%%%%%%%%%%%%%%%%%%%%%%%

In an object of finite size,  the density and pressure 
depend on distance from the center of object.
The dependence can be obtained using the differential (local) form of the 
equality of forces, which is given by the equation of hydrostatic equilibrium.  
For degenerate neutrino gas it is reduced to the 
Lane-Emden (L-E) equation~\cite{chandrasekhar1957introduction}. 
%%The
%%Lane-Emden equation can be derived either from hydrostatic equilibrium,
The L-E equation can be also derived from equations of motion of the fields 
(see Appendix~\ref{sec:Lane-Emden}).

The equation of hydrostatic equilibrium expresses the equality of
the repulsive force due to the degenerate pressure, $F_{{\rm deg}}$,
and the Yukawa attractive force, $F_{\text{Yuk}}$, acting on a
unit of volume at distance $r$ from the center:
\begin{equation}
F_{{\rm deg}}(r) \equiv \frac{d{\cal P}_{{\rm deg}}(r)}{dr} = F_{\text{Yuk}}(r)\thinspace.
\label{eq:hsequil}
\end{equation}
For non-relativistic neutrinos, ${\cal P}_{{\rm deg}}$
is given in Eq.~\eqref{eq:-71}, and using Eq.~\eqref{eq:-12}, ${\cal P}_{{\rm deg}}$
can be rewritten as 
\begin{equation}
{\cal P}_{{\rm deg}} = Kn^{5/3},\, \, \,  \ K\equiv\frac{(6\pi^{2})^{5/3}}{30\pi^{2}m_{\nu}} 
= \frac{(6\pi^{2})^{2/3}}{5m_{\nu}}\thinspace.\label{eq:-109}
\end{equation}
The Yukawa force equals 
\begin{equation}
F_{{\rm Yuk}}(r)=-\frac{y^{2}}{4\pi r^{2}}n(r)N(r)\thinspace, 
\label{eq:-110}
\end{equation}
where 
$$
%\begin{equation}
N(r)=\int_{0}^{r}dr'4\pi r'{}^{2}n(r')
%\label{eq:-111}
%\end{equation}
$$
is the number of neutrinos inside a sphere of radius $r$. 
Inserting  Eqs.~\eqref{eq:-109} 
and \eqref{eq:-110} into Eq.~\eqref{eq:hsequil}
gives the equation for $n(r)$:
$$
%\begin{equation}
\frac{5}{3}Kn^{2/3}\frac{dn}{dr} =
\frac{y^{2}}{r^{2}}n\int_{0}^{r}dr'r'^{2}n(r')\thinspace.
%\label{eq:-112}
%\end{equation}
$$
Dividing this equation by $n/r^{2}$ and then differentiating by $r$, 
we obtain 
$$
%\begin{equation}
\frac{5}{2}K\frac{d}{dr}\left[r^{2}\frac{dn^{2/3}}{dr}\right]=-y^{2}r^{2}n\thinspace, 
%\label{eq:-113}
%\end{equation}
$$
or 
\begin{equation}
\frac{1}{r^{2}}\frac{d}{dr}\left[r^{2}\frac{dn^{2/3}}{dr}\right]=-y^{2}\kappa n\thinspace,
\label{eq:-114}
\end{equation}
where
$$
%\begin{equation}
\kappa\equiv\frac{2m_{\nu}}{(6\pi^{2})^{2/3}}\thinspace.
%\label{eq:-115}
%\end{equation}
$$
Eq.~\eqref{eq:-114} is nothing but the Lane-Emden equation~\cite{chandrasekhar1957introduction} 
for specific 
values of parameters. In particular, $\gamma$, defined as  
the ratio of powers of $n$ in the r.h.s. and the l.h.s. 
of Eq.~\eqref{eq:-114}, equals $\gamma = 1/(2/3) = 3/2$.  
For this value 
the $N$-body system has a finite size $R$ defined by 
$$
%\begin{equation}
n(R)=0.
%\label{eq:-75}
%\end{equation}
$$

\begin{figure}
\centering
\includegraphics[width=0.6\textwidth]{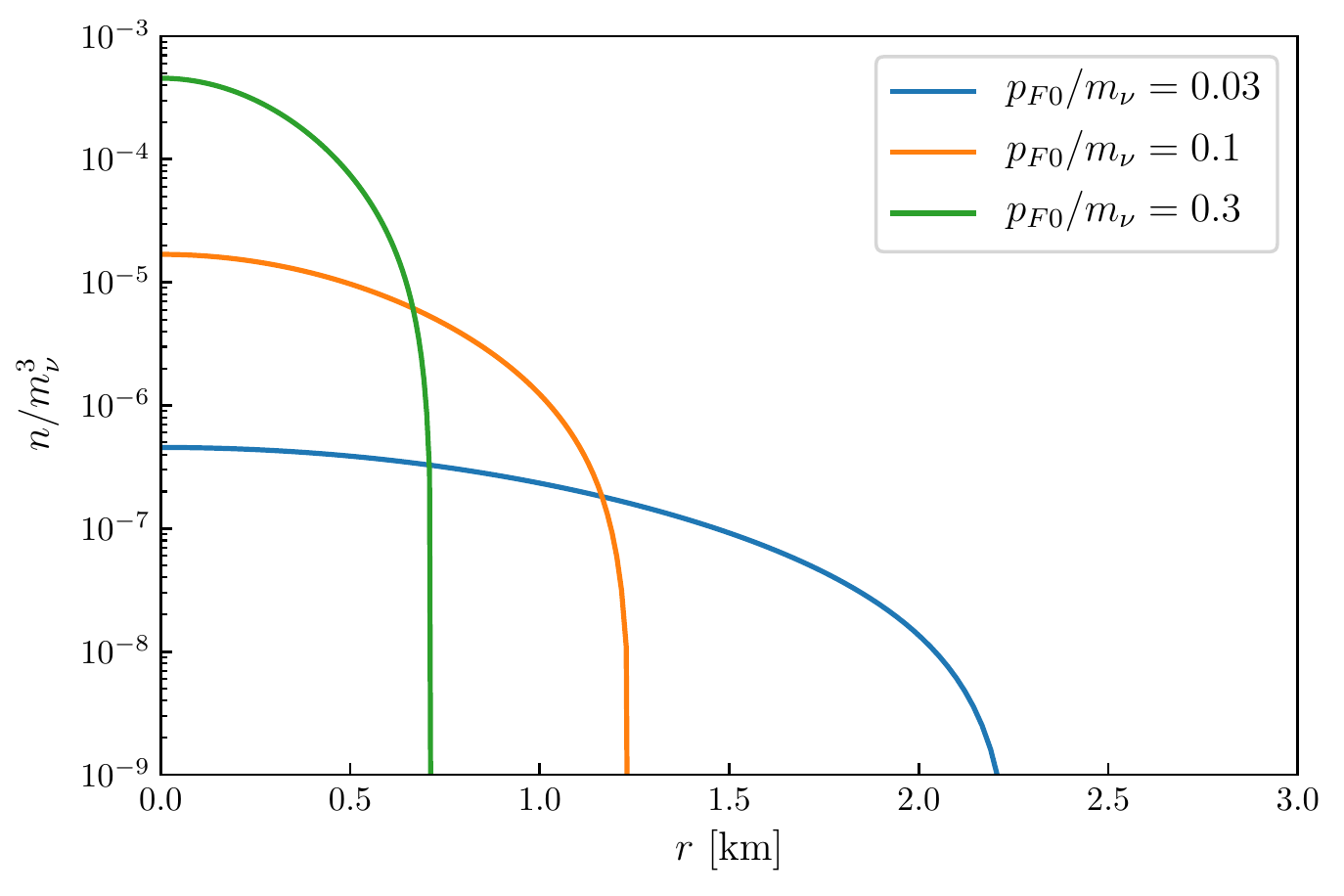}
\caption{\label{fig:distribution} Dependence  of neutrino number density
$n(r)$ on distance from center for different values of central density  
$p_{F0}/m_\nu$. We take $y=10^{-7}$ and  $m_{\nu}=0.1$ eV. 
}
\end{figure}

We solve the L-E equation~\eqref{eq:-114} numerically with the boundary condition 
$$
%\begin{equation}
n(r = 0) = n_0,~~ {\rm or~~ equivalently}, ~~ p_F(r = 0) = p_{F0}.
%\end{equation}  
$$
Fig.~\ref{fig:distribution} shows solutions of   
$n(r)$ for different values of $p_{F0}/m_\nu$.   
As $r$ increases, the number densities drop down to zero at  
\begin{equation}
R\approx\frac{19.9}{y\sqrt{m_{\nu}p_{F0}}}.  
\label{eq:-77}
\end{equation} 
Integrating  $n(r)$ with $r$ ranging from $0$ to $R$
gives the  total number of neutrinos:
\begin{equation}
N \approx\frac{93.6}{y^{3}}
\left(\frac{p_{F0}}{m_{\nu}}\right)^{3/2}\thinspace.
\label{eq:20210526-1}
\end{equation}
Combining Eqs.~\eqref{eq:-77} and \eqref{eq:20210526-1}, we 
obtain expression for the radius
\begin{equation}
R\approx\frac{90.4}{m_{\nu}y^{2}}\left(\frac{1}{N}\right)^{1/3}\thinspace,
\label{eq:20210531}
\end{equation}
which shows that $R$ decreases with $N$, in agreement 
with the qualitative  estimations in %Sec.~ IV.A. 
Sec.~\ref{sub:N-qualitative}.
The results in Eqs.~\eqref{eq:-77}, \eqref{eq:20210526-1} and \eqref{eq:20210531}
have the same parametric dependence as  
Eqs.~\eqref{eq:-73}, \eqref{eq:-154} and \eqref{eq:-34} and differ by numerical factors 
2.5, 2.3 and 3 correspondingly. 

Notice that for the Fermi momentum $p_{F0} \sim m_\nu$ the 
Eq.~\eqref{eq:-77} gives $R = 19.9 /y m_\nu$, 
which practically coincides with
(\ref{eq:nradnew}) obtained from rescaling of
parameters of neutron star.
The coincidence is not accidental since parameters of neutron star
correspond to $p_{F} \sim m_n$.
Other characteristics of a $\nu$-cluster  for $p_{F0} \sim m_\nu$
are $M_\nu = 3 \cdot 10^{-11}$ g (for $y = 10^{-7}$) 
and $n_{0} = 2.6 \cdot 10^{8}$  cm$^{-3}$.
The central density is determined according to  Eq.~\eqref{eq:-12} as 
\begin{equation}
\frac{n_{0}}{m_\nu^3} = \frac{1}{6\pi^2} \left(\frac{p_{F0}}{m_\nu}\right)^3. 
\label{eq:n0mnu}
\end{equation}

%%%%%%%%%%%%%%%%%%%%%%%%ffff2 %%%%%%%%%%%%%%%%%%%%%%%%%%%%%%%%%%%%%%%%%%%%%%%%%%%
%%%%%%%%%%%%%%%%%%%%%%%%%%%%%%%%%%%%%%%%%%%%%%%%%%%%%%%%%%%%%%%%%%%%%%%%%

In Eq. (\ref{eq:-114}) both $y^{2}$ and $\kappa$ can be absorbed by the redefinition of $r$: 
$$
%\begin{equation}
r' = r y \sqrt{\kappa} \propto r y \sqrt{m_\nu}. 
%\label{eq:redef}
%\end{equation}
$$
In terms of $r'$, Eq.~\eqref{eq:-114} can be written as the universal
form 
$$
%\begin{equation}
\frac{1}{r'^{2}}\frac{d}{dr'}\left[r'^{2}\frac{dn^{2/3}}{dr'}\right] = 
- n\thinspace,
%\label{eq:univ}
%\end{equation}
$$
and its solution depends on the boundary condition $n(0)$ which is 
free parameter 
restricted from above by $p_{F0} <  m_\nu$. 
Thus, solution for different values of $y$ and $m_\nu$ can be obtained from rescaling of $r$: 
\begin{equation}
r = r_7 \left(\frac{10^{-7}}{y}\right) 
\left(\frac{0.1 {\rm eV}} {m_\nu}\right)^{1/2}, 
\label{eq:resc}
\end{equation}
where $r_7 \equiv r(y = 10^{-7})$.  The radius of a cluster  is rescaled as in (\ref{eq:resc}). 
For $y = 10^{-14}$ this equation  gives  $R \sim 10^{12}$ cm, 
{\rm i.e.} orders of magnitude smaller than the Earth orbit. 
For the $\nu$-cluster made of $\nu_2$ (in the case of mass hierarchy) 
the radius is 6 times larger. 
The mass of such a cluster equals $3 \cdot 10^{10}$ g. 
The central density, and consequently,  the mass of $\nu$-cluster is an independent 
parameter restricted by  the neutrino mass. 
This consideration matches results in Eqs. 
(\ref{eq:-77}) and  (\ref{eq:20210526-1}).

\subsection{Non-degenerate $\nu$-clusters }
%%%%%%%%%%%%%%%%%%%%%%%%%%%%%%%%%%%%%%%%%%%%%%%%%%%%%%%%%%%%%%%%%

As we will show in Sec.~\ref{sub:energy-loss} 
% sect. VI D,  
the energy loss (cooling) of $\nu$-clusters 
due to $\phi$ bremsstrahlung  is negligible. 
Therefore the final configuration of degenerate neutrino 
gas may not be achieved. 
In this connection we consider the non-degenerate neutrino 
system characterized by a non-zero temperature $T$. 
The number density equals 
$$
%\begin{equation}
n_{T}=  \frac{1}{2\pi^2} \int_{0}^{\infty} \frac{p^2dp}{e^{p/T}+1}
= \frac{I_3}{2 \pi^2} T^3,   
%\label{eq:20210526}
%\end{equation}
$$
where 
\begin{equation}
I_n \equiv \int_{0}^{\infty} \frac{dx x^{n - 1}}{e^x +1}, 
\label{eq:inn}
\end{equation}
and $I_3$ is related to the Riemann zeta function: $I_3 = 3\zeta (3)/2 = 1.803$. 

For simplicity we assume that  neutrinos are distributed uniformly in the
sphere of radius $R_T$. Then 
\begin{equation}
R_T = \left(\frac{3N}{4\pi n_{T}}\right)^{1/3} 
= \left(\frac{3\pi}{ 2I_3}\right)^{1/3} \frac{N^{1/3}}{T}.  
%%\approx 1.7 \frac{N^{1/3}}{\sqrt{\eta} m_\nu}.
%%\frac{7.7}{ym_{\nu}}\sqrt{\frac{p_{F0}}{\eta m_{\nu}}}\thinspace.
\label{eq:20210526-2}
\end{equation}
If neutrinos are non-relativistic, the total kinetic energy can be computed as 
\begin{equation}
E_{K} \approx \frac{4\pi}{3} R_{T}^{3} \frac{1}{2\pi^2} 
\int^{\infty}_0   \frac{p^4 dp}{2m_\nu} \frac{1}{e^{p/ T} + 1}
= \frac{I_5}{2I_3} \frac{T^2}{m_\nu} N,  
\label{eq:20210526-3}
\end{equation}
where $I_5  = 45 \zeta (5)/2 = 23.33 $ [see  \eqref{eq:inn}] and we used Eq.~\eqref{eq:20210526-2} for 
$R_T$.  

The potential energy of a cluster equals
\begin{equation}
E_{V}=- \frac{y^2}{4\pi}  \int_{0}^{R_T}\frac{(4\pi n_Tr^{3}/3)(4\pi n_T r^{2}dr)}{4\pi r}= 
-\frac{4}{15}\pi n_T^2 R_T^5 y^{2}
= - \frac{y^2 N^2}{8\pi R_T},  
\label{eq:20210526-4a}
\end{equation}
or eliminating $R_T$:
\begin{equation}
E_V = - \frac{3}{20 \pi} \left(\frac{2I_3}{3\pi}\right)^{1/3}  y^2 T N^{5/3}.
\label{eq:20210526-4}
\end{equation}
The ratio of the kinetic and  potential energies can be written as  
\begin{equation}
\xi_E  \equiv \frac{E_K}{|E_V|} = 
\frac{10 I_5}{3 (I_3/\pi)^{4/3}} \left(\frac{3}{2}\right)^{1/3}    
~\frac{T}{y^2 m_\nu N^{2/3}}.  
\label{eq:eratio}
\end{equation}
The ratio  $\xi_E$ in Eq.~(\ref{eq:eratio}) depends
on the temperature $T$.
Therefore, given a certain value of $\xi_E$, it allows to determine
$T$: 
$$
%\begin{equation}
T = \frac{1}{5}  
\left(\frac{3}{2}\right)^{2/3} 
\left(\frac{I_3}{\pi}\right)^{4/3} I_5^{-1} y^2 N^{2/3} \xi_E m_\nu. 
%\frac{E_K^{\rm in}}{E_V^{\rm in}} 
%\label{eq:eratio1}
%\end{equation}
$$

For  $T \rightarrow p_F/3 $ it matches Eq. (\ref{eq:20210526-1}),  numerically:
\begin{equation}
T = 0.081 m_\nu  \left(\frac{y}{10^{-7}} \right)^2 
\left(\frac{N}{6 \cdot 10^{22}}\right)^{2/3} \xi_E 
= 0.081 m_\nu  \left(\frac{y^3 N}{60} \right)^{2/3} \xi_E.  
%%\frac{3E_K^{\rm in}}{E_V^{\rm in}}. 
\label{eq:eratio2}
\end{equation}
%%
%%Using expression  \eqref{eq:eratio1} for $\eta_{\rm in}$ we find from 
%%\eqref{eq:20210526-2} 
Then, according to (\ref{eq:20210526-2}) the radius equals 
$$
%\begin{equation}
R_T = \frac{1}{m_\nu} 
\left(\frac{2}{3}\right)^{1/3} 
\frac{5 I_5}{(I_3/\pi)^{5/3}}~ 
\frac{1}{y^2 N^{1/3}\xi_E}, 
%\label{eq:20210526-3a}
%\end{equation}
$$
and numerically 
\begin{equation} 
R_T = 1.33~ {\rm km}
\left(\frac{10^{-7}}{y }\right)^{2}
\left(\frac{6 \cdot 10^{22} }{N }\right)^{1/3}
\left(\frac{1}{\xi_E }\right)
\left(\frac{0.1 {\rm eV}}{m_\nu}\right). 
\label{eq:rin1}
\end{equation}
These equations  match Eq. (\ref{eq:20210531}).

The stable configuration of a cluster corresponds to the hydrostatic 
equilibrium ${\cal P}_T = - {\cal P}_{\rm Yuk}$.
In the non-relativistic case, ${\cal P}_T = (2/3)\rho_T$
and therefore from Eq.~(\ref{eq:20210526-3}) we have
$E_K = 3/2 V {\cal P}_T$. So, ${\cal P}_T = (2/3) E_K /V$.
The Yukawa pressure equals ${\cal P}_{\rm Yuk} = E_V /V$.
From this we find that the  hydrostatic equilibrium
corresponds to $ 2/3 E_K = E_V$ or $\xi_E = 3/2$.

%%In general (for arbitrary $\xi$) we find
%%the radius of $\nu$-star (96).

%%Notice that $R$ decreases with increase of $\xi_E$
%%or kinetic energy, on the contrary to expectations.

The ratio of the degenerate $\nu$-cluster radius in Eq.~\eqref{eq:20210531}, denoted by 
$R_{\rm deg}$, to the radius $R_T$,  equals
\begin{equation}
 \frac{R_{\rm deg}}{R_T} = 0.346 \xi_E = 0.52,
\label{eq:ratio-rad}
\end{equation}
where the last equality is for the hydrostatic equilibrium.
The ratio does not depend on other parameters.
So, if the evolution proceeds via the formation of a cluster 
in hydrostatic equilibrium, achieving final configuration
require that the radius decreases by a factor of two.
Our consideration of the thermal case is oversimplified: 
applying the Hydrostatic equilibrium locally leads to a distribution
of neutrinos in a cluster with a larger radius.

%%2. In the energy balance the energy in the scalar
%%field  should be taken into account.

Therefore, the density of a non-degenerate cluster equals 
$$
%\begin{equation}
n = m_\nu^3 \frac{9 \cdot 10^{-3}}{\pi^6}
~\frac{I_3^5}{I_5^3}~ y^6 N^2\xi_E^3, 
%\label{eq:din-3}
%\end{equation}
$$
and numerically 
$$
%\begin{equation} 
n = 6.2 \cdot 10^6 ~ {\rm cm}^{-3}     
\left(\frac{y}{10^{-7}}\right)^{6}
\left(\frac{N}{6 \cdot 10^{22}}\right)^{2}
\xi_E^3      
\left(\frac{m_\nu}{0.1 {\rm eV}}\right)^{3}. 
%\label{eq:din1}
%\end{equation}
$$

%%For this $N$ the final configuration has central density $n_f (0) = 1.3 \cdot 10^7$  cm$^{-3}$, 
%%and $R_f = 4.2$ km. [[check consistency]]

%%%%%%%%%%%%%%%%%%%%%%%%%%%%%%%%%%%%%%%%%%%%%%%%%%%%%%%%%%%
\section{Relativistic regime  \label{sec:Relativistic}}
%%%%%%%%%%%%%%%%%%%%%%%%%%%%%%%%%%%%%%%%%%%%%%%%%%%%%%%%%%%%

\subsection{Equations for degenerate  neutrino cluster}
%%%%%%%%%%%%%%%%%%%%%%%%%%%%%%%%%%%%%%%%%%%%%%%%%%%%%%%%%%%%%%%%%%%%%%%%%%

For a sufficiently large $N$,
neutrinos become relativistic ($p_{F}\gg \tilde{m}_{\nu}$)
in the center of cluster, while near the surface
they are almost at rest $(p_{F}=0)$.
For  relativistic neutrinos the Yukawa potential 
is suppressed by the factor 
$ \tilde{m}_{\nu}/E_{\mathbf{p}}$ in Eq.~\eqref{eq:-61}. Due to this suppression, 
the pressure of degenerate gas 
is able to equilibrate  
the Yukawa attraction for arbitrarily large $N$, in contrast to the gravity case. 

The system is described by    Eq.~\eqref{eq:-61}
and relativistic generalization of equilibrium condition 
between the Yukawa attraction and the degenerate
neutrino gas repulsion,  Eq.~(\ref{eq:hsequil}). To avoid potential ambiguities
in the definition of forces in the relativistic regime, 
instead of  $F_{{\rm Yuk}}$ and $F_{{\rm deg}}$ 
we will use the derivatives $d\phi/dr$  and $dp_{F}/dr$ which 
represent these two forces and are well-defined quantities
in the relativistic case. The equilibrium condition  can be derived   
from the equations of motion for the fields
(see  Appendix~\ref{sec:chemical-equilibrium}) giving 
\begin{equation}
y(m_{\nu}+y\phi)\frac{d\phi}{dr}=-p_{F}\frac{dp_{F}}{dr}, 
\label{eq:-164}
\end{equation}
($r < R$) which is the relativistic analogy  of equality  
$F_{{\rm Yuk}} = F_{{\rm deg}}$. 
In turn, Eq.~\eqref{eq:-61} for the scalar field 
can be rewritten as 
\begin{equation}
\left[\nabla^{2}-m_{\phi}^{2}\right]y\phi=y^{2}\tilde{n},
\label{eq:-61-1}
\end{equation}
where
\begin{equation}
\tilde{n}=  \frac{\tilde{m}_{\nu}}{2\pi^2}  \int_{0}^{p_{F}}
\frac{p^2dp}{\sqrt{\tilde{m}_{\nu}^2+ p^{2}}} = 
\frac{\tilde{m}_{\nu}^{3}}{4\pi^{2}}\left[\frac{p_{F}}{\tilde{m}_{\nu}}
\sqrt{1+p_{F}^{2}/\tilde{m}_{\nu}^{2}}-\tanh^{-1}
\left(\frac{p_{F}}{\sqrt{p_{F}^{2}+\tilde{m}_{\nu}^{2}}}\right)\right] 
\label{eq:-165}
\end{equation}
is the effective neutrino density and 
% $\tilde{m}_{\nu} = m_\nu + y\phi$ is the   
% the effective neutrino mass in the $\phi$ field  introduced in
$\tilde{m}_{\nu}$ is  introduced in
Eq.~\eqref{eq:-59}. 
%%
%%\begin{equation}
%%\tilde{m}=m_{\nu}+y\phi.\label{eq:-166}
%%\end{equation}
%%is the effective neutrino mass
%%in the $\phi$-field:
Let us underline that since the effective neutrino mass is 
given by $\tilde{m}_{\nu}$, the relativistic regime is determined by the condition 
$p_{F}/ \tilde{m}_{\nu} \gg 1$ which can differ substantially from $p_{F}/ m_\nu \gg 1$.

Using Eq.~\eqref{eq:-165}, we can rewrite the system of  Eqs.~\eqref{eq:-164} and \eqref{eq:-61-1}
in terms of $\tilde{m}_{\nu}$ and $p_{F}$:
\begin{align}
% \left[\nabla^{2}-m_{\phi}^{2}\right]\tilde{m}_{\nu} & = 
\left[\nabla^{2}-m_{\phi}^{2}\right](\tilde{m}_{\nu}-m_{\nu}) & = 
y^{2}\frac{\tilde{m}_{\nu}^{3}}{4\pi^{2}}
\left[\frac{p_{F}}{\tilde{m}_{\nu}}\sqrt{1+p_{F}^{2}/\tilde{m}_{\nu}^{2}} - 
\tanh^{-1}\left(\frac{p_{F}}{\sqrt{p_{F}^{2}+\tilde{m}_{\nu}^{2}}}\right)\right],
\label{eq:-167}\\
\tilde{m}_{\nu}\frac{d\tilde{m}_{\nu}}{dr} &  = -p_{F}\frac{dp_{F}}{dr} ~~~ (r < R).
\label{eq:-168}
\end{align}
% where in the first equation we have neglected the term $m_\nu m_\phi^2$. 
% Otherwise it coincides with the equation used in 
% \cite{Stephenson:1996qj}. 
Eqs.~\eqref{eq:-167} and \eqref{eq:-168} is a system 
of differential equations for $p_{F}$ and $\tilde{m}_{\nu}$ 
with the boundary conditions  
in the center:
\begin{equation}
\tilde{m}_{\nu}(r=0) = \tilde{m}_{0}, 
\,  \,  \, \,  \,   p_{F}(r=0) = p_{F0}.
\label{eq:-170}
\end{equation}
Eq.~\eqref{eq:-168} can be integrated 
from $r = 0$ to a given $r$: 
$$
%\begin{equation}
\tilde{m}_{\nu}^2 =  - p_{F}^2  + \tilde{m}_0^2 + p_{F0}^2. 
%\label{eq:pm-conn}
%\end{equation}
$$
Inserting $\tilde{m}_{\nu}$ from this equation into 
Eq.~\eqref{eq:-167} we obtain a single equation for $p_{F}$, which can be solved numerically\footnote{
 See Appendix~\ref{sec:n-sol}. Our numerical code is available at \url{https://github.com/xunjiexu/neutrino-cluster}. 
%     To improve the stability 
% of the numerical method when $y$ is very
% small, we absorb $y$ into $\nabla^{2}$ and $m_{\phi}^{2}$ by redefining
% $\nabla'=\nabla/y$ and $m'_{\phi}=m_{\phi}/y$.
} and using   
$n(r)= p_{F}^{3}(r)/ 6\pi^2$ (which still holds for relativistic
neutrinos) to compute the number density.

In the differential equations, the evolution of 
$\tilde{m}_{\nu}$ represents the evolution of the scalar field while $p_F$  determines the neutrino density. 
Therefore the boundary condition for  $\tilde{m}_{\nu}$ is 
essentially the condition for the
field $\phi$. 
Note  that $p_{F0}/\tilde{m}_0$ is the measure of relativistic 
character of the system. 
Hence $p_{F0}/\tilde{m}_0$ is used as the main input condition which determines the rest, 
while $\tilde{m}_{0}$ is determined by a self-consistent condition below.

As in the case of the  non-relativistic Lane-Emden equation, 
at certain  $r=R$, the Fermi momentum
$p_{F}$ (and hence the density) drops down  to zero: 
$$
%\begin{equation}
p_{F}(r = R) = 0,\,  \, 
%%\,  p_{F}(r<R)>0.
%\label{eq:-171}
%\end{equation}
$$
and this defines the radius of a cluster.  
Since the field $\phi$ extends beyond a star radius, $r > R$,   
the effective mass $\tilde{m}_{\nu}$ differs from $m_\nu$ at $r > R$. 
But it should coincide with the vacuum mass $m_\nu$ at $r\rightarrow\infty$.  
To obtain a self-consistent solution, we need to tune $\tilde{m}_{0}$ 
to match $\tilde{m}_{\nu} (r\rightarrow\infty)$ with $m_\nu$, i.e.~$\tilde{m}_{0}$ 
is not a free parameter but a parameter determined by the neutrino mass at $r\rightarrow\infty$.

%{\color{red}
	The system of Eq.~\eqref{eq:-167} and \eqref{eq:-168} is similar to that for nuggets in 
	\cite{Gresham:2017zqi}.
	Indeed, differentiating Eq.~(5) of
	\cite{Gresham:2017zqi}  with respect to $r$
	gives exactly Eq.~\eqref{eq:-168}.
	Eq.~\eqref{eq:-167} corresponds (up to a factor of $1/\pi$)
	to Eq.~(6) in \cite{Gresham:2017zqi}.
	However,  the boundary conditions (7) there  differs from
	our boundary conditions determined by Eq.~\eqref{eq:-170} and by the aforementioned matching of $\tilde{m}_{\nu} (r\rightarrow\infty)$.
	The boundary condition
	in the relativistic regime together with the matching is non-trivial 
	as already noticed in \cite{Stephenson:1996qj}.
	This condition can affect the existence of solutions.
%}

Eqs.~\eqref{eq:-167} and \eqref{eq:-168} 
can be reduced to the Lane-Emden equation in the non-relativistic limit. 
Indeed, in this limit $\tilde{n} \approx n$ and for  $m_\phi = 0$ we obtain from 
\eqref{eq:-167}
\begin{equation}
\frac{1}{r^2} \frac{d}{dr} \left[r^2 \frac{d (y \phi)}{dr} \right] = - y^2 n, 
\label{eq:nonrel1}
\end{equation}
which is the spherically symmetric case of \eqref{eq:-167}.
From \eqref{eq:-164}, neglecting  $V = y \phi$ in comparison to $m_\nu$ we find  
$$
%\begin{equation}
\frac{d(y\phi)}{dr}= - \frac{1}{2m_\nu} \frac{dp_{F}^2}{dr} = 
- \frac{(6\pi^2)^{2/3} }{2m_\nu} \frac{d n^{2/3}}{dr}.
%\label{eq:seconde}
%\end{equation}
$$
Inserting this expression for the derivative into 
Eq.~\eqref{eq:nonrel1} gives  Eq.~\eqref{eq:-114}.

\subsection{Solution for massless mediator}
%%%%%%%%%%%%%%%%%%%%%%%%%%%%%%%%%%%%%%%%%%%%%%%%%%%%%%%%%%%%%%%%%%%%%%%%%

%%%%%%%%%%%%%%%%%%%%%%%%%%%%%%%%%ffff3  %%%%%%%%%%%%%%%%%%%%%%%%%%%%%%%%%%%%%
\begin{figure}
    \centering
    \includegraphics[width=0.6\textwidth]{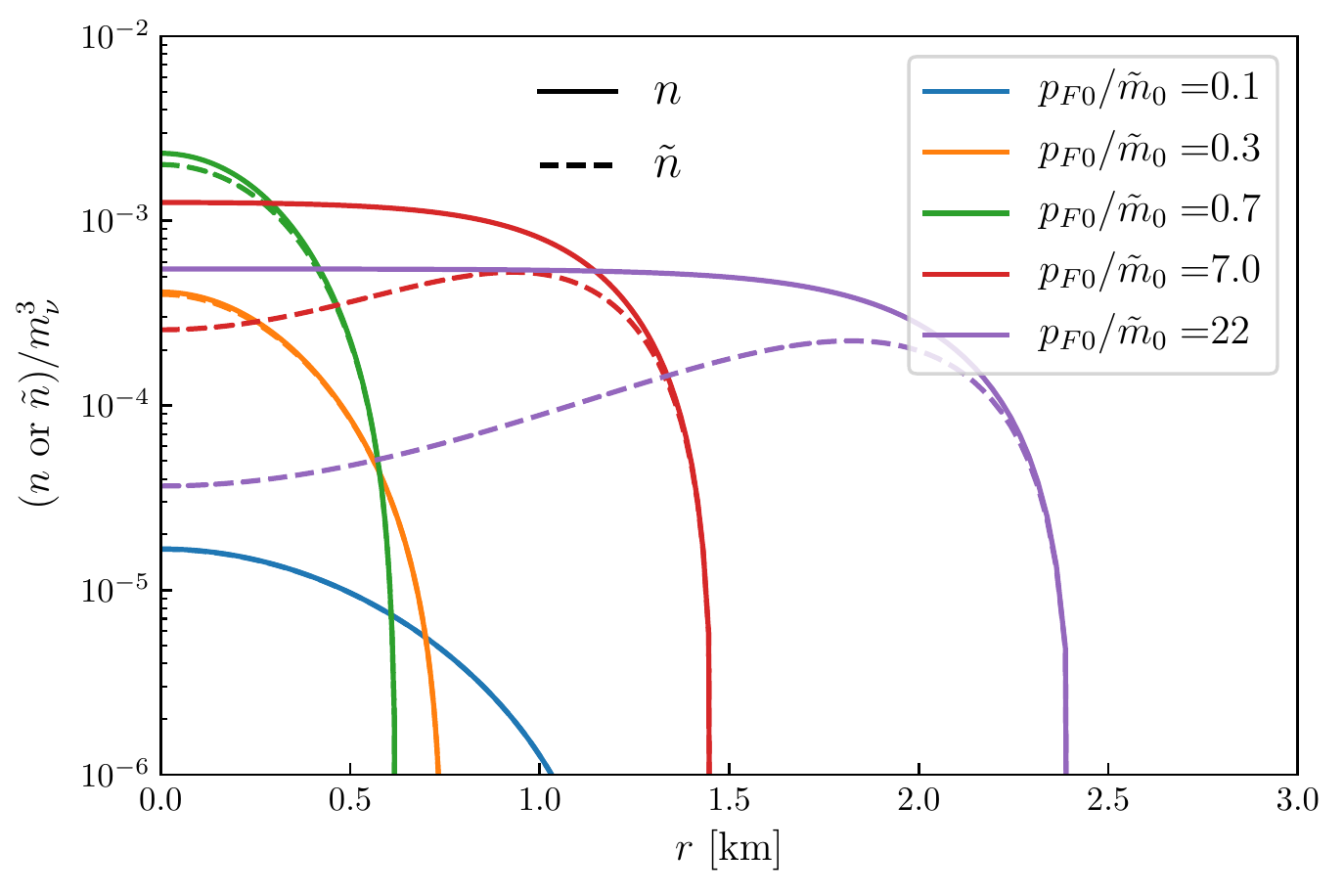}
    \caption{\label{fig:distribution-1} The same as in  Fig.~\ref{fig:distribution}
    but for the relativistic case.  
    Shown are the density (solid) and effective density (dashed) profiles 
    for different values of  $p_{F0}/\tilde{m}_{0}$. 
    }
\end{figure}
%%%%%%%%%%%%%%%%%%%%%%%%%%%%%%%%%%%%%%%%%%%%%%%%%%%%%%%%%%%%%%%

Following the above setup, we first solve the differential equations \eqref{eq:-167} and \eqref{eq:-168}  with $m_\phi=0$. 
In Fig.~\ref{fig:distribution-1}, we show  the obtained  
density $n(r)$ and the effective density $\tilde{n}(r)$ 
profiles for several values of 
$p_{F0}/\tilde{m}_{0}$. Recall that the effective density 
is the number density 
that includes the suppression factor $ \tilde{m}_{\nu}/E_{\mathbf{p}}$. 
This factor depends on $p_F$,  and   decreases toward the center.  
As a result, suppression in the center is stronger 
and the maximum of $\tilde{n}(r)$ occurs
at certain point away from $r=0$.  This effect becomes stronger 
with the increase of $p_{F0}/\tilde{m}_{0}$. 

Using the obtained density profiles we computed various characteristics of 
the neutrino clusters for $y = 10^{-7}$ and $m_\phi = 0$ 
(see   Tab.~\ref{tab:star}). 
The correlations of characteristic parameters ($R$, $N$, $p_{F0}$) are shown 
by blue lines in Figs.~\ref{fig:R-N} and \ref{fig:pf-N}. 
For a given value of $p_{F0}/\tilde{m}_{0}$
 (second line in Tab.~\ref{tab:star})  we determined $\tilde{m}_{0}/m_\nu$  
(the third line). 
The total numbers of neutrinos $N$ (the first line) was computed according to 
$$
%\begin{equation}
N  = \int_{0}^{R}n(r)4\pi r^{2}dr. 
%\label{eq:-173}
%\end{equation}
$$
Using the numbers in the second and  third lines we find  $p_{F0}/m_\nu$ (the 4th line),  
which in turn, gives the central density of a $\nu$-cluster:  
$$
%\begin{equation}
n_{0} = 2.1 \cdot 10^9~{\rm cm^{-3}} ~\left(\frac{p_{F0}}{m_\nu}\right)^3  
\left(\frac{m_\nu}{0.1~{\rm eV}}\right)^3\,.
%\end{equation} 
$$
As $N$ increases, the central density $n_0$ (also $p_{F0}$) first increases, 
reaches maximum at $N y^3 = 1.5 \cdot 10^2$ or $N = 1.5 \cdot 10^{23}$  
and then it decreases. The maximum  corresponds to $p_{F0} \simeq 0.6~m_\nu$, 
that is, to the point 
of transition between the non-relativistic and relativistic regimes. 
The maximal density 
is determined by the mass of neutrino only:   
$$
%\begin{equation}
n_\nu^{\max}(0) = 4.3 \cdot 10^{8}~{\rm cm^{-3}} 
\left(\frac{m_\nu}{ 0.1 {\rm eV}}\right)^3.   
%\end{equation}
$$
The corresponding values of radius of a cluster are shown in the last line of Tab.~\ref{tab:star}.
On the contrary,  
as  $N$ increases, 
%%from value that correspond to non-relativistic case  
the radius $R$ first (in non-relativistic range) decreases,  reaches the minimum 
\begin{equation}
R_{\min}\approx0.62 ~{\rm km} \left(\frac{10^{-7}}{y}\right) 
\left(\frac{0.1\ {\rm eV}}{m_{\nu}}\right){\rm km}\thinspace 
\label{eq:-179}
\end{equation}
at $N \approx 1.5 \cdot 10^{23}$ and then increases, as shown in Fig.~\ref{fig:R-N}. 
The minimal radius \eqref{eq:-179} is about 20\% smaller than the non-relativistic
result in Eq.~\eqref{eq:-77} for $p_{F}=0.3m_{\nu}$,  
which can still be accurately computed using the non-relativistic 
approximation according to the difference between the dashed and solid curves in Fig.~\ref{fig:distribution-1}.

%{\color{red}
	Notice that dependence $n(r)$ substantially differs from $n(r) = 
	p_F^3(r)/6\pi^2$
	with $p_F(r)$ taken from the ansatzes of \cite{Wise:2014ola}.
	For some distances $r$, the difference is given by  a factor of 5 to 6.
	At the same time there is good agreement of our $n(r)$ shown in Fig.~\ref{fig:distribution-1}
	with that obtained in \cite{Gresham:2017zqi}.
%}

%%%%%%%%%%%%%%%%%%%table1%%%%%%%%%%%%%%%%%%%%%%%%%%%%%%%%%%%%%%%%%%%%
\begin{table*}
    \caption{\label{tab:star}Characteristics of final (degenerate) 
    states of neutrino clusters for $y = 10^{-7}$ and $m_\phi = 0$.}
    \begin{ruledtabular}
    \begin{tabular}{lccccc}
    %\hline
    $N$  & $2.96\cdot 10^{21}$ & $1.63\cdot 10^{22}$ & $5.96\cdot 10^{22}$ & $9.35 \cdot 10^{23}$ & 
    $2.34\cdot 10^{24}$ \tabularnewline
    \hline
    $p_{F0}/\tilde{m}_{0}$ & 0.10   & 0.31   & 0.75   & 7.0    & 22      \tabularnewline
    $\tilde{m}_{0}/m_\nu$  & 0.991  & 0.922  & 0.688  & 0.060  & 0.014   \tabularnewline
    $p_{F0}/m_\nu$         & 0.099  & 0.286  & 0.561  & 0.420  & 0.308   \tabularnewline
    $n_0$ [cm$^{-3}$]   & $2.0\cdot 10^6$  & $4.9\cdot 10^7$  & $3.7\cdot 10^8$ & $1.5\cdot 10^8$  & 
    $6.1 \cdot 10^7$ 
    \tabularnewline
    $R$ [km]                & 1.25   &  0.75  & 0.62   & 1.46   &  2.41  \tabularnewline
    \end{tabular}
\end{ruledtabular}
\end{table*}
%%%%%%%%%%%%%%%%%%%%%%%%%%%%%%%%%%%%%%%%%%%%%%%%%%%%%%%%%%%%%%%%%%%%%%%%%%%%%%%%%%%%%%%%%%%%%%%%%%%%%%%%%%%%%

The effective neutrino mass at the border of the cluster can be obtained as 
$\tilde{m}_{\nu}(R) = m_\nu  + V(R)$ with $V(R)$ defined in Eq.~\eqref{eq:-169}: 
$$
%\begin{equation}
\tilde{m}_{\nu}(R)  = m_{\nu} 
- \frac{y^{2}}{m_{\phi}R}e^{-m_{\phi}R}\int_{0}^{R}r\tilde{n}(r)\sinh(m_{\phi}r)dr. 
%\label{eq:-174}
%\end{equation}
$$
Here 
$\tilde{n}(r)$ should be computed according to Eq.~\eqref{eq:-165}.
For $m_{\phi}=0$ the expression simplifies to 
$$
%\begin{equation}
\tilde{m}_{\nu}(R) =  m_{\nu}
 -  \frac{y^{2}}{R}\int_{0}^{R}\tilde{n}(r)r^{2}dr.
%\label{eq:-177}
%\end{equation}
$$

Dependence of $N$ and $R$ on $y$ can be obtained 
using rescaling: $ r\to \overline{r} = r y,  m_\phi \to \overline{m}_\phi= m_\phi/y$\,---\,see Eq.~\eqref{eq:redef2} 
in Appendix~\ref{sec:n-sol}.
% Indeed, $y$  can be eliminated from 
% the equations by redefinition    
% \begin{equation}
% r' = r y, ~~~~ m_\phi' = m_\phi/y,   
% \label{eq:redef2}
% \end{equation}
% so that  $n  = n(r', m_\phi')$. 
% Thus, with decrease of $y$ the radius increases while the mass of $\phi$ decreases. 
The rescaling means that  relative effects of the density gradient and mass of $\phi$ 
in the equation for $\tilde{m}_{\nu}$ do not change with $y$, if $ m_\phi/y$ is given at a fixed value.
We can write 
\begin{equation}
R(y) = R_7 \left(\frac{10^{-7}}{y}\right), ~~~ N(y) = N_7 \left(\frac{10^{-7}}{y}\right)^3,
\label{eq:nandr}
\end{equation}
where $R_7$ and  $N_7$ are the radius and density at $y = 10^{-7}$  
given in the Table I. 
From Eq.~\eqref{eq:nandr}, we find e.g.~that at $y = 10^{-14}$ 
and parameters in the last column of the Table I, 
$R = 2.4 \cdot 10^{12}$ cm and $N  = 2.3 \cdot 10^{45}$,  
which corresponds to  $M = 3.3 \cdot 10^{12}$ g. 
For $m_\phi = 0$, the distribution in $r$  
simply dilates 
with  $y$, and $N$ changes correspondingly.

According to Tab.~\ref{tab:bounds}, for small $N$ the field $\phi$ 
in a cluster is weak, therefore $\tilde{m}_{\nu} \approx m_\nu$ and  
$p_{F0}/\tilde{m}_0 \approx p_{F0}/m_\nu $. 
With increase of $N$ the field increases, $\tilde{m}_0$ decreases, {\it i.e.},  
the medium suppresses the effective neutrino mass. 
$N_7  = 1.5 \cdot 10^{23}$ is critical value between 
the non-relativistic and relativistic cases. 
Further increasing $N$, the dependence of $R$ and $n_0$ on $N$ becomes opposite: $R$ increases, while $n_0$ decreases.

\subsection{Neutrino clusters for nonzero $m_\phi$}
%%%%%%%%%%%%%%%%%%%%%%%%%%%%%%%%%%%%%%%%%%%%%%%%%%%%%%%%%%%%%%%%%%%%%%%%%%%%%%%%%%%%%%%%%

The dependence of properties of a $\nu$-cluster on $m_\phi$ is rather 
complicated.
% since it is not reduced to rescaling of $r$ and  other parameters.
With increase of $m_\phi$  the radius of scalar interaction becomes smaller
and hence the interactions become weaker. Therefore, in general, one expects that
the binding effect weakens and the system becomes less compact for a fixed $N$.
As a result, the radius $R$ increases with increase of $m_\phi$,
which indeed is realized in the non-relativistic case.

%%%%%%%%%%%%%%%%%%%%%%%%%%%ffff4 %%%%%%%%%%%%%%%%%%%%%%%%%%%%%%%%%%%%%%%%%%%%%%%%%%
\begin{figure}
\centering
\includegraphics[width=0.6\textwidth]{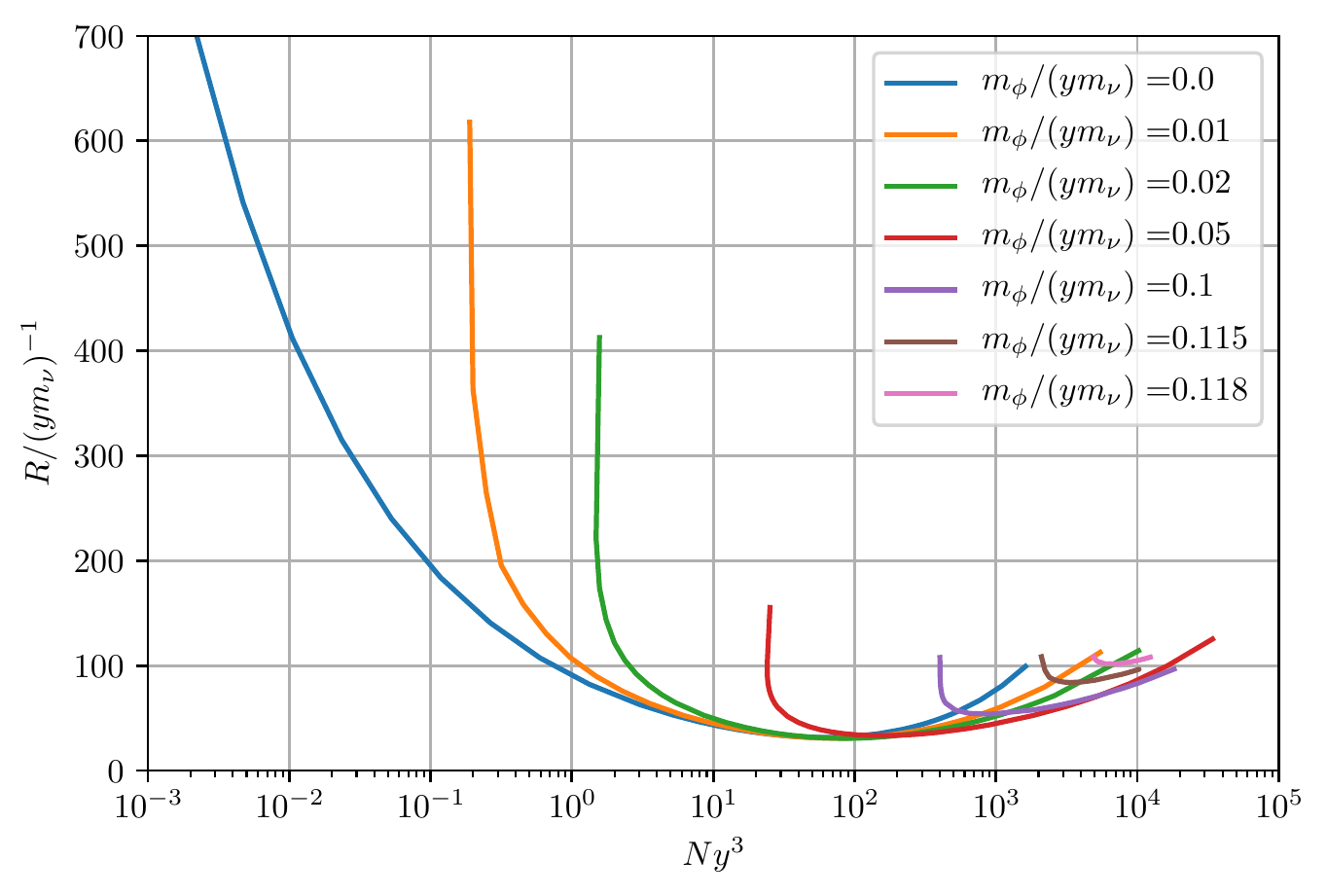}
\caption{\label{fig:R-N} Dependence of the radius $R$
of a neutrino degenerate cluster in the unit of $y m_\nu$  on $Ny^3$ 
for different values of $m_\phi$.
}
\end{figure}
%%%%%%%%%%%%%%%%%%%%%%%%%%%%%%%%%%%%%%%%%%%%%%%%%%%%%%%%%%%%%%%%%%%%%%%%%%%%%%%

In Fig.~\ref{fig:R-N}  we show the dependence of the radius $R$ on $N$ for
different values of $m_\phi$. 
For a given  $m_\phi$ the dependence has a minimum $R_{\min}$.  
%%which corresponds to transition from
%%non-relativistic (NR) to ultrarelativistic (UR) relativistic regimes 
%%at $p_{F0} \sim m_\nu$  [[if so it should not shift]].
As in the massless case, with increase of $N$ the radius 
first decreases, reaches minimum
and then increases in the relativistic regime. 
This behavior is explained by the fact that 
in the relativistic case the attractive force is suppressed
by $\tilde m_\nu/ E$, and
with increase of $N$ the effective mass $\tilde{m}_{\nu}$ decreases. 

The main features of dependence of $R$ on $N$ are discussed below. \\

(i) Existence of a lower bound on $N$. 
 At certain $N$
the lines become vertical.  
The bound is absent for $m_\phi = 0$. 
For each nonzero $m_\phi$, there is a lower bound, $N_{\min}$,  below which it is impossible
% For $N \rightarrow N_{\min}$, the radius diverges, $R \rightarrow  \infty$,  
% which means that neutrinos are not confined.
to make the neutrinos confined.
Since the potential energy (attraction) increases
with $N$,  to support bound system  one should
have large enough  number of neutrinos. 
$N_{\min}$ increases with increase of $m_\phi$,  that
is, with  decrease of radius of interactions of individual
neutrinos $1/m_\phi$.  The smaller the radius, the larger $N$ should be
to keep the system bounded.

%{\color{red}
	The dependence of $R$ on $N$  in Fig. \ref{fig:R-N} differs from that
in Fig. 2 of \cite{Gresham:2017zqi}.
	In the region of $N$ around  minimal radius and above that for $m_\phi = 0$ 
radius in Fig. \ref{fig:R-N}  is larger than  the radius in \cite{Wise:2014ola} 
and \cite{Gresham:2017zqi} and the difference 
increases with decrease of $N$. 
%The difference is related 
%probably with different boundary conditions, 
%which are a rather subtle issue. 
The difference is due to that \cite{Gresham:2017zqi} used an analytical solution 
for $\phi$ which is only valid when $\tilde{n}$ is a constant.
	
Furthermore, the curves $R = R(N, m_\phi)$ of \cite{Gresham:2017zqi} do  not extend 
far below $N$ that correspods to the
	minimum of $R$, since the number $N$ becomes small.
Indeed, in \cite{Gresham:2017zqi} the constant $\alpha_\phi \sim (0.01 - 0.1)$ corresponds to
$y^3 = (0.045 - 1.4)$. For $N \geq 10^2$ this gives $N y^3 \geq (4.5 - 140)$.
Our results extend  to much smaller values of $N y^3$, where important feature of
existence of minimal value of $N$ for $m_\phi \neq 0$ is realized.
	
The dependence of $R$ on $m_\phi$ for fixed $N$ is  also similar here to those  in \cite{Gresham:2017zqi}.
In the relativistic branch  the radius decreases with the increase of  $m_\phi$.
%}

For $m_\phi = 0$ (blue line) the dependence
of $R$ on $N$ is well reproduced by Eq.~(\ref{eq:20210531})
in the non-relativistic case. Notice that in this case 
the scale of $R$ is related to the neutrino mass.   \\

(ii) Existence of a minimal radius for a given $m_\phi$. 
The minimum  corresponds to $p_{F0} \sim \tilde{m}_{\nu}$ ({\it i.e.} to the effective mass of neutrinos which 
can be much smaller than the vacuum mass).
This minimal radius, $R_{\min}$, in the $m_\phi=0$ 
case is determined by Eq.~\eqref{eq:-179}. For nonzero $m_{\phi}$, 
it becomes larger because nonzero $m_\phi$ suppresses the Yukawa potential exponentially.

With  increase of $m_\phi$ the minimum $R_{\min}$ shifts to larger $N$. 
%%Empirically $N \propto m_\phi^\alpha$ with $\alpha \sim 1$.
The value $R_{\min}$ itself changes weakly.  
For large $m_\phi$ the minimum  of $R(N)$  does not correspond
to $\tilde{m}_\nu \sim p_{F0}$, but still occurs close to 
transition from NR to UR  regimes.
Because of the shift of the curves for fixed $N$  in the NR regime
the radius increases with increase of $m_\phi$, while in the
UR case it decreases. The shift can be interpreted in the following way: 
For fixed $R$
the volume per neutrino is determined by $m_\phi$:
$V_1 \propto (1/m_\phi)^\kappa$ ($\kappa \sim 3$). Therefore
with increase of $m_\phi$, the volume per neutrino 
decreases and the total number of $N$ in
a cluster increases.

There is maximal value of $m_\phi$ for which solution
with finite radius exists. 
%%We find that
%%\begin{equation}
%%m_\phi^{\max} = 0.... y m_\nu. 
%%\label{eq:nf-lim}
%%\end{equation}
With approaching of $m_\phi$ to this value the curve further shifts
to larger $N$ and minimum moves up.
For $m_\phi > m_\phi^{\max}$ the radius of interaction
becomes too short to bound the neutrino system and the system disintegrates.
The upper bound on $m_\phi$ corresponds to the lower bound on
strength of interaction:
\begin{equation}
S_\phi  \equiv \frac{y^2 m_\nu^2}{m_\phi^2}  \gtrsim    S_\phi ^{\min} = 70\, , 
\label{eq:b-strength}
\end{equation}
which can be compared  with  the bound  
$G_\phi  > 3.27$ found in \cite{Stephenson:1996qj}
from the condition of existence of minimum of total
energy per neutrino for infinite medium. Notice that our bound is obtained from 
condition of stability of finite configuration. 

%%Inverse of maximal mass $1/m_\phi^{\max}$ can be compared to the
%%radius of a star for zero $m_\phi$, $R_0$,  given in \ref{}:
%%For $p_{F0}/m_\nu = (0.3 - 0.8)$ we obtain
%%$$
%%1/m_\phi^{\max} > (0.5 - 0.7) R_0.
%%$$
%%So, the radius of interactions should be bigger than the radius
%%of star in the limit $m_\phi \rightarrow  0$.

Let us  compare the radius $R$ with the radius of interactions 
$R_\phi = 1/m_\phi$. For small $m_\phi$ there is the range of values of $N$
in which $R < 1/m_\phi$. It is possible to obtain $R = R_\phi$ at two different values of $N$. 
For $m_\phi = 0.03 y m_\nu$ the equality $R = R_\phi$
is fulfilled at definite value $N = 10^2/y^3$.
For $m_\phi > 0.03y m_\nu$, we have $R > R_\phi$ for all $N$
and the difference between them increases.
$R$ can be orders of magnitude larger than $R_\phi$
and therefore in general $1/m_\phi$ does not determine the
size of the cloud and fragmentation scale.

We denote by $X$ and $Y$ the values at 
horizontal and vertical axes of Fig.~\ref{fig:R-N} correspondingly: 
$$
X \equiv Ny^3~~ {\rm  and}~~~  Y \equiv R y m_\nu. 
$$
The coordinate  $Y$ can be written as 
$Y = R/R_\phi S_\phi^{-1/2}$.
Therefore the radius of a cluster over the radius of interactions equals 
\begin{equation}
\frac{R}{R_\phi} = Y S_\phi^{-1/2}.
\label{eq:radrat}
\end{equation}
For $S_\phi^{-1/2} = 0.01$ the radius (in units
$y m_\nu$) is in the interval $Y = (30 - 180)$ (see Fig. 4)
 which gives according to Eq.(\ref{eq:radrat})
$R/R_\phi = 0.3 - 1.8$. For $S_\phi^{-1/2} = 0.02$
the corresponding numbers are  $Y = (30 - 170)$ and
$R/R_\phi = (0.6 - 3.4)$. For $S_\phi^{-1/2} = 0.1$
we have $Y = (55 - 110)$ and $R/R_\phi = 5 - 11$.
Thus, with decrease of strength the radius increases and becomes
substantially larger than $R_\phi$. 
%{\color{red}
	This regime 
was observed in \cite{Gresham:2017zqi} and called 
``saturation''. 
%}
Still $1/m_\phi$ determines
the scale of sizes of $\nu$-clusters within an order of magnitude.

%%%%%%%%%%%%%%%%%%%%%%%%%%%ffff5 %%%%%%%%%%%%%%%%%%%%%%%%%%%%%%%%%%%%%%%%%%%%%%%%%%
\begin{figure}
\centering
\includegraphics[width=0.6\textwidth]{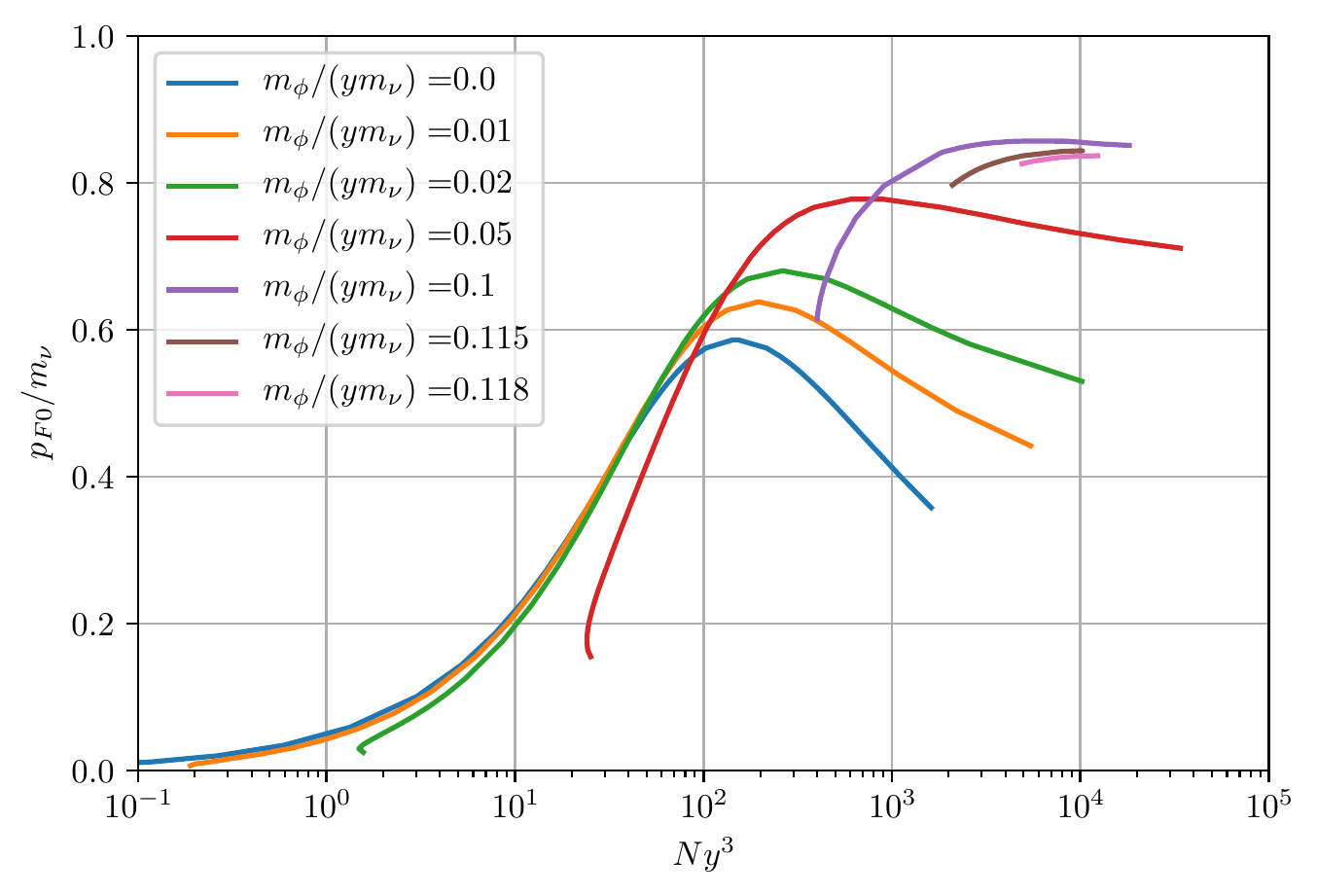}
\caption{\label{fig:pf-N} 
Dependence of the Fermi momentum in the center of $\nu$-cluster,  
$p_{F0}$,  on $Ny^3$ for different values of $m_\phi$ ($m_\phi/y m_\nu$).
% Dependence of the Fermi momentum $p_{F0}$ in the center 
% of the $\nu$-cluster in the unit of $m_\nu$ on $Ny^3$ for 
%%different values of $m_\phi$ in the unit of $y m_\nu$.
}
\end{figure}
%%%%%%%%%%%%%%%%%%%%%%%%%%%%%%%%%%%%%%%%%%%%%%%%%%%%%%%%%%%%%%%%%%%%%%%%%%%%%%%

In Fig. \ref{fig:pf-N}  we show the dependence of $p_{F0}$ (and consequently, 
the central density) on $N$ for different values of $m_\phi$.
For fixed $m_\phi$,  $p_{F0}$ as a function of $N$ has a maximum. 
With increase of $m_\phi$, this maximum 
shifts to larger $N$ and increases. 
However, this increase slows down and 
is restricted by $p_{F0}/m_\nu \lesssim 0.9$. 
Explicitly, for this value   
\begin{equation}
n_0^{\max} = 1.5 \cdot 10^9 {\rm cm^{-3}} \left( \frac{m_\nu}{0.1 {\rm eV}} \right)^3. 
\label{eq:nmax}
\end{equation}
With increase of $m_\phi$ the central density 
decreases at small $N$ (the NR regime) and it increases
at larger $N$ (UR regime). This anti-correlates with change of
radius  (for fixed $N$) in Fig.~\ref{fig:R-N}. 
The maximal density (corresponding to the maximal $p_{F0}$) increases
with $m_\phi$ and shifts to larger $N$.
The anticorrelation is due to  
%%along the line of constant 
%%$S$ [[everywhere?]] 
the relation
\begin{equation}
N \sim b \frac{4\pi}{3} R^3 n_{F0} =
b \frac{2}{9} \pi R^3 p_{F0}^3, 
\label{eq:rela2}
\end{equation}
where $b \approx 0.2 - 0.5$ is smaller than 1,  since the density distribution is not uniform 
and the density is substantially smaller than $n_{F0}$
in peripheral regions.

Thus, the overall change of characteristics
of cluster (for fixed $N$) with $m_\phi$
is the following: In the NR case central density decreases
and radius increases (cluster becomes less compact).
In the UR case, the radius varies non-monotonically  
with $m_\phi$, as shown by the relativistic branches of the curves in Fig.~\ref{fig:R-N}. 
These characteristics could be important for cluster formation processes.

%{\color{red}
	For fixed $m_\phi$ the bound system exists
	for $N$ above certain minimal value.
	With the increase of $N$  the central density
	(and $p_{F0}$) increases  quickly and therefore
	the radius of cluster decreases. Then the density reaches its maximum and 
	with further increase of $N$
	the density decreases. Correspondingly the
	radius of cluster increases. 
%}

Notice that strengths of interactions in the case of
two neutrino bound states $\lambda$ [see Eq. (\ref{eq:-14})]
and for the bound neutrino system [Eq.~(\ref{eq:b-strength})] have
different functional dependence on the parameters.
%%[[why?]].
From Eqs.~(\ref{eq:-14}) and (\ref{eq:b-strength}) we obtain the following relation between
the two strengths:
$$
\lambda^2 = \left(\frac{y}{8 \pi}\right)^2 S_\phi.
$$
For $y \leq 10^{-7}$ and $\lambda^2 \geq 0.7$ required to have
$2\nu$-bound state we find $S_\phi > 6.3 \cdot 10^{16}$. That is, if
$2\nu$-bound states exist also $\nu$-clusters should exist.
The opposite does not work: For $S_\phi \sim 10^2$
we have $\lambda \sim 1.6 \cdot 10^{-15}$, and so the $2\nu$ bound
states are not possible.
Both the bound states and bound systems exist for very small
masses of scalars.

\subsection{Neutrino bound systems for given $y$ and $m_\phi$}
%%%%%%%%%%%%%%%%%%%%%%%%%%%%%%%%%%%%%%%%%%%%%%%%%%%%%%%%%%%%%

In the previous sections we focused on properties of final
stable configurations of neutrino clusters.
Let us comment on implications of the obtained results.

From the particle physics perspective  the input parameters are  $y$ and $m_\phi$,  
and the questions are whether bound systems for given  $y$ and $m_\phi$ exist, 
and what the characteristic parameters of these systems are. The parameters
$y$ and $m_\phi$ determine the interaction strength $S_\phi$ defined in Eq.~\eqref{eq:b-strength}.
The main condition for existence of bound system is that the strength $S_\phi$ is large enough,
satisfying the lower bound (\ref{eq:b-strength})
or $S_\phi^{-1/2} \leq 0.12$.
%% as we have found in  sect.
For a given $y$, the condition on the strength
%%Eq. (\ref{eq:strengthbound})
gives an upper bound on $m_\phi$.
So, essentially $y$ and $S_\phi$ determine all other characteristics.

Let us  consider properties of clusters at the critical (minimal) value of strength.
According to Fig.~\ref{fig:R-N}, $S_\phi^{\min}$ is realized at
$$
%\begin{equation}
X^{\rm cr} \approx 10^4, \, \, \, Y^{\rm cr} \approx 10^2
%\label{eq:extr}
%\end{equation}
$$
with very small spread. As follows from Fig.~\ref{fig:pf-N}, it corresponds to
maximal $p_{F0}/m_\nu = 0.86$ and therefore maximal central
density of the cluster. 
%%(It is in the transition region
%%between the relativistic and non-relativistic cases).
%%
With these specific values of $X$ and $Y$, the parameters of cluster are determined
by $y$. Taking $y = 10^{-7}$, 
we obtain
$$
m_\phi^{\rm cr} \approx 10^{-8} m_\nu = 10^{-9}\, {\rm eV}, 
$$
and 
$$
N^{\rm cr} = X^{\rm cr} y^{-3} \approx 10^{25}, ~~~
R^{\rm cr} =  2 \cdot 10^{-4} {\rm cm} ~ Y^{\rm cr} y^{-1}
\approx  2 \cdot 10^{5} {\rm cm}.
%\label{eq:extr1}
%\end{equation}
$$

If $m_\phi < m_\phi^{\rm cr}$ (for fixed $y$),
the  strength increases and according to Fig.~\ref{fig:R-N},
the ranges of $X$ and $Y$ expand. Correspondingly,
$$
%\begin{equation}
N = X y^{-3} = 10^{25} X/X^{\rm cr},
%\label{eq:extr12}
%\end{equation}
$$
and therefore clusters with much smaller number
of neutrinos are possible. E.g. for $S_\phi = 10^4$,
$X$ can be as small as 0.2, and therefore $N = 2 \cdot 10^{20}$.
%%[[even smaller than in the Table I for massless $\phi$]]
$X$ becomes a free parameter in the interval $0.2 - 10^{4}$.
The allowed range  of $Y$ expands in both directions:
$Y$ can be as small as $0.3 Y^{\rm cr}$, and therefore the radius
reduces down to $0.7$ km. The maximal value of $Y$ can be much bigger than
$Y^{\rm cr}$ and it increases with increase of strength (decrease of
$m_\phi$).
$$
%\begin{equation}
R = 2 \cdot 10^{-4} {\rm cm} ~Y y^{-1} =
2 \cdot 10^{5} {\rm cm} ~Y/Y^{\rm cr}.
%\label{eq:r12}
%\end{equation}
$$
For $S_\phi = 10^4$ (orange line in Fig.~\ref{fig:R-N})
$Y = (30 - 380)$ and correspondingly, $R= (0.6 - 7.6)$ km.
If also $y$ decreases and  $m_\phi$ decreases,
so that $S_\phi$ does not change, the line $Y = Y(X)$
is the same, but parameters of cluster increase as
$N \propto y^{-3}$, $R \propto y^{-1}$. \\

For fixed $y$ and $m_\phi$  stable bound systems 
are situated along the corresponding lines
$S_\phi ={\rm const}$ in Figs.~\ref{fig:R-N} and \ref{fig:pf-N}. 
This, in turn, fixes the interval of $X$ and $Y$ (hence $N$ and $R$):
\begin{equation}
N  \in \frac{1}{y^{3}} \left[ X_{{\rm min}}, \,  X_{{\rm max}}\right] \thinspace, \, \, \, \, \, 
R \in \frac{1}{ym_{\nu}} [ Y_{{\rm min}}, \,  Y_{{\rm max}}]\thinspace.
\label{n-and-r}
\end{equation}
As an  example, for $y = 10^{-20}$ and
$m_\phi = 10^{-23}$ eV we obtain $S_{\phi}^{-1/2} = 0.01$ (orange line),  
and according to Fig.~\ref{fig:R-N} we have  $X \in [0.2,\ 5 \cdot 10^{3}]$.
For chosen $y$, this gives 
$N \in [2 \cdot 10^{59},\ 5 \cdot 10^{63}]$. Depending on $N$,  the central density 
changes (see Fig. \ref{fig:pf-N})  
following the relation (\ref{eq:rela2}).  The interval
$Y \in [30,\ 370]$ (Fig.~\ref{fig:R-N}) corresponds to  the interval of values 
$R \in  [6 \cdot 10^{17},\ 7.4 \cdot 10^{19}]$ cm
and  $R/R_\phi \in [0.3,\ 3]$.
With increase of $y$  for fixed strength, $N$ and $R$ change as in Eq.~\eqref{n-and-r}.
With decrease of strength, the intervals of $X$ and $Y$ shrink. 
Thus for $S_{\phi}^{-1/2} = 0.1$,  we have $X \in [400,\  2\cdot 10^{4}] $ 
and  $Y \in [50,\ 110]$. If $y = 10^{-20}$ and
$m_\phi = 10^{-22}$ eV, we obtain $N \in [4 \cdot 10^{62}, \ 
2 \cdot 10^{64}]$  and $R \in [10^{18},\  2.3 \cdot 10^{18}]$ cm.

From the cosmological perspective the key input parameter
is $N$. It is this  quantity that determines
in the nearly uniform background evolution and formation of clusters.
If $N$ is conserved (no merging of clusters, no particle
loss, {\it etc.}), this quantity determines the final radius  and
density of the cluster. From Eq. (\ref{eq:20210531}) we obtain
$$
%\begin{equation}
R = 1.8~ {\rm km} ~\left(\frac{10^{-7}}{y}\right)^2
\left(\frac{10^{21}}{N}\right)^{1/3},
%\label{eq:r12N}
%\end{equation}
$$
and from Eqs.~(\ref{eq:20210526-1}) and (\ref{eq:n0mnu}):
$$
%\begin{equation}
n_0 = 2.25 \cdot 10^8 ~{\rm cm^{-3}} (y^3 N)^2.
%\label{eq:noN}
%\end{equation}
$$
These relations are valid for $m_\phi \ll 1/R$.
With decrease of $y$ (and fixed $N$), 
$R$ increases as  $1/y$ and the density decreases as $y^{3/2}$.

%%Notice that stable configurations of neutrino clusters 
%%do not ensure that they are realized in the real world:
%%additional restrictions can come from formation of clusters
%%in the  expanding Universe.

%%%%%%%%%%%%%%%%%%%%%%%%%%%%%%%%%%%%%%%%%%%%%%%%%%%%%%%%%%%%%%%%%%%%%%%%%%%%%%
\section{Formation of neutrino clusters \label{sec:Formation-and-lifetime}}
%%%%%%%%%%%%%%%%%%%%%%%%%%%%%%%%%%%%%%%%%%%%%%%%%%%%%%%%%%%%%%%%%%%%%%%%%%%%%%

Let us consider the formation of neutrino clusters from 
the uniform background of relic neutrinos in  
the course of cooling and expansion  of the Universe.

We find that for the allowed values of $y$ 
and $m_\nu$, the typical time  of cooling via $\phi$
bremsstrahlung and via annihilation  $ \nu \bar{\nu} \rightarrow  \phi \phi$ 
are much larger than the age of the Universe (see Appendices \ref{sub:energy-loss} and \ref{subsec:Rates}). 
Therefore the formation of neutrino clusters differs from the formation of usual stars.\footnote{
	Notice that the formation of  neutrino
	clusters is very different 
	from the formation of nuggets of ADM
	\cite{Gresham:2018anj}. The latter proceeds via fusion of particles of 
	DM and requires C-asymmetry to avoid complete annihilation.} 
As we will show in subsections~\ref{sub:dip} and \ref{sub:frag}, for the interaction strength above a certain critical value it has a character 
of phase transition at which the kinetic energy 
is transformed into the increase of  the effective neutrino mass.
The mechanism was first suggested in \cite{Stephenson:1996qj} and here we  explore it further.

The phase transition leads to fragmentation of the cosmic neutrino background. Here we present some simplified arguments and analytic estimations. 
A complete solution of the formation problem  would require numerical simulation 
of evolution of the $\nu$ background. Such a simulation  is beyond the scope of this paper.

%%%%%%%%%%%%%%%%%%%%%%%%%%%%%%%%%%%%%%%%%%%%%%%%%%%%%%%%%%%%%%%%%%%%%%%%%%
\subsection{Maximal density and fragmentation of the $\nu$ background\label{sub:fragmentation}}
%%%%%%%%%%%%%%%%%%%%%%%%%%%%%%%%%%%%%%%%%%%%%%%%%%%%%%%%%%%%%%%%%%%%%

The key parameter that determines formation of  $\nu$-clusters 
is the central density given by Eq.~(\ref{eq:nmax}). 
According to Fig.~\ref{fig:pf-N},  there is a maximal value 
of $p_{F0}$,  and consequently, a maximal value of the
density   $n_{0}^{\rm max}$  which is determined by the mass of neutrino in Eq.~(\ref{eq:nmax}). 
After the fragmentation of a uniform relic 
$\nu$-background (see below), the central density of a cluster  stopped decreasing. 
Consequently, at the epoch of fragmentation the density  was 
$$
%\begin{equation}
n_{\rm frag} \leq  n_0^{\max}.  
%\label{eq:nmax1}
%\end{equation}
$$
Thus, we can take  
$$
%\begin{equation}
n_{\rm frag} \leq 10^9~ {\rm cm^{-3}}. 
%\end{equation} 
$$
In turn,  this density determines the epoch of fragmentation: 
\begin{equation}
(z_f+1) = \left( \frac{n_{\rm frag}}{n_{\rm rel}} \right)^{1/3} = 215,  
\label{eq:zfrag}
\end{equation}
where $n_{\rm rel} = 113$ is the number density 
of one neutrino mass state that would be in the present epoch 
in the absence of clustering. That is, formation of $\nu$-clusters  starts  
after the epoch of recombination ($z=1100$, $T=0.1$ eV). 
At this epoch ($z_f+1=215$) the neutrino average 
momentum was $p \sim 0.02$ eV. 
So, fragmentation could start when $p \sim m_\nu$, 
which is not accidental since $n_0^{\max}$ is determined by $m_\nu$ only. 

%%%%%%%%%%%%%%%%%%%%%%%%%%%%%%%%%%%%%%%%%%%%%%%%%%%%%%%%
\subsection{Evolution of the effective neutrino mass }
%%%%%%%%%%%%%%%%%%%%%%%%%%%%%%%%%%%%%%%%%%%%%%%%%%%%%%%%%

Dependence of the effective mass $\tilde{m}_{\nu}$ on $T$ plays crucial role in the formation 
of neutrino bound systems. 
% In the absence of other factors like gravitational perturbations 
%  related to DM halos ...
%%Dynamically, the fragmentation of the uniform neutrino sea 
%%may start when in the expanding Universe 
%%the system of neutrinos interacting via light mediators  
%%becomes unstable.  
%%The expansion of the Universe can trigger fragmentation
%%if at some temperature (density) 
%%the energy of the system reaches minimum and further 
%%expansion would require increase (injection) of energy from outside. 
%%In this consideration we follow to some extent 
%%the analysis in \cite{Stephenson:1996qj}. 
%%We assume that there is no initial density of $\phi$
 If neutrinos are the only source of the field $\phi$,  
for the allowed coupling constant 
%the production of $\phi$ is negligible and there is no $\phi$ particles in the Universe.
$\phi$ cannot be thermalized in the early Universe and the amount of $\phi$ particles is negligible.
Evolution of the scalar field in expanding universe is described by 
Eq.~(\ref{eq:-40}) 
with an additional expansion term  $3 H(t) d\phi/dt $, 
where the Hubble constant depends on time as $H(t) = 1/2t$ in the radiation-dominated epoch
and as $H(t) = 2/3t$ in the matter-dominated epoch.

Notice that in general $m_\phi \neq 0$ regularizes the procedure 
of computations. Otherwise, it is regularized by the Hubble expansion rate $ H$. Indeed, 
from the equation of motion (\ref{eq:-40}) for uniform  distribution,   assuming 
$\dot{\phi}  = a H \phi$, 
%,  where $a$ is [[]], 
where $a$ is the scale factor in the Friedmann-Robertson-Walker (FRW) metric, 
we obtain
\begin{equation}
\phi  = - \frac{y}{a^2 H^2 + m_\phi^2}\langle \bar{\nu}\nu\rangle,  
\label{eq:phiev}
\end{equation} 
and  $\langle \bar{\nu}\nu \rangle$ is given in (\ref{eq:-57}).
Thus, the expansion term regularizes the solution for $m_\phi \rightarrow 0$.
The energy density in the field:  
\begin{equation}
\rho_\phi  = \frac{y^2}{2} \frac{m_\phi^2}{(a^2H^2 + m_\phi^2)^2}
|\langle \bar{\nu}\nu \rangle|^2. 
\label{eq:rhophi}
\end{equation}
In what follows we assume that $H \ll m_\phi$, 
so that expansion can be treated as slow 
change of density and temperature in the solution of equation 
without expansion. 

Neglecting the Hubble expansion term, Eq.~(\ref{eq:phiev}) gives for uniform medium 
\begin{equation}
\phi  = - \frac{y}{m_\phi^2}\langle \bar{\nu}\nu \rangle.
\label{eq:phiev1}
\end{equation}

From Eq.~(\ref{eq:phiev1}), we obtain the  equation for the effective mass of neutrino: 
$$
%\begin{equation}
\tilde{m}_{\nu} = m_\nu - \frac{y^2}{m_\phi^2} 
\langle \bar{\nu}\nu\rangle,  
%\label{eq:tildemeq}
%\end{equation} 
$$
where [see Eq. (\ref{eq:-57})]
$$
%\begin{equation}
\langle \bar{\nu}\nu\rangle  = 
\frac{\tilde{m}_{\nu}}{2 \pi^2} \int_0^{\infty} 
\frac{dp~p^2}{\sqrt{p^2 + \tilde{m}_{\nu}^2}} f(p,T), 
%\label{eq:matr2}
%\end{equation} 
$$
and  $f(p,T)$ is the  distribution of neutrinos 
(thermal or degenerate\footnote{Notice that results for the degenerate case
can be obtained by substituting $ I_3 T^3 \rightarrow \frac{p_F^3}{3} $ 
in the results for the thermal case.}).  
Explicitly the equation for $\tilde{m}_{\nu}$ can be written as 
$$
%\begin{equation}
\tilde{m}_{\nu} = m_\nu - \frac{y^2}{m_\phi^2} \frac{\tilde{m}_{\nu}}{2 \pi^2} 
\int_0^{\infty} \frac{dp~p^2}{\sqrt{p^2 + \tilde{m}_{\nu}^2}} f(p,T). 
%\label{eq:tildemeq1}
%\end{equation}  
$$
Here we use the thermal distribution with $T \neq 0$, which 
is relevant for $T > m_\nu$, while in  \cite{Stephenson:1996qj} 
the distribution of degenerate Fermi gas was used. Qualitatively, the results are similar.  
%[[and below $m_\nu$???]]

%%%%%%%%%%%%%%%%%%%%%%%%%%%ffff6 %%%%%%%%%%%%%%%%%%%%%%%%%%%%%%%%%%%%%%%%%%%%%%%%%%
\begin{figure}
\centering
\includegraphics[width=0.6\textwidth]{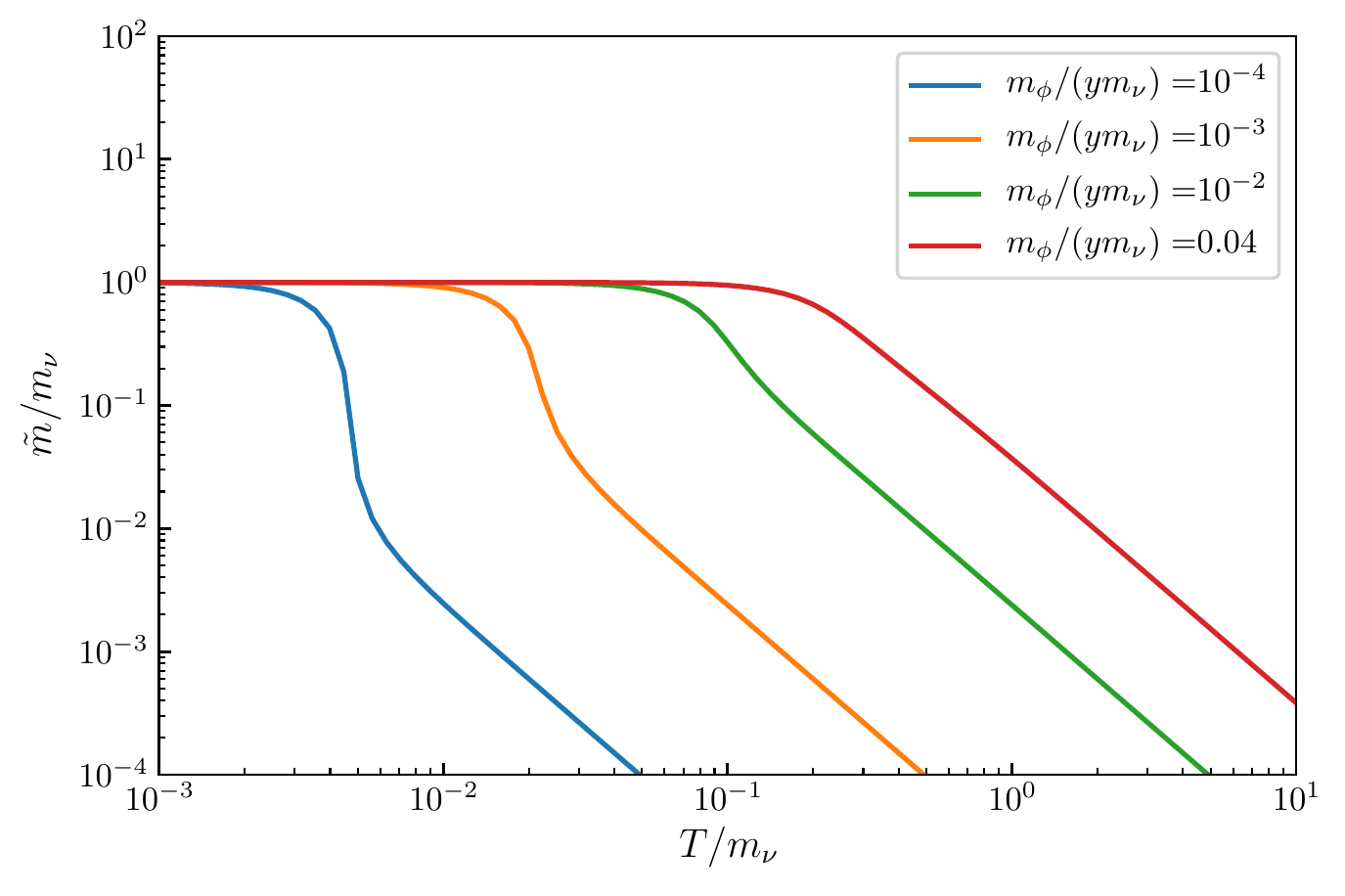}
\caption{\label{fig:tm-T} 
Dependence of the effective neutrino mass $\tilde{m}_{\nu}$ 
% in the unit of $m_\nu$ 
on $T/m_\nu$ for different values 
of $m_\phi$.
%in the units of $y m_\nu$
% (inverse of square root of strength of interaction). 
}
\end{figure}
%%%%%%%%%%%%%%%%%%%%%%%%%%%%%%%%%%%%%%%%%%%%%%%%%%%%%%%%%%%%%%%%%%%%%%%%%%%%%%%

In Fig.~\ref{fig:tm-T}, we show dependence of $\tilde{m}_{\nu}$ 
on $T$ for different values of 
strength of interaction.  In the  ultra-relativistic limit,  
$T \gg \tilde{m}$,  we obtain
\begin{equation}
\tilde{m}_{\nu} = m_\nu - \frac{y^2 \tilde{m}_{\nu} T^2}{m_\phi^2}
\frac{I_2}{2 \pi^2},
\label{eq:tildemrel}
\end{equation}
where 
%%$I_2  = \pi^2/12$, and 
in general, 
the integrals $I_n$ have been introduced in Eq. (\ref{eq:inn}).  
From Eq.~(\ref{eq:tildemrel}) we find
\begin{equation}
\tilde{m}_{\nu} = \frac{m_\nu}{1 + \frac{y^2 T^2 I_2}{2 \pi^2 m_\phi^2}}
\approx m_\nu \frac{2 \pi^2 m_\phi^2}{y^2 T^2 I_2} = 
m_\nu \frac{24 m_\phi^2}{y^2 T^2} = 
m_\nu \frac{24}{S_\phi} \left(\frac{m_\nu}{T} \right)^2 \equiv \tilde{m}_{\nu}^{\rm rel}.
\label{eq:tildemrel2}
\end{equation}
That is, the effective mass $\tilde{m}_{\nu} \propto T^{-2}$  
decreases as $T$ increases and becomes negligible in the  early Universe.  

In the non-relativistic case, $T \ll \tilde{m}_{\nu}$, the solutions is 
\begin{equation}
\tilde{m}_{\nu} = m_\nu - \frac{y^2 T^3}{m_\phi^2} \frac{I_3}{2 \pi^2} \equiv \tilde{m}_{\nu}^{\rm nr}.
\label{eq:tildemnonr}
\end{equation}
That is, as $T\to 0$, $\tilde{m}_{\nu} \rightarrow m_\nu$.
These analytic results agree well with results of numerical computations shown in Fig. \ref{fig:tm-T}.

Note that the relativistic regime is determined by $\tilde{m}_{\nu}$ 
and not $m_\nu$: $T > \tilde{m}_{\nu}$. 
The effective mass $\tilde{m}_{\nu}$ 
increases as $1/T^2$ till 
$T \approx \tilde{m}_{\nu}$. According to 
Eq.~(\ref{eq:tildemrel2})  this equality gives 
\begin{equation}
\frac{T_{\rm rel}}{m_\nu} = \left(\frac{24}{S_\phi}\right)^{1/3}.
\label{eq:Trel1}
\end{equation}
Below this temperature the growth of $\tilde m$ is first faster 
and then slower than 
$1/T^2$. The change of derivative occurs at maximum of the relative 
energy in the scalar field (see below). $\tilde m (T)$ 
converges to the non-relativistic regime 
at $\tilde{m}_{\nu}^{\rm rel} (T)  = m_\nu$. 
This gives, according to Eq.~(\ref{eq:tildemrel2}), 
%\begin{equation}
$$
\frac{T_{\rm nr}}{m_\nu} = \left(\frac{24}{S_\phi}\right)^{1/2}. 
%\end{equation}
$$

\subsection{Energy of the $\nu$-$\phi$ system and the dip \label{sub:dip}}
%%%%%%%%%%%%%%%%%%%%%%%%%%%%%%%%%%%%%%%%%%%%%%%%%%%%%%%%%%%%%%%%%%%%%%

Let us compute the energy density in the scalar field, 
$\rho_\phi$ , in neutrinos, $\rho_\nu$, and the  
total energy density defined as 
\begin{equation}
\rho_{\rm tot}   = \rho_\nu   + \rho_\phi  
\label{eq:rhotot}
\end{equation}
as a function of $T$.  
Expansion of the Universe can be accounted for by considering the average energy per neutrino: 
\begin{equation}
\epsilon_i  \equiv \frac{\rho_i}{n_\nu}, ~~~~i = \{\nu, ~~ \phi, ~~ {\rm tot} \}, 
\label{eq:rhotot}
\end{equation}
where the number density of neutrinos equals
$$
%\begin{equation}
n_\nu =  \frac{1}{2 \pi^2} \int_0^{\infty} dp~ p^2 f(p,T) 
= \frac{I_3}{2 \pi^2} T^3.
%\label{eq:ndens}
%\end{equation} 
$$

The energy density in the scalar field is given by 
\begin{equation}
\rho_\phi  = \frac{1}{2} m_\phi^2 \phi^2. 
\label{eq:scdens}
\end{equation} 
Here $\phi$ can be expressed via the effective mass, $\phi = (\tilde{m}_{\nu} - m_\nu)/y$, and therefore  
\begin{equation}
\rho_\phi(\tilde{m}_{\nu})  = \frac{m_\phi^2}{2 y^2}(\tilde{m}_{\nu} - m_\nu)^2.  
\label{eq:scdens1}
\end{equation} 
So, $\rho_\phi$ is directly determined by the deviation of the effective mass of neutrino from 
the vacuum mass. 
The energy per neutrino equals
\begin{equation}
\epsilon_\phi  =  m_\nu  \frac{\pi^2}{I_3 S_\phi}
\left(\frac{\tilde{m}_{\nu}}{m_\nu} - 1\right)^2 \left(\frac{m_\nu}{T}\right)^3. 
\label{eq:scdenseff}
\end{equation}
% $$
% G_\phi \equiv \frac{g^2 m_\nu^2}{m_\phi^2}. 
% $$
% is the strength of interaction. 

Using the known form of $\tilde{m}_{\nu}(T)$, Eq.~(\ref{eq:tildemrel}), 
we compute the energy density $\epsilon_\phi$. 
In the ultra-relativistic limit,  according to Eq.~(\ref{eq:scdens1}),  
the energy density in $\phi$ converges to the constant:
$$
%\begin{equation}
\rho_\phi^{\rm rel}  = \frac{m_\phi^2}{2 y^2} m_\nu^2. 
%\label{eq:scdens3}
%\end{equation}
$$
Therefore
$$
\epsilon_\phi^{\rm rel} = \frac{m_\nu}{\chi},  
$$
where 
$$
%\begin{equation}
\chi  \equiv \frac{S_\phi I_3}{\pi^2} \left(\frac{T}{m_\nu}\right)^3.  
%\label{eq:xidef}
%\end{equation}
$$

In the non-relativistic limit 
$$
%\begin{equation}
\rho_\phi^{\rm nr} = \frac{1}{8\pi^4}
\frac{y^2}{m_\phi^2} I_3^2 T^6.
%\label{eq:rhophinra}
%\end{equation}
$$
and 
\begin{equation}
\epsilon_\phi^{\rm nr} = 
m_\nu \frac{S_\phi I_3}{4 \pi^2} \left(\frac{T}{m_\nu}\right)^3 = 
\frac{m_\nu}{4} \chi.   
\label{eq:relphinr}
\end{equation}
According to Eq.~(\ref{eq:relphinr}), as the Universe 
cools down, $\epsilon_\phi$ first increases and then decreases. 
The maximum of $\epsilon_\phi$ is achieved when  
$\epsilon_\phi^{\rm rel} \approx \epsilon_\phi^{\rm nr}$, which  happens  at 
$\chi = 2$, or 
$$
%\begin{equation}
\frac{T^{\max}}{m_\nu} = \left(\frac{2\pi^2}{S_\phi I_3} \right)^{1/3}. 
%\label{eq:tmax}
%\end{equation}
$$
Hence the maximum is $\epsilon^{\max} = m_\nu/2$.  
When $S_\phi$ increases (correspondingly 
$m_\phi$ decreases), the maximum $\epsilon^{\max}$  
does not change significantly,  but shifts to smaller $T$.

%%%%%%%%%%%%%%%%%%%%%%%%%%%ffff7%%%%%%%%%%%%%%%%%%%%%%%%%%%%%%%%%%%%%%%%%%%%%%%%%%
\begin{figure}
\centering
\includegraphics[width=0.6\textwidth]{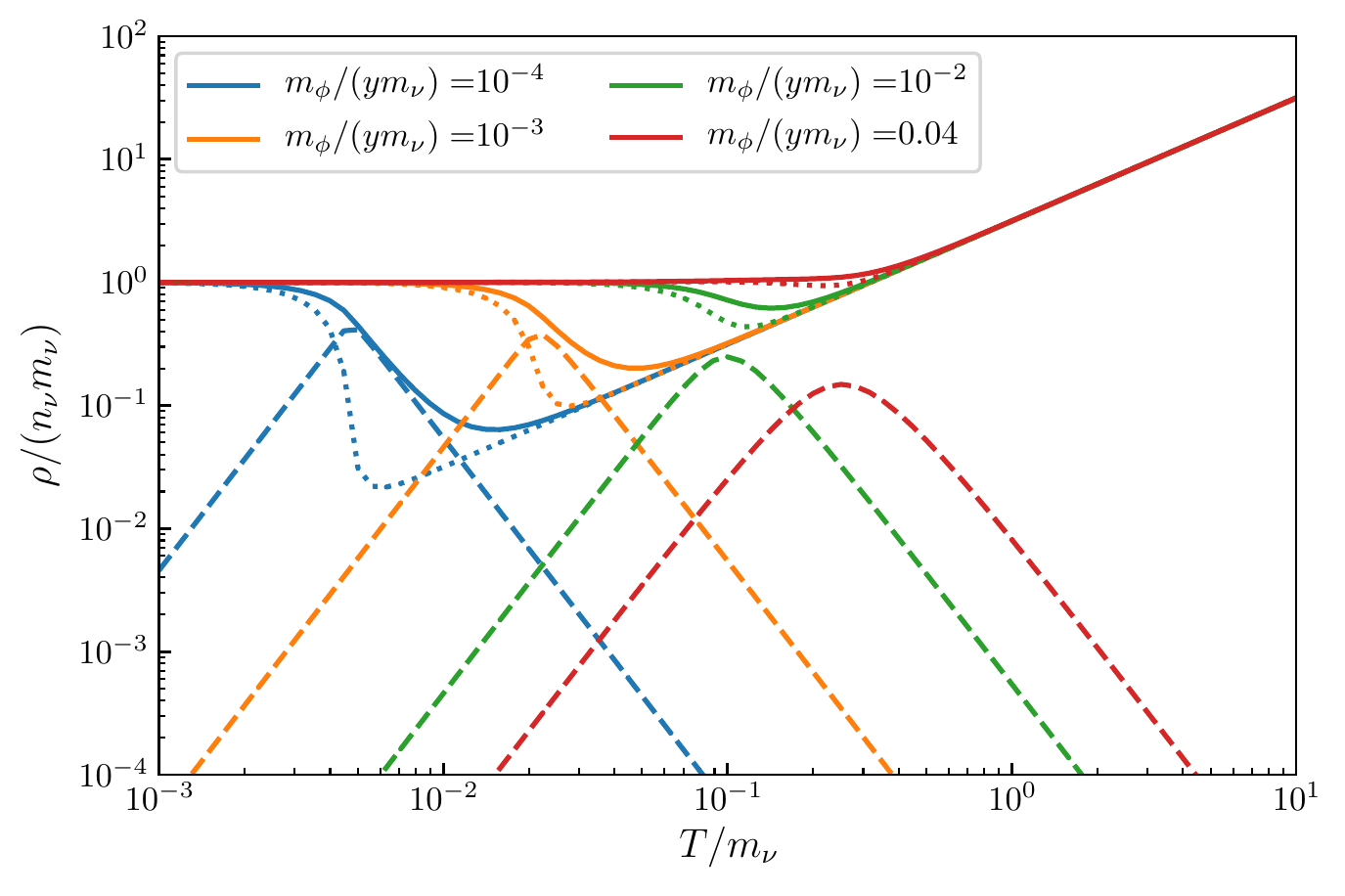}
\caption{\label{fig:eps-T} 
Dependence of comoving energy densities
% different energies  per neutrino in the units of $m_\nu$ 
on $T/m_\nu$ for different values of $m_\phi/(y m_\nu)$. The  
dashed lines corresponds to the energy of the scalar field,
the dotted lines to the energy of neutrinos, and the solid lines to the  
total energy. 
}
\end{figure}
%%%%%%%%%%%%%%%%%%%%%%%%%%%%%%%%%%%%%%%%%%%%%%%%%%%%%%%%%%%%%%%%%%%%%%%%%%%%%%%

In Fig.~\ref{fig:eps-T}, we show the dependence of $\epsilon_\phi/m_\nu$  on $T$ (dashed lines). 
As can be easily checked, the analytic expressions above agrees well with
the numerical results. 
% The energy density $\rho_\nu$ is constant at high temperatures. 

The energy density of neutrinos is given by 
$$
%\begin{equation}
\rho_\nu (\tilde{m}_{\nu}) = 
\frac{1}{2 \pi^2} \int_0^{\infty} dp~p^2 
\sqrt{p^2 + \tilde{m}_{\nu}^2} f(p,T). 
%\label{eq:rhonu}
%\end{equation} 
$$
It can be rewritten as 
\begin{equation}
\rho_\nu (\tilde{m}_{\nu}) = 
\frac{1}{2 \pi^2} 
\left[\int_0^{\infty} \frac{dp~p^4}{\sqrt{p^2 + \tilde{m}_{\nu}^2}} f(p,T) +  
\tilde{m}_{\nu}^2\int_0^{\infty} \frac{dp~p^2}{\sqrt{p^2 + \tilde{m}_{\nu}^2}} f(p,T)
\right].
\label{eq:rhonu1}
\end{equation}
Here the second integral coincides with integral 
in $\langle \bar{\nu}\nu \rangle$ and $\phi$.
%%The integrals can be computed explicitly for the degenerate distribution.  
The dependence of $\rho_\nu (\tilde{m}_{\nu})$ on $T$ can be found explicitly.

In the non-relativistic case we obtain from Eq.~(\ref{eq:rhonu1})
$$
%\begin{equation}
\rho_\nu^{\rm nr} =
\frac{1}{2\pi^2}
\left[\tilde{m}_{\nu} I_3 T^3+ \frac{1}{\tilde{m}_{\nu}} I_5 T^5
\right] \approx  \frac{1}{2\pi^2} m_\nu I_3 T^3, 
%\label{eq:rhonunon}
%\end{equation}
$$
and consequently,  $\rho_\nu^{\rm nr} = m_\nu n_\nu$, or 
$$
%\begin{equation}
\epsilon_\nu^{\rm nr} = m_\nu.
%\label{eq:relnunr}
%\end{equation}
$$
In the relativistic case, we have
$$
%\begin{equation}
\rho_\nu^{\rm rel} =
\frac{1}{2\pi^2}
\left[I_4 T^4 + \tilde{m}_{\nu}^2 I_2 T^2
\right] \approx  \frac{I_4}{2\pi^2} T^4   \approx \frac{7\pi^2}{240} T^4,
%\label{eq:rhonurel}
%\end{equation}
$$
and
$$
%\begin{equation}
\epsilon_\nu^{\rm rel} = \frac{I_4}{I_3} T \approx 3.15 T,
%\label{eq:relnunr}
%\end{equation}
$$
as expected. 
The dependence of $\epsilon_\nu$ on $T$ is shown in 
Fig.~\ref{fig:eps-T} (dotted lines).  
%%The key feature of the dependence is appearance 
%%of the dip at $T/m_\nu <   m_\nu/3$. 

Solid lines in Fig. \ref{fig:eps-T} correspond 
to the total energy (\ref{eq:rhotot}). 
Scalar contributions  shift the minimum of $\epsilon_{\rm tot}$ to slightly larger $T$
with respect to the minimum of $\epsilon_\nu$.

The key feature of the $\epsilon_{\rm tot}$  dependence  on $T$ is development of
the dip at some temperature below $m_{\nu}$,  denoted by $T_{\rm dip}$.

According to Fig.~\ref{fig:eps-T}, $\epsilon_{\rm tot}$ becomes smaller than $m_\nu$  when $T \simeq m_\nu/3$. 
%The dip has a minimum at $T_{\rm min}$. 
As $T$ decreases below $T_{\rm dip}$,   the energy  $\epsilon_{\rm tot}$ converges to $m_\nu$
from below. The dip essentially coincides
with the transition region for $\tilde{m}_{\nu}$ (Fig. \ref{fig:tm-T}). 

Without the Yukawa interaction (or if the Yukawa interaction is too weak, as illustrated by the red line for $S_{\phi}=0.04^{-2}=625$ 
in Fig.~\ref{fig:eps-T}), $\epsilon_{\rm tot}$
would decrease as $\epsilon_{\rm tot}\propto T$  in the relativistic  case; 
at $T \simeq  m_\nu/3$, the curve flattens and
$\epsilon_{\rm tot}$ converges to $m_\nu$.
For the entire range it would always be $\epsilon_{\rm tot} > m_\nu$ which corresponds to
unbounded neutrinos. 
For larger $S_{\phi}$ (other curves in Fig.~\ref{fig:eps-T}), the dip appears. 
The appearance of the dip with $\epsilon_{\rm tot} < m_\nu$
means that neutrinos can form bound systems with the average 
binding energy equal to $m_\nu - \epsilon_{\rm tot}$.
As the interaction strength increases, 
the dip shifts to lower temperatures
and becomes deeper.
Correspondingly the binding energy becomes stronger.

The critical value of the strength needed for existence of  the dip, 
$S_\phi^c$, is determined by the condition that 
temperature of beginning of deviation of $\epsilon_\nu^{\rm rel}$ from linear decrease
equals $T_{\rm rel} \approx m_\nu/3$. 
This temperature coincides with $T_{\rm rel}$ of deviation of $\tilde{m}_{\nu}$ 
from its $1/T^2$ behavior. 
The condition (\ref{eq:Trel1}) can be rewritten as 
$$
%\begin{equation}
S_\phi^{c} \approx 24 \left(\frac{m_\nu}{T_{\rm rel}}\right)^3.
%\end{equation}
$$
Then for $T_{\rm rel}/m_\nu \approx  1/3$ it gives $S_\phi^{c}  = 580$, which is 
close to $S_{\phi}=625$ for the red curve in Fig.~\ref{fig:eps-T}. In actual numerical calculations, we find that the dip appears roughly at $S_{\phi}^c\approx 700$.

\subsection{Fragmentation \label{sub:frag}}
%%%%%%%%%%%%%%%%%%%%%%%%%%%%%%%%%%%%%%%%%%%%%%%%%%%%%%%%%%%%%

Here we consider $S_\phi > S_\phi^c$ so that  the dip occurs. 
Above $T_{\rm dip}$, cosmic neutrinos evolve as a uniform 
background with decreasing density and effective temperature.
For $T < T_{\rm dip}$, further cosmological expansion assuming the uniform neutrino background would imply
that the energy of the system increases.
Since there is no injection of energy into the  $\nu$-$\phi$ system from outside,  
it is energetically profitable that the uniform neutrino 
background fragments into parts (clusters) 
with  the temperature  $ T \approx T_{\rm dip}$, which determines also the corresponding number density. Further evolution of fragments  will proceed, subject to the dynamics of $\nu$ clusters. 
The expansion of the Universe then increases the distance between fragments.

The highest temperature of fragmentation corresponds to  
$T_f \sim 0.3 m_\nu \sim 0.03$ eV. This temperature is achieved at redshift $z_f = 200$ when
the number density of neutrinos was
\begin{equation}
n_f = 9 \cdot 10^8 \, \, {\rm cm}^{-3}.
\label{eq:dens2}
\end{equation}
In the process of formation of final configuration
this density could further increase in the central parts of objects.
This value agrees well with the one in Sec.~\ref{sub:fragmentation}.

The biggest possible object after fragmentation, which satisfies
minimal (free) energy condition, would have 
the size  $D_0 \leq  1/2 D_U (z_f)$ in one dimension,
where $D_U (z_f)$ is the size of horizon in the epoch of fragmentation,
$D_U (0) = 4.2$ Gpc.  Therefore in three dimensions one would expect that
$2^3 = 8$ such objects could be formed, 
and for estimations we will use  $ \sim 10$ objects.
At $z_f = 200$ we find $D_0 = 10$ Mpc.
In the present epoch distance between the objects 
is $z_f$ times larger, that is $\sim 2000$ Mpc. 

%%We assume that the size of the object does not change 
%%substantially during further evolution. 
Using the number density density (\ref{eq:dens2}) and 
the radius $R_f = 5$ Mpc we obtain the   
total number of neutrinos in a cluster 
\begin{equation}
N_f = \frac{4\pi}{3} R_f^3 n_f = 1.2 \cdot 10^{85}. 
\label{eq:nf}
\end{equation}
Correspondingly, the mass is $M = 4 \cdot 10^{17} M_{\odot}$. 
%%This quantity does not depend on $z$. 
These biggest structures can be realized for 
$m_\phi =  4 \cdot 10^{-30} {\rm eV}$ and $y = 4 \cdot  10^{-27}$. 

%%It can be reduced if we increase the number of objects.\\ 

If $m_\phi$ and $y$ are much larger,  
these structures do not satisfy conditions for stable bound systems. 
The number of neutrinos  and radii should be much smaller. 
%(See an example for $y = 10^{20}$ in sect V.D). 
%%
%%Since  formation of $\nu$ clusters proceeds in the non-relativistic regime, the relativistic branch 
%%in Figs. \ref{fig:R-N}, \ref{fig:pf-N} does not work.  For instance,  for $S_{\phi}^{-1/2} = 0.01$
%%an additional bound $X < 10^2$ appears, which further restricts $N$ as 
%%$N  <  10^{62}$, while interval of radiuses do not change. 
%%
One can explore two possible ways of formation of small scale structures: 

1. The neutrino background first fragments into the biggest structures, 
thus satisfying the energy requirement. These structures are unstable 
and further fragment into smaller parts. The secondary fragmentation 
can be triggered by density fluctuation 
of the neutrino background itself or via gravitational interactions 
with already existing structures like the Dark Matter halos.

If the secondary fragmentation is slow, the overall final structure 
could consist of superclusters (originated from primary fragmentation)
and clusters within superclusters.

2. The uniform neutrino background may immediately fragment into small structures 
(close to final stable bound states)
with $R = {\cal O} (1/m_\phi)$.  These structures may interact among themselves.\\

As we established, for given $m_\phi$ and $y$ ({\rm i.e.} fixed strength) 
there is a range  of stable configurations with different $N$ and $R$.
E.g. for $S_{\phi}^{-1/2} = 0.01$,  $N$  can be within  4 orders of magnitude. 
What is the distribution of clusters with respect to $N$?  The answer depends on details of formation, 
in particular, on the surface effects. 
One can consider several options:

(i) a flat distribution within the allowed interval;

(ii) a peak at $N = X(n_{F0})/y^3$,  
where $n_{F0}$ is the density at the beginning of fragmentation  
which, in turn, is determined by the strength, {\it etc}.  \\

%%%%%%%%%%%%%%%%%%%%%%%%%%%%%%%%%%%%%%%%%%%%%%%%%%%%%%%%%%%%%%%%%%%%
Considering $T_{\min}$ as the temperature of the beginning of
fragmentation we can find using 
Figs. \ref{fig:tm-T} and \ref{fig:eps-T} the physical conditions 
of medium which determine the initial states of pre-clusters.
The kinetic energy per neutrino is (in the unit of $m_\nu$)
$\epsilon^{\rm kin} = 3.14T_{\min}/m_\nu$. The kinetic energy turns out to be much  
bigger than $\tilde {m} (T_{\min}) $:
$$
\epsilon^{\rm kin} \gg \tilde {m} (T_{\min}).
$$
So neutrinos are ultra-relativistic and therefore our usage of the relativistic 
distribution is justified. Correspondingly,  the total energy in neutrinos
nearly coincides with the kinetic energy: $\epsilon_\nu \approx \epsilon^{\rm kin}$.

Thus, for $S_{\phi}^{-1/2} = 10^{-3}$ we obtain  from Figs. \ref{fig:tm-T} and \ref{fig:eps-T}:
$T_{\min}/m_\nu = 5 \cdot 10^{-2}$,
$\epsilon^{\rm kin} = 0.15$, and $\tilde {m}/m_\nu = 0.01$.
The energy in the scalar field is substantially smaller
than in neutrinos:  $\epsilon_\phi = 0.05$. 
However it increases quickly  with the temperature further decreasing. 
The total energy in the system is $\epsilon_{\rm tot} = 0.2$. 

The number density of neutrinos is given by $n_0^{\min} = 2 \cdot 10^6$ cm$^{-3}$
which would correspond to the Fermi momentum $p_{F}/m_\nu = 0.1$.
This momentum  is smaller
than the kinetic energy, so the cooling is needed to reach
degenerate configuration.

%Notice that according to Fig. 7 the increase of
%$\epsilon_{tot}$ is due to increase of the energy
%of the scalar field. 
%%Due to fragmentation that would probably not possible.

%%In the infinite size uniform medium (as used in the figures)  

Further results and statements  should be considered as a possible trend since to construct 
the figures we used approximation of uniform infinite medium and 
relativistic distributions of neutrinos over momenta.  
Below $T_{\rm dip}$, the dynamics of finite size objects should be included.

As  $T$ deceases below $T_{\min}$,   the kinetic energy of neutrinos decreases while the effective mass $\tilde {m}$ increases. 
This means that the kinetic energy transforms into the 
increase of the mass as well as into the energy of scalar field.
Thus, for $T/m_\nu = 0.03$  we obtain  $\epsilon^{\rm kin} = 0.09$,
$\tilde {m}/m_\nu = 0.03$, {\it i.e.} the ratio $\epsilon^{\rm kin}
m_\nu / \tilde {m} = 3$;  for $T/m_\nu = 0.02$ the ratio becomes 0.3.

According to Figs.~\ref{fig:tm-T} and \ref{fig:eps-T},  
the minimum of total  neutrino energy, $\epsilon_\nu$, is achieved at 
a temperature lower than  $T_{\rm dip}$.  
In the minimum, for $S_{\phi}^{-1/2} = 10^{-3}$, $\epsilon_\nu = 0.1$ and $\tilde m/m_\nu = 0.07$, 
{\it i.e.} neutrinos become non-relativistic.

As $T$ further decreases,  both $\epsilon_\nu$ and $\tilde m$
quickly increase with $\tilde m \rightarrow \epsilon_\nu$. This indicates that the kinetic energy 
is transformed into $\tilde m$ as well as into $\epsilon_\phi$. 
One can guess that the phase transition ends when $\epsilon_\phi$ reaches maximum 
which corresponds to the maximal potential energy. This happens at a slightly lower  temperature
than the temperature of $\epsilon_\nu$ reaching its minimum.

%%Similar results can be obtained for other values of strength of interaction
%%Thus, for $S_{\phi}^{-1/2} = 10^{-2}$ we have  $T_{min}/m_\nu = 0.15$,
%%$\epsilon^{kin} = 0.47$ and $\tilde {m}/m_\nu = 0.1$.
%%For scalar field: $\epsilon_\phi = 0.14$.
%%The Fermi momentum $p_{F}/m_\nu = 0.3$ turns out to be
%%closer to $\epsilon^{kin}$ then in the case of bigger strength.

The period of (primary) fragmentation can be estimated as the   
time interval which corresponds to the width of the dip: 
%%$$
%%\Delta t \sim \left(\frac{dT}{dt}\right)^{-1} \Delta T. 
%%$$
%%This gives $\Delta t \sim t$,  e.g. the age of the Universe in epoch 
%%of fragmentation.{\color{blue} [I think it is more reasonable to consider 
%%that the fragmentation time is 
\begin{equation}
\Delta t = t(T_{1})-t(T_{2}), \label{eq:t_frag}
\end{equation}
where   $t\approx\frac{2}{3H(T)}$ (for matter 
dominated era) and $H$ is roughly proportional to $T^2$. The specific form  can be determined by 
$H^{2}=8\pi\rho_{{\rm total}}/(3m_{{\rm pl}}^{2})$ where $\rho_{{\rm total}}$
is the total energy density of the Universe and $m_{{\rm pl}}$ is
the Planck mass. The two temperatures $T_1$ and $T_2$ correspond to  the maximum of the dashed curve in Fig.~\ref{fig:eps-T} and the minimum of the solid curve respectively.  
This  gives e.g.  $\Delta t = 0.014 \tau_U$ for $S_\phi^{-1/2} = 10^{-3}$ and  
$\Delta t = 0.24 \tau_U$ for $S_\phi^{-1/2} = 10^{-4}$, where $\tau_U$ is the present age of the Universe. 
Note that the estimations based on Fig.~\ref{fig:eps-T} may not be reliable 
since in the  region of temperatures below $T_{\rm dip}$ 
neutrinos become non-relativistic and the dynamics of clusters starts to dominate.\\

%%$t\sim1/H$ ($t\approx\frac{1}{2H(T)}$
%%	for radiation dominated era and 

%%During fragmentation the energy balance 
%%(the kinetic energy and level of degeneracy)  
%%is settled down in the clusters.

%
%[[Fragmentation starts at $T \sim m_\nu/3$ when deviation from linear dependence 
%$\epsilon_{\rm tot} \propto T$ or at $T_{\rm min}$ ? ]] \\

According fo Fig.~\ref{fig:eps-T},  
$T_{\rm dip}$ decreases as   
$S_\phi$ increases. 
This means that for larger $S_{\phi}$, fragmentation starts at later epochs and lower densities. 
For $S_{\phi}^{-1/2} =(0.12,~  10^{-2}, ~10^{-3})$, 
 we find $T_{\rm min}/m_\nu = (0.3,~ 0.12, ~0.04)$,  
$n_0 = (9 \cdot 10^8, ~ 5.7 \cdot 10^7, ~ 2.1 \cdot 10^6)$ cm$^{-3}$
and the redshift at fragmentation 
$(z_f + 1) =  (200, ~ 80, ~ 27)$. 
With the increase of strength, $n_0$ and $p_{F0}$ decrease. 
Then according to Figs.~\ref{fig:pf-N} and \ref{fig:R-N}, $N$ and $R$ decrease. \\

Let us underline that 
results on the very fact of formation based on the energy consideration  
and properties of final configurations are robust. 
Details and features of fragmentation (the way of fragmentation, the distribution of clusters as a function of $N$, 
the level of degeneracy in the final states of bound systems) require 
further studies, and in general, extensive  numerical simulations.

\subsection{Virialization \label{sub:vir} } 
%%%%%%%%%%%%%%%%%%%%%%%%%%%%%%%%%%%%%%%%%%%%%%%%%%%%%%%%%%%%%%%%%%%%%%%%%

%%For $S_{\phi} \gtrsim 600$, we believe that cosmic neutrinos have experienced 
%%the phase transition and fragmentation. 
%%After fragmentation,  the homogeneous cosmic neutrino background breaks 
%%into fragments which will then go through virialization to form spherically symmetric neutrino clusters.

%%For $70 \lesssim S \lesssim 600$,  fragmentation is absent but the growth of 
%%local over-density, when becomes non-linear,  might go also through virialization 
%%to form neutrino clusters. similar to the formation of DM halos. 

%%Here we focus on the former case since whether local over-density 
%%of neutrinos can grow to virialized neutrino clusters remains 
%%an open question which probably can only be addressed by simulations.

%%Virialization of DM halos is purely a gravitational process, 
%%which re-distributes the kinetic and potential energies so that the virial equilibrium is reached,  
%%without energy loss. Here we apply virialization to the Yukawa force which 

In the range of $S_{\phi} = 70 - 700$, the dip disappears and the phase transition is absent. Such interaction strengths still allow for stable bound systems of neutrinos.
In this case, $\nu$ clusters might be formed from  the growth of local over-density to the non-linear regime and then go also through virialization, similar to the formation of dark matter halos.

Quantitatively, results for Yukawa forces (if $m_{\phi}$
is sufficiently small) can be obtained from the gravitational 
results by replacing 
\begin{equation}
\frac{GM^{2}}{r}\rightarrow\frac{(yN)^{2}}{4\pi r}\thinspace,
\label{eq:-107}
\end{equation}
where $G$ is the gravitational constant and $M = m_\nu N$ is the total mass. 
The key difference from the gravitational case is that the strength of
Yukawa interaction is many orders of magnitude
larger than the strength of gravity.
%%As a result, the scalar field itself is strong and plays an important role 
%%in the dynamics of $\nu$-clusters. 
%As we saw, the contribution
%of the field to the neutrino mass can be comparable to the mass: 
%$V \sim m_\nu$, so that the effective mass can differ 
%strongly from the vacuum mass (in contrast to the WIMP case).\\

With this analogy we can consider the following picture of structure formation.

1. The initial state is the neutrino background with
some primordial or initial density perturbations
(primordial clouds) with $\Delta n_p/ n_p \ll 1$
and sizes $R_p$. These perturbations may be related to
the density  perturbation of DM in the Universe, since neutrino
structures are formed latter.

2. As the Universe expands, the size of clouds and
distance between them increase simultaneously as
the scale factor $R \propto d \propto a$. Correspondingly, the
density decreases as $a^{-3}$ and $\Delta n/ n$ is constant.
This is the linear regime.

3. At certain epoch, 
%$a_{NL}$ 
when $T < 0.3 m_\nu$
the evolution becomes non-linear: the 
attraction becomes more efficient  than expansion and the increase
of the size of clouds becomes slower than the expansion of the
Universe. This happens provided that the potential energy
of the cloud is substantially larger
than the total kinetic energy.
Then the distance between two clouds increases faster that size of clouds.
Also the decrease of the density slows down.
4. At some point (for which we denote the scale factor by $a_{\max}$) the density stops decreasing,
and after that ($a > a_{\max}$),  the system starts collapsing.
In this way a bound system is formed and its evolution is largely
determined by internal factors. The distance between two
clouds continues increasing.
%[[remove]]

Since the system is collisionless and there is no energy loss
(emission of particles or and  waves) the contraction
has a character of virialization\footnote{A system of collisionless 
({\it i.e.} no additional interactions except
for gravity) particles can collapse from a large cold distribution 
(or more strictly, a distribution with low mean kinetic energy since 
it is usually not a thermal distribution) 
 to a smaller and hotter one. This processes is known as virialization.}, 
which starts from the stage that kinetic energy of the neutrino system  
becomes significantly smaller than the potential energy: 
\begin{equation}
\frac{E_K}{|E_V|} \equiv \xi_E \ll 1.
\label{eq:befr}
\end{equation} 
During virialization,
the kinetic energy increases (the effective
$T$ increases, although the distribution may not be thermal).
The process stops at $a = a_{\rm vir}$ when  {\it virial equilibrium}
is achieved, which also corresponds to hydrostatic equilibrium.
The radius decreases by a factor of 2 so that the density
increases by a factor of 8. Approaching the equilibrium the cluster may oscillate. 
At this point, the formation of the cluster is accomplished.

In what follows we will make some estimations and
check how this picture matches the final configurations of
clusters we obtained before. We will discuss bounds
on initial parameters of the perturbations ,
formation procedure, epochs of
different phases, etc, which follow from this matching.

The total energy is conserved during virialization. When the system reaches  
virial equilibrium, the virial theorem implies that 
\begin{equation} 
2E_K^{\rm vir} + E_V^{\rm vir} = 0 ~~~{\rm or}~~~
\frac{E_{K}^{\rm vir}}{E_{V}^{\rm vir}} = \frac{1}{2}\thinspace,
\label{eq:-122}
\end{equation}
Here $E_{K}^{\rm vir}$ and $E_{V}^{\rm vir}$ are the total
kinetic and potential energies after virialization. Thus,  the ratio of energies increases.

Although quantitative studies of virialization involve $N$-body
simulation, the orders of magnitude of various characteristics 
can be obtained from simplified considerations. For gravitational virialization,  
it is known that\footnote{See, e.g., \url{http://www.astro.yale.edu/vdbosch/astro610_lecture8.pdf}
or Ref.~\cite{Mota:2004pa}.}  the virialization
time (roughly the time of collapse) driven by gravity equals  
\begin{equation}
t_{{\rm vir}} \simeq 2\pi GM\left|\frac{M}{2(E_{K}+E_{V})}\right|^{3/2},
\label{eq:-108}
\end{equation}
where $E_{K}$ and $E_{V}$ are the total kinetic and potential energies
($E_{K}+E_{V}<0$) of the system.

%%In analogy with gravity due to the Yukawa 
%%attraction the system starts to contract and its virialization (the  $m_{\phi}$-driven virialization) occurs. 
The time of $m_{\phi}$-driven virialization can be obtained from Eq.~\eqref{eq:-108}
by making the  replacement~\eqref{eq:-107}:  
\begin{equation}
t_{{\rm vir}} \simeq \frac{\sqrt{m_\nu}N^{5/2}y^{2}}{4\sqrt{2}|E_{K}+E_{V}|^{3/2}}\thinspace.  
\label{eq:-116}
\end{equation}

%%Let us estimate the energies $E_{K}$ and  $E_{V}$. 

The contraction stops when virial equilibrium 
is achieved and we assume 
that the equilibrium has the same form as 
in the gravity case~\eqref{eq:-122}. 
Since the total energy is conserved during virialization, 
$$
%\begin{equation}
E_{K}^{{\rm vir}}+E_{V}^{{\rm vir}}=E_{K}+E_{V}\thinspace,
%\label{eq:-123}
%\end{equation}
$$
we obtain $E_{K}^{\rm vir}=|E_{K}+E_{V}|$ and $E_{V}^{\rm vir}=-2|E_{K}+E_{V}|$.

The total energy in the beginning of virialization can be written as 
$$
%\begin{equation}
| E_K + E_V| =  |E_V| (1 - \xi_E) = \frac{2}{3} |E_V|,  
%\end{equation}
$$
where in the last equality we have taken $\xi_E = 1/3$. 
Inserting this total energy with $|E_V|$ given in  \eqref{eq:20210526-4}
into  Eq.~\eqref{eq:-116}, we obtain  the time of virialization 
$$
%\begin{equation}
t_{{\rm vir}}  \simeq  
\sqrt{3} \left(\frac{5}{2} \right)^{3/2} 
\frac{\pi^2}{I_3^{1/2}}\frac{1}{y m_{\nu}} \left(\frac{m_\nu}{T}\right)^{3/2}, 
%\label{eq:20210526-5}
%\end{equation}
$$
which does not depend on $N$. Numerically, we get
\begin{equation}
t_{{\rm vir}}  \simeq 3.4 \cdot 10^{-6} {\rm sec}
\left(\frac{10^{-7}}{y }\right)
\left(\frac{0.1 {\rm eV}}{m_\nu}\right)
\left(\frac{m_\nu}{T}\right)^{3/2},
\label{eq:tvirn}
\end{equation} 
which implies that for $T/m_\nu = 1$, the virialization takes
$t_{{\rm vir}}  \simeq 3.4 \cdot 10^{-6}$ sec.

As aforementioned, after the virialization process,  the radius decreases 
by a factor of 2 and  correspondingly, the density increases by a factor of 8.  
Hence in the example we discussed at the end of the previous subsection, 
this gives  $R_{\rm vir} = 1.3$ km,  and $n_{\rm vir} = 7 \cdot  10^{10}$ cm$^{-3}$.  

As a result of virialization, the distribution  reaches virial equilibrium
with the virialized energies given by 
\begin{equation}
E_{K}^{({\rm vir})}=-\frac{1}{2}E_{V}^{({\rm vir})} = E_{K}^{\rm in}  + E_{V}^{\rm in}. 
%%\approx\frac{4.1m_{\nu}}{y^{3}}\left(\frac{p_{F0}}{m_{\nu}}\right)^{7/2}\text{eV}\thinspace.
\label{eq:-124}
\end{equation}
A halo of neutrinos is formed. In
the absence of energy loss, its number density remains constant without
being diluted by $a^{-3}$ ($a$ is the scale factor) 
due to the cosmological expansion.  That is, a halo of high-density
cosmic neutrinos would be   frozen till today. 

Let us estimate temperatures of the $\nu$-clusters at different phases of
formation. We can assume that deviation from the linear regime
occurs at $T \sim m_\nu$.  Also we assume that before collapse the state of the system can
be approximately described
by  thermal equilibrium and the border effects can be neglected.
Therefore, we can use results obtained in sect. IV C.
The expression for temperature (\ref{eq:eratio2}) can be rewritten as
$$
\frac{T}{m_\nu} = 5.3 \cdot 10^{-3} (y^3 N)^{2/3} \xi_E.
$$
Let us take $y^3 N = 1$ (see implications in the Fig. 4, 5) then
$T/m_\nu = 1$ (the scale factor $a_{NL}$) corresponds to $\xi_E = 189$.
For $\xi_E = 1/3$ we obtain $T/m_\nu = 1.8 \cdot 10^{-3}$. 
For  $y^3 N = 32$,  $T/m_\nu = 1$ corresponds to
$\xi_E = 19$ and $\xi_E = 1/3$ is achieved at
$T/m_\nu = 1.8 \cdot 10^{-2}$. If $y^3 N = 10^3$,
$T/m_\nu = 1$ is realized at $\xi_E = 1.9$, and
$\xi_E = 1/3$ at $T/m_\nu = 0.18$.

%%According to  Eq.~\eqref{eq:eratio2}, systems with larger 
%%numbers of neutrinos start 
%%to collapse at higher temperatures ($\eta$) and therefore earlier. 
%%Probably maximal $T$ corresponds to $T/m_\nu = 1$.
%%For $\eta = 1$ we obtain $N = 1.3 \cdot 10^{25}$,
%%consequently, the initial density equals 
%%$n_{\rm in} = 1.1\cdot  10^{10}$ cm$^{-3}$, 
%%and the initial radius $R_{\rm in} = 0.65$ km. 

%%[[ Neutrinos become non-relativistic at $z = 1000$, 
%%when their number  density was $10^{11}$ cm$^{-3}$   ]]

\subsection{Neutrino structure of the Universe}
%%%%%%%%%%%%%%%%%%%%%%%%%%%%%%%%%%%%%%%%%%%%%%%%%%%%%%%%%%%%%%%%%%%%%%%%%%

The formation of the neutrino clusters leads to neutrino 
structure in the Universe with overdense areas and voids. 
For simplicity we consider that all the clouds have the same number of neutrinos $N$. \\

In the present epoch the distance between neutrinos without 
clustering equals
$$
%\begin{equation}
d_0 = (n_\nu^0)^{1/3} = 0.2 ~ {\rm cm}.
%\end{equation}
$$
If neutrinos cluster into clouds with $N$ neutrino in each,
the distance between (centers of) these clusters is
$$
%\begin{equation}
d = d_0 N^{1/3}.
%\end{equation}
$$
Taking, e.g., $N = 6 \cdot 10^{22}$, we obtain $d = 78$ km 
which is much larger than the radius of each cloud, $R = 0.62$ km given by Eq.~\eqref{eq:-179}. \\

The existence of such a structure could have an impact on the direct detection 
of relic neutrinos in future experiments such as the PTOLEMY~\cite{PTOLEMY:2018jst,PTOLEMY:2019hkd}. 
The result also depends on the motion of the clusters with respect to the Earth.
The motion of the Earth and the solar system will average the effect of these small structures.  
According to Eq.~(\ref{eq:20210531}), in the non-relativistic regime the radius of  cluster
decreases with increase of $N$ as $N^{-1/3}$.
Therefore the ratio of spacing (distance) and radius increases
as $\propto N^{2/3}$:
$$
%\begin{equation}
\frac{d}{R} \approx  1.1 \cdot 10^{-2}~ d_0 m_\nu y^2 N^{2/3}.
%\label{eq:drratio}
%\end{equation}
$$
Taking the above example ($d = 78$ km, $R=0.62$ km), we have $d/R = 126$. So, voids occupy most of the space.
With decrease of $y$, the radius of cluster increases and $N$ also increases
(for fixed maximal density).  As a result,  $d/R =$ const
 which is determined by the neutrino mass.\\

% In the relativistic case  according to the Table I, the radius increases as
% $N^{1/3}$ [[not exactly]],  and therefore with increase of $N$ 
% $d/R$ approaches constant  [[dependence on $y$?]]. 

Let us estimate how small scale clusters may affect the direct detection
of relic neutrinos assuming that clouds reached their
final configurations.  The distance needed for
a detector to travel through to meet a $\nu$-cluster, in average,  can be estimated as
$L = (\sigma_c n_c)^{-1}$,  where $\sigma_c \approx \pi R^2$
is the cross section of the cluster and $n_c = d^{-3}$
is the number density of clusters. Consequently,
\begin{equation}
L = \frac{d}{\pi} \left(\frac{d}{R}\right)^2.
\label{eq:lexpr}
\end{equation}
Numerically, in our example, we obtain from (\ref{eq:lexpr}) $L = 4 \cdot 10^5$ km.
Since the circumference of the Earth is  $4 \cdot 10^4$ km,
the Earth detector will cross the cluster every 10 days in average.
Here we have not taken into account motion of clusters.

If the time of observation is much longer than 10 days, the
total rate of events should be the same  with and without clustering
of neutrinos. This can be seen by noticing that the interaction rate 
when the detector is in the cloud is enhanced by a factor of $ n_c/ n_0$ 
while the rate of the detector being in a sphere of neutrinos 
is reduced by a factor of $ n_c/ n_0$. Hence the time-averaged 
result will be the same as in the non-clustered case.
% case.
% Indeed, 
% let us consider
%  $\sigma_{\nu D } n$
% where $\sigma_{\nu D }$ is the cross section of a single neutrino with a scatter (e.g. tritium nucleus in the PTOLEMY experiment)
% and $n$ is the number of scatterers,  which determine
% the rate of events. We compare this product
% for both cases. In the case of unclustered neutrinos
% $\sigma_{\nu D } = \sigma_0 (\nu)$\,\textemdash\,the cross section of neutrino interaction with
% particles in a detector and $n = n_0$ is the number density of
% relic neutrinos, so that the product equals $\sigma_0 (\nu)
% n_0$. In the case of clustering, $n = n_c = n_0/N $ and
% $\sigma_{\nu D } = \sigma_0 (\nu) N$ since in each scatterer we have
% $N$ neutrinos. The product will be the same as in the non-clustered
% case.\\

%%[[inderplay with gravitational clustering \textemdash for future]]

If the clusters are very big 
(e.g. comparable with the solar system or more) which is realized for
very small  coupling  constant $y$, 
the rate will be modified if the detector is in a void
or in the overdense region of cluster.

\section{Conclusions \label{sec:Conclusion}}
%%%%%%%%%%%%%%%%%%%%%%%%%%%%%%%%%%%%%%%%%%%%%%%%%%%%%%%%%%%%%%%%%%%%%%%%%

Yukawa interactions of neutrinos with a new light scalar boson $\phi$ are widely
considered  recently. One  of the generic consequences
of such interactions is the potential existence of stable bound states
and bound systems of many neutrinos ($\nu$-clusters). 
Our study addresses these questions: whether and when do bound systems exist? what are the characteristic parameters
of these systems and how are they determined? what are the observational (cosmological, laboratory, {\it etc.}) consequences of $\nu$-clusters?\\

We find that for two-neutrino bound states 
the Yukawa coupling $y$ and the mass of the scalar $m_{\phi}$ should satisfy 
the bound $\lambda \equiv y^2 m_\nu /( 8\pi m_\phi) >0.84$.
Eigenstates and eigenvalues of  a
$2\nu$ bound system are obtained by solving 
the Schr\"odinger equation numerically.
Analytic results (including the critical value
of $\lambda$ as well as the binding energy and the radius of the system) can be obtained approximately using the variation principle.
%The variation principle can
%be used to obtain approximate results, including the critical value
%of $\lambda$ as well as the binding energy and the radius of the system. 
%%---see Eqs.~\eqref{eq:-152}, \eqref{eq:-144-1} and \eqref{eq:-145-1}.
The binding energy in a two-neutrino state is always small
compared to the neutrino mass due to smallness of $y$.   
For allowed values of couplings $y$ and masses of scalar $m_\phi$,
the bound state of two neutrinos would have the size greater than 
$10^{12}$ cm. Bound states of sub-cm size are possible for 10 keV sterile neutrinos with coupling $y > 10^{-4}$. 
As long as the Yukawa coupling is below the perturbativity bound, 
neutrinos in a two-neutrino bound state are  non-relativistic. 
%%, which justifies the use of
%%the Schr\"odinger equation in this case.\\

For $N$-neutrino bound systems, the ground state corresponds  
to a system of degenerate Fermi gas in which  the Yukawa attraction is balanced
by the Fermi gas pressure. 
For non-relativistic neutrinos 
the density distribution in the $\nu$-cluster is described by 
the Lane-Emden equation. We re-derive this equation  
and solve it numerically. 
As a feature of the Lane-Emden equation, the solution always has a finite radius $R$. 
The radius $R$ decreases as $N$ increases in the non-relativistic regime.

Unlike $2\nu$ bound states, the  $N$-neutrino bound system can enter the
relativistic regime with sufficiently large $N$. 
We have derived a set of equations to describe the  $\nu$-clusters  
in relativistic regime.  
In the non-relativistic limit they reduce to the  L-E equation.
In the relativistic
regime, the Yukawa attraction acquires the $m_\nu/E$ suppression which  
leads to a number of interesting consequences: 

\begin{itemize}

\item
The effective number density of neutrinos, $\tilde{n}$,   
is suppressed. It deviates from the conventional number density $n$ significantly when
the neutrino momentum is large;

\item
The Fermi degenerate pressure is  able to balance the attraction for any $N$, 
thus preventing the  collapse of the system in contrast to gravity;

\item
The radius of system increases with $N$, which implies that 
relativistic $N$-body bound states cannot be much more compact 
than the non-relativistic ones.

\item As $N$ increases the radius of a $\nu$-cluster first decreases
(in the non-relativistic regime) and then increases (in the relativistic
regime), attaining a minimum when $p_{F0}$ is comparable to the effective neutrino mass.

\item There is a maximal central density $\sim 10^9$ cm$^{-3}$, 
which is determined by the neutrino mass. 

\item For a given $m_\phi$ there is a minimal value of $Ny^3$
for which stable configurations can be formed.
This minimal value increases as the interaction strength   $S_{\phi} \equiv \frac{y^2 m_\nu^2}{m_\phi^2} $ decreases. 

\item For a given interaction strength, there is a minimal radius of the cluster. 

\item The minimal interaction strength (maximal $m_\phi$ for fixed $y$) required to bind neutrinos in a $\nu$-cluster is $S_{\phi} \gtrsim 70$.

\end{itemize}

We have investigated possible formation of the $\nu$-clusters from the uniform relic neutrino background as the Universe cools down.
\begin{itemize}
\item For the interaction strength $S_{\phi} \gtrsim 700$, a dip (see Fig.~\ref{fig:eps-T}) appears in the evolution of the total energy of the $\nu$-$\phi$ 
system  at some temperature below $m_{\nu}$. In this case, the formation of
neutrino structures has a character of phase transition. The dip leads to the instability of the uniform $\nu$ background, causing its fragmentation when $T$ further decreases. 
The typical size of final fragments is $R \sim  (0.1 - 10)/ m_\phi$.
A detailed picture of fragmentation should be obtained
by numerical simulations, which is beyond the scope of this work.

\item For $70 \lesssim  S_{\phi} \lesssim 700$,  the dip is absent, and hence 
fragmentation does not occur, though stable neutrino clusters can exist. 
In this case  neutrino clusters might be formed via the growth of local density perturbations  
followed by virialization, in a way similar to the formation of DM halos, though the energy loss of neutrino clusters due to 
$\phi$ emission and neutrino annihilation  is negligible. In this case, a new cooling mechanism would be required.

\item For $S_{\phi} \lesssim 70$, stable neutrino clusters cannot exist.

\end{itemize}

%These clusters, once formed, may or may not be in degenerate configurations, depending on whether the kinetic energy is fully transformed into effective neutrino masses at the phase transition.

%The energy loss of neutrino clusters due to 
%$\phi$ emission as well as due to neutrino annihilation 
%is negligible.  Therefore transition to degenerate configuration 
%could be due to transformation of the kinetic energy into 
%growing effective neutrino mass. \\

If clusters are formed at redshift $z_f$ (which can be as large as 200) and retain the same size until today,
the neutrino structure of the universe would show up as $\nu$-clusters which are $z_f^3$ times denser than the homogeneous neutrino background and 
voids which are $z_f$ times larger than clusters. 
This leads to important implications for the detection of 
relic neutrinos. 
%% and might possibly affect the observation of the future PTOLEMY experiment.

While conclusion of formation of the neutrino clusters via fragmentation at $S_{\phi} \gtrsim 700$  is 
robust, details of fragmentation, the distribution of clusters with respect to $N$, and the evolution of clusters 
require further studies.

\appendix

\section{Calculation of $\langle\overline{\psi}\psi\rangle$ and $\langle\psi^{\dagger}\psi\rangle$
\label{sec:QFT}}
%%%%%%%%%%%%%%%%%%%%%%%%%%%%%%%%%%%%%%%%%%%%%%%%%%%%%%%%%%%%%%%%%%%%%%%%%%%%%%%%%%%%%%%%%%%%%%%%%%%

\noindent The quantization of $\psi$ is formulated as follows:
\begin{equation}
\psi(x)=\int\frac{d^{3}\mathbf{p}}{(2\pi)^{3}}\frac{1}{\sqrt{2E_{\mathbf{p}}}}
\sum_{s}\left[a_{\mathbf{p}}^{s}u^{s}(p)e^{-ip\cdot x}+b_{\mathbf{p}}^{s\dagger}v^{s}(p)e^{ip\cdot x}\right].
\label{eq:-41}
\end{equation}
Now consider a single particle state which has been modulated by a
wave-packet function $w(\mathbf{p})$:
\begin{equation}
|w\rangle=\int\frac{d^{3}\mathbf{p}}{(2\pi)^{3}}\frac{w(\mathbf{p})}{\sqrt{2E_{\mathbf{p}}}}|\mathbf{p},s\rangle,\ 
|\mathbf{p},s\rangle=\sqrt{2E_{\mathbf{p}}}a_{\mathbf{p}}^{s\dagger}|0\rangle,\label{eq:-42}
\end{equation}
so that it is localized in both the coordinate space and the momentum space.
More explicitly, for
\begin{equation}
\Psi(\mathbf{x})\equiv\int\frac{d^{3}\mathbf{p}}{(2\pi)^{3}}\frac{w(\mathbf{p})}{\sqrt{2E_{\mathbf{p}}}}u^{s}(p) 
e^{i\mathbf{p}\cdot\mathbf{x}},\label{eq:-43}
\end{equation}
 the probability of this particle appearing at position $\mathbf{x}$
is ${\cal P}(\mathbf{x})\equiv|\Psi(\mathbf{x})|^{2}$ and we have
chosen $w(\mathbf{p})$ in a way that 
\begin{equation}
{\cal P}(\mathbf{x})\approx\begin{cases}
{\rm finite} & \text{\ensuremath{\mathbf{x}}}\in[\mathbf{x}_{0}-\Delta\mathbf{x},\ \mathbf{x}_{0}+\Delta\mathbf{x}]\\
0 & {\rm otherwise}
\end{cases},\ w(\mathbf{p})\approx\begin{cases}
{\rm finite} & \mathbf{p}\in[\mathbf{p}_{0}-\Delta\mathbf{p},\ \mathbf{p}_{0}+\Delta\mathbf{p}]\\
0 & {\rm otherwise}
\end{cases},\label{eq:-44}
\end{equation}
which is always possible as long as $\Delta\mathbf{x}\Delta\mathbf{p}\gg1$. 
In this way, we can say that the particle is located approximately at $\mathbf{x}_{0}$ with an approximate momentum $\mathbf{p}_{0}$.

% Here  $w(\mathbf{p})$ is normalized by
Next, let us consider the normalization:
\begin{equation}
1=\int{\cal P}(\mathbf{x})d^{3}\mathbf{x}.\label{eq:-47}
\end{equation}
Substituting Eq.~\eqref{eq:-43} in ${\cal P}(\mathbf{x})\equiv|\Psi(\mathbf{x})|^{2}$
and then in Eq.~\eqref{eq:-47}, we have
\begin{align}
1 & =\int d^{3}x\frac{d^{3}\mathbf{\mathbf{k}}}{(2\pi)^{3}}\frac{d^{3}\mathbf{p}}{(2\pi)^{3}}
\frac{w^{*}(\mathbf{k})w(\mathbf{p})}{\sqrt{2E_{\mathbf{k}}}\sqrt{2E_{\mathbf{p}}}}u^{s\dagger}(k)u^{s}(p)
e^{i(\mathbf{p}-\mathbf{k})\cdot\mathbf{x}}\nonumber \\
 & =\int\frac{d^{3}\mathbf{k}}{(2\pi)^{3}}\frac{d^{3}\mathbf{p}}{(2\pi)^{3}}(2\pi)^{3}\delta^{3}(\mathbf{p}-\mathbf{k})
\frac{w^{*}(\mathbf{k})w(\mathbf{p})}{\sqrt{2E_{\mathbf{k}}}\sqrt{2E_{\mathbf{p}}}}u^{s\dagger}(k)u^{s}(p)\nonumber \\
 & =\int\frac{d^{3}\mathbf{p}}{(2\pi)^{3}}|w(\mathbf{p})|^{2},\label{eq:ww1}
\end{align}
where it is useful to note that 
\begin{equation}
u^{s\dagger}(p)u^{s}(p)=2E_{\mathbf{p}},\ \overline{u^{s}(p)}u^{s}(p)=2m_{\psi}.\label{eq:-45}
\end{equation}

With the above setup, we can evaluate the mean value of $\overline{\psi}\psi$
and $\psi^{\dagger}\psi$ in the presence of the single particle state
$|w\rangle$. First, note that
\begin{align}
a_{\mathbf{p}}^{s}|w\rangle & =a_{\mathbf{p}}^{s}\int\frac{d^{3}p'}{(2\pi)^{3}}w(\mathbf{p}')a_{\mathbf{p}'}^{s\dagger}|0\rangle\nonumber \\
 & =\int d^{3}\mathbf{p}'\thinspace w(\mathbf{p}')\delta^{3}(\mathbf{p}-\mathbf{p}')|0\rangle\nonumber \\
 & =w(\mathbf{p})|0\rangle.\label{eq:-46}
\end{align}
Hence for $\langle w|\psi^{\dagger}\psi|w\rangle$, using Eq.~\eqref{eq:-41}
and Eq.~\eqref{eq:-46}, we have
\begin{align}
\langle w|\psi^{\dagger}(x)\psi(x)|w\rangle & =\int\frac{d^{3}\mathbf{k}}{(2\pi)^{3}}\frac{d^{3}\mathbf{p}}{(2\pi)^{3}}
\frac{1}{\sqrt{2E_{\mathbf{k}}}\sqrt{2E_{\mathbf{p}}}}\langle w|a_{\mathbf{k}}^{s\dagger}u^{s\dagger}(k)e^{ik\cdot x}a_{\mathbf{p}}^{s}
u^{s}(p)e^{-ip\cdot x}|w\rangle\nonumber \\
 & =\int\frac{d^{3}\mathbf{k}}{(2\pi)^{3}}\frac{d^{3}\mathbf{p}}{(2\pi)^{3}}
\frac{u^{s\dagger}(k)u^{s}(p)}{\sqrt{2E_{\mathbf{k}}}\sqrt{2E_{\mathbf{p}}}}
e^{i(k-p)\cdot x}\langle0|w(\mathbf{k})^{*}w(\mathbf{p})|0\rangle\nonumber \\
 & =\Psi^{\dagger}(x)\Psi(x),\label{eq:-50}
\end{align}
where the spacetime-generalized wavefunction, $\Psi(x)$, is defined similar to Eq.~\eqref{eq:-43} except
that $e^{i\mathbf{p}\cdot\mathbf{x}}$ is replaced by $e^{-ip\cdot x}$. 

Similarly, for $\langle w|\overline{\psi}\psi|w\rangle$, we obtain
\begin{equation}
\langle w|\overline{\psi}(x)\psi(x)|w\rangle=\overline{\Psi}(x)\Psi(x).\label{eq:-49}
\end{equation}

As has been formulated in Eq.~\eqref{eq:-44}, the particle at $t=0$
is localized in the region $\mathbf{x}\in[\mathbf{x}_{0}-\Delta\mathbf{x},\ \mathbf{x}_{0}+\Delta\mathbf{x}]$.
So it is useful to inspect the average value of $\langle w|\overline{\psi}\psi|w\rangle$
and $\langle w|\psi^{\dagger}\psi|w\rangle$. At $t=0$, we have 
\begin{align}
\langle w|\psi^{\dagger}\psi|w\rangle\xrightarrow{t=0,\ \mathbf{x}\ {\rm averaged\ locally}} & 
\frac{1}{\Delta x^{3}}\int_{\mathbf{x}_{0}-\Delta\mathbf{x}}^{\mathbf{x}_{0}-\Delta\mathbf{x}}d^{3}\mathbf{x}
\langle w|\psi^{\dagger}\psi|w\rangle\nonumber \\
= & \frac{1}{\Delta x^{3}}\int_{-\infty}^{+\infty}d^{3}\mathbf{x}\Psi^{\dagger}(\mathbf{x})\Psi(\mathbf{x})\nonumber \\
= & \frac{1}{\Delta x^{3}}\int\frac{d^{3}\mathbf{k}}{(2\pi)^{3}}\frac{d^{3}\mathbf{p}}{(2\pi)^{3}}\frac{u^{s\dagger}(k)
u^{s}(p)}{\sqrt{2E_{\mathbf{k}}}\sqrt{2E_{\mathbf{p}}}}w(\mathbf{k})^{*}w(\mathbf{p})\nonumber \\
 & \times\left[\int_{-\infty}^{+\infty}d^{3}\mathbf{x}e^{i(\mathbf{p}-\mathbf{k})\cdot\mathbf{x}}\right]\nonumber \\
= & \frac{1}{\Delta x^{3}}\int\frac{d^{3}\mathbf{p}}{(2\pi)^{3}}
\frac{u^{s\dagger}(p)u^{s}(p)}{2E_{\mathbf{p}}}w(\mathbf{p})^{*}w(\mathbf{p})\nonumber \\
= & \frac{1}{\Delta x^{3}},\label{eq:-51}
\end{align}
which is exactly the particle number density of a single particle
distributed in the small volume $\Delta x^{3}$. 

If $\psi^{\dagger}$ is replaced with $\overline{\psi}$ in Eq.~\eqref{eq:-51},
at the last step we would have $\overline{u^{s}(p)}u^{s}(p)\rightarrow2m_{\psi}$
{[}see Eq.~\eqref{eq:-45}{]} and hence
\begin{equation}
\langle w|\psi^{\dagger}\psi|w\rangle\xrightarrow{t=0,\ \mathbf{x}\ {\rm averaged\ locally}}\frac{1}{\Delta x^{3}}
\int\frac{d^{3}\mathbf{p}}{(2\pi)^{3}}\frac{m_{\psi}}{E_{\mathbf{p}}}|w(\mathbf{p})|^{2}.\label{eq:-52}
\end{equation}
Since $|w(\mathbf{p})|^{2}$ is mostly concentrated near $\mathbf{p}_{0}$
{[}see Eq.~\eqref{eq:-44}{]}, it can be approximately reduced to
$\frac{1}{\Delta x^{3}}\frac{m_{\psi}}{E_{\mathbf{p}_{0}}}$, which
in the non-relativistic regime is equivalent to the number density
but in the relativistic regime is suppressed.

The above calculation can be straightforwardly generalized to $N$
particles with different momentum and difference spins. Each of them
has an independent wave-packet function $w_{i}(\mathbf{p})$. In this
case, we sum over the contributions together:
\begin{equation}
\langle\overline{\psi}\psi\rangle=\sum_{i=1}^{N}\langle w_{i}|\overline{\psi}\psi|w_{i}\rangle,\ \ 
\langle\psi^{\dagger}\psi\rangle=\sum_{i=1}^{N}\langle w_{i}|\psi^{\dagger}\psi|w_{i}\rangle.\label{eq:-53}
\end{equation}
Denote
\begin{equation}
f(\mathbf{p})=\sum_{i=1}^{N}\frac{1}{\Delta x^{3}}|w_{i}(\mathbf{p})|^{2},\label{eq:-54}
\end{equation}
which can be regarded as the momentum distribution function of the
$N$ particles. Then the summation gives
\begin{align}
\langle\psi^{\dagger}\psi\rangle & =\int\frac{d^{3}\mathbf{p}}{(2\pi)^{3}}f(\mathbf{p}),\label{eq:-55}\\
\langle\overline{\psi}\psi\rangle & =\int\frac{d^{3}\mathbf{p}}{(2\pi)^{3}}\frac{m_{\psi}}{E_{\mathbf{p}}}f(\mathbf{p}).\label{eq:-56}
\end{align}

\section{Potential issues on the binding energy and the field energy\label{sec:energy-issues}}
%%%%%%%%%%%%%%%%%%%%%%%%%%%%%%%%%%%%%%%%%%%%%%%%%%%%%%%%%%%%%%%%%%%%%%%%%%%%%%%%%%%%%%%%%%%%%%%%%%%

\noindent 
There is a potential issue regarding the binding energy of
two particles 
in Sec.~\ref{sec:Yukawa-potential}: for two particles, the binding energy does not contain a factor of two.
To address this issue, 
here we discuss the two-body problem from both the
classical and modern (field theory) perspectives. 

In the classical picture, we assume that each particle can be infinitely
split to smaller particles. Denote the two particles as $A$ and $B$.
If we could remove an infinitely small percentage ($\epsilon$, $\epsilon\ll1$)
of particle $B$ to $r=\infty$, the energy cost would be $\frac{y^{2}}{4\pi}\frac{1}{r}e^{-rm_{\phi}}\epsilon$
(due to the attraction of $A$) plus the energy cost to overcome the
attraction of $B$. The latter part will be canceled out when all
the small fractions of $B$ are re-assembled at $r=\infty$. So the
total energy cost (i.e. binding energy) to separate $A$ and $B$
is 
\begin{equation}
E_{{\rm binding}}=\frac{y^{2}}{4\pi}\frac{1}{r}e^{-rm_{\phi}}\sum\epsilon=\frac{y^{2}}{4\pi}\frac{1}{r}e^{-rm_{\phi}}.\label{eq:-142}
\end{equation}
Hence we conclude that the total binding energy for a two-body system
should take $N=1$ in Eq.~\eqref{eq:V_NR}.

In addition to the classical picture, from the perspective of field
theory, an equivalent interpretation is that the binding energy is
stored in the field $\phi$. According to Eq.~\eqref{eq:L}, the Hamiltonian
for static $\phi$ is 
\begin{equation}
H=\int\left[\frac{1}{2}(\nabla\phi)^{2}+\frac{1}{2}m_{\phi}^{2}\phi^{2}\right]d^{3}\mathbf{x}.\label{eq:-63}
\end{equation}
For the case of two particles, we have 
\begin{align}
n(\mathbf{x}) & =\delta^{3}(\mathbf{x}-\mathbf{x}_{A})+\delta^{3}(\mathbf{x}-\mathbf{x}_{B}),\label{eq:-64}\\
\phi(\mathbf{x}) & =-\frac{y}{4\pi}\left(\frac{1}{r_{A}}e^{-r_{A}m_{\phi}}+\frac{1}{r_{B}}e^{-r_{B}m_{\phi}}\right),\label{eq:-65}
\end{align}
 where $\mathbf{x}_{A}$ and $\mathbf{x}_{B}$ are coordinates of
$A$ and $B$, and $r_{A,B}=\sqrt{|\mathbf{x}|^{2}-|\mathbf{x}_{A,B}|^{2}}$.
Although $H$ computed in this way is divergent, by comparing two
cases with $|\mathbf{x}_{A}-\mathbf{x}_{B}|=r$ and $|\mathbf{x}_{A}-\mathbf{x}_{B}|=\infty$,
the difference of $H$ is finite. Plugging Eq.~\eqref{eq:-65} into
Eq.~\eqref{eq:-63} and using $\int(\nabla\phi)^{2}d^{3}\mathbf{x}=-\int\phi\nabla^{2}\phi d^{3}\mathbf{x}$,
it is straightforward to obtain the energy difference:
\begin{equation}
V=\Delta H=-\frac{y^{2}}{4\pi}\frac{1}{r}e^{-rm_{\phi}}.\label{eq:V_NR2}
\end{equation}
This is consistent with Eq.~\eqref{eq:-142} in the classical picture.

\section{Numerical method to solve the two-particle Schr\"odinger equation
\label{sec:two-body-sch}}

\noindent For a quantum system containing two interacting particles,
the Schr\"odinger equation reads
\begin{equation}
i\frac{\partial}{\partial t}\Psi(\mathbf{r}_{1},\ \mathbf{r}_{2})=H\Psi(\mathbf{r}_{1},\ \mathbf{r}_{2}),\label{eq:-82}
\end{equation}
where $\Psi(\mathbf{r}_{1},\ \mathbf{r}_{2})$ is the wave function
with $\mathbf{r}_{1}$ and $\mathbf{r}_{2}$ being the coordinates
of the two particles, and $H$ is the Hamiltonian:
\begin{equation}
H=-\frac{1}{2m_{1}}\nabla_{1}^{2}-\frac{1}{2m_{2}}\nabla_{2}^{2}+V(\mathbf{r}_{1},\ \mathbf{r}_{2}).\label{eq:-83}
\end{equation}
Here $m_{1}$ and $m_{2}$ are the masses of the two particles and
$V$ is the potential. 

Define the reduced mass of the two particles:
\begin{equation}
\mu\equiv\frac{m_{1}m_{2}}{m_{1}+m_{2}},\label{eq:-84}
\end{equation}
and
\begin{equation}
\mathbf{r}\equiv\mathbf{r}_{1}-\mathbf{r}_{2},\ \mathbf{R}\equiv\frac{m_{1}\mathbf{r}_{1}+m_{2}\mathbf{r}_{2}}{m_{1}+m_{2}},\label{eq:-85}
\end{equation}
where $\mathbf{R}$ is the coordinate of the center of mass. Then
$\nabla_{1}$ and $\nabla_{2}$ can be replaced by the derivatives
with respect to $\mathbf{r}$ and $\mathbf{R}$ (denoted as $\nabla_{R}$
and $\nabla_{r}$ below):
\begin{equation}
\nabla_{1}=\frac{\mu}{m_{2}}\nabla_{R}+\nabla_{r},\ \nabla_{2}=\frac{\mu}{m_{1}}\nabla_{R}-\nabla_{r},\label{eq:-86}
\end{equation}
So the Hamiltonian can be transformed to
\begin{equation}
H=H_{R}+H_{r},\label{eq:-88}
\end{equation}
where
\begin{equation}
H_{R}\equiv-\frac{1}{2\left(m_{1}+m_{2}\right)}\nabla_{R}^{2},\ \ H_{r}\equiv-\frac{1}{2\mu}\nabla_{r}^{2}+V(r).\label{eq:-87}
\end{equation}
Note that here we assume that $V$ depends on $r$ only. Eqs.~\eqref{eq:-88}
and \eqref{eq:-87} allows us to separate the motion of $\text{\ensuremath{\mathbf{R}}}$
with that of $\text{\ensuremath{\mathbf{r}}}$:
\begin{equation}
\Psi(\mathbf{r}_{1},\ \mathbf{r}_{2})=\psi(\mathbf{r})\tilde{\psi}(\mathbf{R}).\label{eq:-89}
\end{equation}
In this work we are only concerned with the part of $\psi(\mathbf{r})$,
which can be expanded using spherical harmonics. Plugging Eq.~\eqref{eq:psi_sol}
to 
\begin{equation}
H_{r}\psi(\mathbf{r})=E\psi(\mathbf{r}),\label{eq:-90}
\end{equation}
we obtain 
\begin{equation}
-\frac{1}{2\mu}\frac{d^{2}u(r)}{dr^{2}}+\left[V(r)+\frac{1}{2\mu}\frac{l(l+1)}{r^{2}}\right]u(r)=Eu(r).\label{eq:-4-1}
\end{equation}

Define
\begin{equation}
\tilde{V}\equiv2\mu V(r)+\frac{l(l+1)}{r^{2}},\ \tilde{E}\equiv2\mu E.\label{eq:-7}
\end{equation}
Then the equation is simplified to
\begin{equation}
-u''(r)+\tilde{V}(r)u(r)=\tilde{E}u(r).\label{eq:-5-1}
\end{equation}

Given any form of $\tilde{V}(r)$, in principle Eq.~(\ref{eq:-5-1})
can be numerically solved. Since $r$ varies from zero to infinity,  %$u(r)$ is a function $[0,\ \infty)$,
it is actually more convenient to make the following variable transformation:
\begin{equation}
v(\omega)=u(r)\cos\omega,\ \omega\equiv\arctan\frac{r}{R},\label{eq:-6}
\end{equation}
 so that $v(\omega)$ is a function on a finite interval $\omega\in[0,\pi/2)$.
Here $R$ in principle can be an arbitrary length scale but in order
to improve the efficiency of the numerical method it should be set
at  the typical length scale of the wave function that is being inspected. 

After the transformation, Eq.~(\ref{eq:-5-1}) becomes
\begin{equation}
-\frac{\cos^{4}\omega}{R^{2}}\frac{d^{2}}{d\omega^{2}}v(\omega)+\left[\tilde{V}(\omega) - 
\frac{\cos^{4}\omega}{R^{2}}\right]v(\omega)=\tilde{E}v(\omega).\label{eq:v_eq}
\end{equation}
This equation can be solved by identifying it as the problem of find
eigenvalues and eigenvectors of a large matrix. If we  divide the
interval $[0,\pi/2)$ into $N$ segments and $\omega$ takes one of
these values:
\begin{equation}
\left(0,\ d\omega,\ 2d\omega,\ \cdots,\ \frac{\pi}{2}-d\omega,\ \frac{\pi}{2}\right),\label{eq:-8}
\end{equation}
where $d\omega=\frac{\pi}{2N}$, then $\frac{d^{2}}{d\omega^{2}}$
can be regarded as an operator in this discrete space, represented
by the $(N+1)\times(N+1)$ matrix below:
\begin{equation}
\frac{d^{2}}{d\omega^{2}}\rightarrow\frac{1}{(\frac{\pi}{2N})^{2}}\left[\begin{array}{cccccc}
-2 & 1\\
1 & -2 & 1\\
 & 1 & -2 & 1\\
 &  &  &  & \cdots\\
 &  &  & 1 & -2 & 1\\
 &  &  &  & 1 & -2
\end{array}\right]_{(N+1)\times(N+1)}.\label{eq:-9}
\end{equation}
And $\tilde{V}(\omega)$ and $\cos^{4}\omega$ are  represented by
diagonal matrices:
\begin{equation}
\tilde{V}(\omega)\rightarrow{\rm diag}\left[\tilde{V}(0),\ \tilde{V}(d\omega),\ 
\tilde{V}(2d\omega),\ \cdots,\ \tilde{V}(\frac{\pi}{2})\right],\label{eq:-93}
\end{equation}
\begin{equation}
\cos^{4}\omega\rightarrow{\rm diag}\left[\cos^{4}(0),\ \cos^{4}(d\omega),\ \cos^{4}(2d\omega),\ \cdots,\ 
\cos^{4}(\frac{\pi}{2})\right].\label{eq:-92}
\end{equation}
In the large-$N$ limit, by solving  eigenvalues and eigenvectors
of $\tilde{H}$ defined below 
\begin{equation}
\tilde{H}\equiv-\frac{\cos^{4}\omega}{R^{2}}\frac{d^{2}}{d\omega^{2}}+\left[\tilde{V}(\omega) - 
\frac{\cos^{4}\omega}{R^{2}}\right],\label{eq:-91}
\end{equation}
we can obtain energy levels and wave functions.

\section{Known constraints on the yukawa coupling\label{sec:Known-constraints}}

\noindent There have been a variety of experimental bounds on a light
scalar coupled to neutrinos. Here we compile known results (as summarized in Tab.~\ref{tab:bounds}) in the
literature and comment on their relevance to the scenario considered
in this work.

%%%%%%%%%%%%%%%%%%%table2%%%%%%%%%%%%%%%%%%%%%%%%%%%%%%%%%%%%%%%%%%%%
\begin{table*}[h]
\caption{\label{tab:bounds}A summary of known bounds on the neutrinophilic yukawa
 coupling $y$, assuming $m_{\phi}$ is well below the neutrino mass. }

\begin{ruledtabular}
\begin{tabular}{cccc}
processes & flavor dependence & bounds & Ref.\phantom{xx}\tabularnewline
\hline 
$\pi^{\pm}$ decay & $\nu_{e}$ & $y<1.3\times10^{-2}$ & \cite{Berryman:2018ogk}\tabularnewline
$K^{\pm}$ decay & $\nu_{e}$, $\nu_{\mu}$ & $y<1.4\times10^{-2}$ ($\nu_{e}$) or $<3\times10^{-3}$ ($\nu_{\mu}$) & \cite{Berryman:2018ogk}\tabularnewline
$\beta\beta$ decay & $\nu_{e}$ & $y<3.4\times10^{-5}$ & \cite{Agostini:2015nwa}\tabularnewline
$Z$ decay & all flavors & $y<0.3$ & \cite{Brdar:2020nbj}\tabularnewline
BBN & all flavors & $y<4.6\times10^{-6}$ & \cite{Huang:2017egl}\tabularnewline
CMB & all flavors & $y<8.2\times10^{-7}$ & \cite{Forastieri:2015paa}\tabularnewline
SN1987A (energy loss)\phantom{xxxx} & all flavors & $y<3\times10^{-7}$ or $2\times10^{-5}<y<3\times10^{-4}$ & \cite{Kachelriess:2000qc}\tabularnewline
SN1987A (deleptonization) & $\nu_{e}$ & $y<2\times10^{-6}$ & \cite{Farzan:2002wx}\tabularnewline
\end{tabular}
\end{ruledtabular}

\end{table*}
%%%%%%%%%%%%%%%%%%%%%%%%%%%%%%%%%%%%%%%%%%%%%%%%%%%%%%%%%%%%%%%%%%%%%%%%%%%%%%%%%%%%%%%%%%%%%%%%%%%%%%%%%%%%%%%%%%%%%%%%%%%%%%%%%%%%%%%%%%%%%%

In laboratory experiments, a light scalar that exclusively couples
to neutrinos can be constrained by particle physics processes such
as meson decay ($\pi^{\pm}$, $K^{\pm}$)~\cite{Barger:1981vd,Lessa:2007up,Pasquini:2015fjv,Berryman:2018ogk},
$Z$ invisible decay~\cite{Brdar:2020nbj}, and  double beta ($\beta\beta$)
decay~\cite{Burgess:1992dt,Burgess:1993xh,Blum:2018ljv,Cepedello:2018zvr,Brune:2018sab,Deppisch:2020sqh}.
Since the typical momentum transfer scales of these processes varying
from $10^{2}$ GeV to MeV are well above the mass of $\phi$ considered
in this work, the constraints from these processes on the coupling
$y$ are almost independent of $m_{\phi}$ when $m_{\phi}$ is lighter
than $m_{\nu}$.  Most of these particle decay bounds only apply
to the electron and/or muon flavor, except for $Z$ invisible decay
which is relevant to all flavors (For couplings to $\nu_{\tau}$,
$\tau$ decay is less restrictive than $Z$ decay \cite{Brdar:2020nbj}).
In addition, the $\beta\beta$ decay bound is only valid when $\phi$
is coupled to two electron neutrinos instead of one neutrino and one
anti-neutrino. The specific values of aforementioned bounds are summarized
in Tab.~\ref{tab:bounds}.  High precision measurements of neutrino
scattering in principle could be sensitive to such interactions via
bremsstrahlung or loop processes, which compared to the standard processes
are typically suppressed by a factor of $y^{2}/16\pi^{2}\sim10^{-2}y^{2}$.
Current neutrino scattering experiments can only measure the cross
sections at percent-level precision~\cite{Bilmis:2015lja,Lindner:2018kjo},
corresponding to an upper limit of $y\lesssim1$, which is negligibly
weak.  Long-range force searches (also referred to as the fifth force
searches in the literature) based on precision test of gravitational
laws (e.g. the inverse-square law, weak equivalent principle) are
sensitive to light mediators coupled to normal matter (electrons or
baryons)~\cite{Adelberger:2006dh,Adelberger:2009zz,Wagner:2012ui}.
Although in the presence of $\phi$-neutrino couplings, $\phi$-electron
couplings can be induced at the one-loop level~\cite{Xu:2020qek,Chauhan:2020mgv},
such loop-induced couplings are typically highly suppressed by the
large hierarchy between neutrino masses and the electroweak scale.
Hence bounds from long-range force searches can be neglected in this
work.

Astrophysical and cosmological bounds, by contrast, are much more
restrictive.  Observations of neutrinos from the supernova event
SN1987A have been employed in the literature to set various constraints
from  different considerations. The most commonly considered effect
is energy loss, which leads to the constraint $y<3\times10^{-7}$
or $2\times10^{-5}<y<3\times10^{-4}$~\cite{Kachelriess:2000qc}.
If the new interaction is able to convert $\nu_{e}$ to $\overline{\nu}_{e}$
 or to neutrinos of other flavors, then another effect known as deleptonization
can also put a strong bound, $y<2\times10^{-6}$~\cite{Farzan:2002wx}.
Other supernova bounds~\cite{Dent:2012mx,Kazanas:2014mca,Heurtier:2016otg,Dev:2020eam}
assuming the presence of quark couplings, heavy masses, or electromagnetic
decays do not apply here.  In addition, for $y>4.6\times10^{-6}$
the scalar could be thermalized in the early universe, causing observable
effects in the Big Bang Nucleosynthesis (BBN)~\cite{Huang:2017egl}.
For smaller $y$, the scalar cannot be  thermalized but could affect
the evolution of cosmological perturbations and hence be constrained
by CMB data~\cite{Forastieri:2015paa}. Theses bounds are also summarized
in Tab.~\ref{tab:bounds}.

%%%%%%%%%%%%%%%%%%%%%%%%%%%%%%%%%%%%%%%%%%%%%%%%%%%%%%%%%%%%%%%%%%%%%%%%%%%%%%%%%%%%%%%%%
\section{Equivalence between chemical equilibrium and force balance 
\label{sec:chemical-equilibrium}}
%%%%%%%%%%%%%%%%%%%%%%%%%%%%%%%%%%%%%%%%%%%%%%%%%%%%%%%%%%%%%%%%

\noindent In this appendix, we show that the balance of the two forces,  $F_{{\rm Yuk}} = F_{{\rm deg}}$ 
in Eq.~(\ref{eq:hsequil}), is equivalent
to chemical equilibrium of the system  (equilibrium of particle numbers
in the presence of external forces). 
%%\begin{equation}
%%F_{{\rm degenerate}}(r)=F_{\text{Yukawa}}(r)\Longleftrightarrow
%%\frac{d}{dr}\left[(m_{\nu}+y\phi)^{2}+p_{F}^{2}\right]=0,
%%\label{eq:-155}
%%\end{equation}
%%where $\phi=\phi(r)$ and $p_{F}=p_{F}(r)$ are $r$-dependent. 
Indeed, the chemical equilibrium of system  means that 
\begin{equation}
\frac{d\mu}{dr}=0,
\label{eq:-156}
\end{equation}
 where $\mu$ is the chemical potential in the Fermi-Dirac distribution
$f=[e^{(E-\mu)/T}+1]^{-1}$ with the neutrino energy $E$ modified
by $\phi$:
\begin{equation}
E=\sqrt{(m_{\nu}+y\phi)^{2}+p^{2}}.
\label{eq:-157}
\end{equation}
The Fermi energy, $E_F$, is defined as Eq.~\eqref{eq:-157} with $p=p_F$.
Since it is the maximal energy of a particle in the distribution, $E_F = \mu$, 
Eq.~\eqref{eq:-156}
is equivalent to  $dE_F/dr = 0$.  So we have 
\begin{equation}
\frac{d}{dr}\left[(m_{\nu}+y\phi)^{2}+p_{F}^{2}\right]=0,\label{eq:-158}
\end{equation}
or
\begin{equation}
y(m_{\nu}+y\phi)\frac{d\phi}{dr}+p_{F}\frac{dp_{F}}{dr}=0.\label{eq:-159}
\end{equation}

Let us show that  in nonrelativistic limit  $\frac{d\phi}{dr}$ and $\frac{dp_{F}}{dr}$ 
can be related to
the Yukawa force and the degenerate pressure respectively.

To make use of the classic concept of forces, we need to take the
non-relativistic limit in which $y\phi\ll m_{\nu}$ so that $y\phi$  can be neglected
in the first term of \eqref{eq:-159}. The gradient $\frac{d\phi}{dr}$ is related to 
the Yukawa force as 
\begin{equation}
F_{{\rm Yuk}}(r)\equiv-yn\frac{d\phi}{dr}.
\label{eq:-160}
\end{equation}
Therefore  from Eq.~\eqref{eq:-159} we have 
\begin{equation}
F_{{\rm Yuk}}(r)=n\frac{p_{F}}{m_{\nu}}\frac{dp_{F}}{dr}.
\label{eq:-162}
\end{equation}
According to Eq.~\eqref{eq:-109}, it is straightforward to see that in the non-relativistic case  
\begin{equation}
\frac{d{\cal P}_{{\rm deg}}}{dr}=n\frac{p_{F}}{m_{\nu}}\frac{dp_{F}}{dr}.\label{eq:-163}
\end{equation}
Thus, in the non-relativistic regime, Eq.~\eqref{eq:-156} and  Eq.~(\ref{eq:hsequil})
are equivalent to each other.

% [[the force acting on the unit volume: $F = dV/dr = y d\phi/dr$]]

%%%%%%%%%%%%%%%%%%%%%%%%%%%%%%%%%%%%%%%%%%%%%%%%%%%%%%%%%%%%%%%%%%%%%%%
\section{Derivation of  the Lane-Emden equation from equations of motion
\label{sec:Lane-Emden}}
%%%%%%%%%%%%%%%%%%%%%%%%%%%%%%%%%%%%%%%%%%%%%%%%%%%%%%%%%%%%%%%%%%%%%%
\noindent For non-relativistic neutrinos without interactions, we
have $E=\sqrt{m_{\nu}^{2}+p^{2}}\approx m_{\nu}+\frac{p^{2}}{2m_{\nu}}$
and hence $E-\mu\leq0$ corresponds to $m_{\nu}+\frac{p^{2}}{2m_{\nu}}-\mu\lesssim0$.
When $p=p_{F}$, $E-\mu\leq0$ is saturated. In the presence of the
$\phi$-induced potential $V\ll m_{\nu}$, according to Eq.~\eqref{eq:-59},
 $E-\mu\leq0$ corresponds to $m_{\nu}+\frac{p^{2}}{2m_{\nu}}+V-\mu\lesssim0$.
Hence it is convenient to define an effective chemical potential to
absorb $m_{\nu}$ and $V$:
\begin{equation}
\mu_{{\rm eff}}\equiv-m_{\nu}-V+\mu\thinspace.\label{eq:-66}
\end{equation}
When $\mu_{{\rm eff}}$ is positive, we have $p_{F}$ determined by
\begin{equation}
\frac{p_{F}^{2}}{2m_{\nu}}-\mu_{{\rm eff}}\approx0\thinspace.\label{eq:-67}
\end{equation}
 Note that $\mu_{{\rm eff}}$ sometimes can be negative, for which
we should take $p_{F}=0$:
\begin{equation}
p_{F}=\begin{cases}
\sqrt{2m_{\nu}\mu_{{\rm eff}}} & \text{if }\mu_{{\rm eff}}>0\\
0 & \text{if }\mu_{{\rm eff}}<0
\end{cases}\thinspace.\label{eq:-68}
\end{equation}
Consider Eq.~\eqref{eq:-60} where the r.h.s depends on the local
number density $n(\mathbf{x})$, which according to Eq.~\eqref{eq:-12}
is determined by the local Fermi momentum $p_{F}(\mathbf{x})$, which
according to Eq.~\eqref{eq:-67} or \eqref{eq:-68} depends on $V$.
So eventually the r.h.s of Eq.~\eqref{eq:-60}, which as a source
term generates the potential $V$, is determined by $V$ itself: 
\begin{equation}
\left[\nabla^{2}-m_{\phi}^{2}\right]V=\frac{y^{2}}{6\pi^{2}}\left[2m_{\nu}(\mu-m_{\nu}-V)\right]^{3/2}.\label{eq:-74}
\end{equation}
It is more convenient to express Eq.~\eqref{eq:-74} in terms of $\mu_{{\rm eff}}$
using Eq.~\eqref{eq:-66}:
\begin{equation}
\nabla^{2}\mu_{{\rm eff}}+m_{\phi}^{2}(\mu-\mu_{{\rm eff}}-m_{\nu})=-\frac{y^{2}(2m_{\nu})^{3/2}}{6\pi^{2}}\mu_{{\rm eff}}^{3/2}.\label{eq:-76}
\end{equation}
For spherically symmetric distributions and $m_{\phi}=0$, it becomes
\begin{equation}
\frac{1}{r^{2}}\frac{d}{dr}\left[r^{2}\frac{d\mu_{{\rm eff}}(r)}{dr}\right]=-\tilde{\kappa}\thinspace\left[\mu_{{\rm eff}}(r)\right]^{\gamma},\label{eq:-69}
\end{equation}
where
\begin{equation}
\tilde{\kappa}\equiv\frac{y^{2}(2m_{\nu})^{3/2}}{6\pi^{2}},\ \ \gamma=\frac{3}{2}.\label{eq:-70}
\end{equation}
When $\mu_{{\rm eff}}$ is expressed in terms of $n$, one obtains
Eq.~\eqref{eq:-114}. 

\section{Numerical details of solving Eqs.~\eqref{eq:-167} and \eqref{eq:-168}\label{sec:n-sol}}
%%%%%%%%%%%%%%%%%%%%%%%%%%%%%%%%%%%%%%%%%%%%%%%%%%%%%%%%%%%%%%%%%%%%%%%%%%%%%%%%%%%%%%%%%%%%%%%%%%%%%%%%%%

%In Eq.~\eqref{eq:-167}, to avoid instability of ODE solver caused by small $y$,
It is useful to note that we can remove the dependence of the differential equation on $y$ by defining
\begin{equation}
\overline{\nabla}\equiv\nabla/y,\ \ \overline{r}\equiv yr,\ \  \overline{m}_{\phi}\equiv m_{\phi}/y,
\label{eq:redef2}
\end{equation} 
and rewrite Eq.~\eqref{eq:-167} as 
\begin{align}
\overline{\nabla}^{2}\tilde{m}-\overline{m}_{\phi}^{2}(\tilde{m}-m_{\nu}) & =\tilde{n},\label{eq:-167-1}
\end{align}
where 
\begin{equation}
\overline{\nabla}^{2}\tilde{m}=\left[\frac{d^{2}}{d\overline{r}^{2}}+2\frac{d}{d\overline{r}}\right]\tilde{m}.\label{eq:-189}
\end{equation}
 Eq.~\eqref{eq:-168} is equivalent to
\begin{equation}
\tilde{m}^{2}(\overline{r})+p_{F}^{2}(\overline{r})=\tilde{m}_{R}^{2}\thinspace,\label{eq:-168-1}
\end{equation}
where $\tilde{m}_{R}$ is $\tilde{m}_{\nu}$ at $r=R$ (or $\overline{r}=\overline{R}\equiv Ry$).
This allows us to substitute $\tilde{m}_{\nu}(\overline{r})\rightarrow\sqrt{\tilde{m}_{R}^{2}-p_{F}^{2}(\overline{r})}$
in Eqs.~\eqref{eq:-167-1} and express it as an equation of $p_{F}(\overline{r})$. 

The initial values in Eq.~\eqref{eq:-170} are determined as follows. 
We note that $\tilde{m}_{0}$ cannot be set arbitrarily because at $r\to \infty$ it should match $m_{\nu}$. 
Technically, for an arbitrarily $\tilde{m}_{0}$,  quite often one gets a solutions with $p_F(r)$  monotonically increasing or oscillating as $r$ increases, which is certainly not a physical solution. We find that in practice it is more convenient to use $\tilde{m}_{R}$ instead of $\tilde{m}_{0}$.
% \begin{equation}
% \tilde{m}_{0}\equiv\tilde{m}(r=0),\ \ p_{F0}\equiv p_{F}(r=0).\label{eq:-170-1}
% \end{equation}
% They are determined as follows. 
Hence, in our code implementation, we first set values of $(\tilde{m}_{R},\ p_{F0})$ and then use $\tilde{m}_{0}=\sqrt{\tilde{m}_{R}^{2}-p_{F0}^{2}}$ to obtain $(\tilde{m}_{0},\ p_{F0})$.
After solving the differential equations for $0<\overline{r}<\overline{R}$,
 one can obtain $\overline{N}\equiv N y^3$ and $\tilde{m}_{\nu}(r\to \infty)$ using to the following integrals:
\begin{align}
    \overline{N} & =\int_{0}^{\overline{R}}n(\overline{r})4\pi\overline{r}^{2}d\overline{r}\thinspace,\label{eq:-173-1}\\
    \tilde{m}_{\nu}(r\to \infty) & =\tilde{m}_{R}+e^{-\overline{m}_{\phi}\overline{R}}\int_{0}^{\overline{R}}\overline{r}\tilde{n}(\overline{r})\frac{\sinh(\overline{m}_{\phi}\overline{r})}{\overline{m}_{\phi}\overline{R}}d\overline{r}\thinspace.\label{eq:-174-1}
\end{align}

With the above setup, we implement the code in the following way:
\begin{itemize}
\item For each pair of the input parameter $(\tilde{m}_{R},\ p_{F0})$, 
the code obtains a curve of $p_F(\overline{r})$ which may or may not drop to zero. 

\item By scanning over a range of $\tilde{m}_{R}$, it can find the 
interval in which the former happens. In this interval, each solution has a radius $\overline{R}$ determined by the zero point of  $p_F(\overline{r})$. 

\item Correspondingly, $\tilde{m}_{\nu}(r\to \infty)$ can be computed using Eq.~\eqref{eq:-174-1}. And one obtains a curve of  $\tilde{m}_{\nu}(r\to \infty)$ that varies with respect to $\tilde{m}_{R}$.
 
\item With this curve, the code solves the  equation $\tilde{m}_{\nu}(r\to \infty)=m_{\nu}$  with respect to $\tilde{m}_{R}$. Depending on $\overline{m}_{\phi}$ and $p_{F0}$, there can be two, one, or zero solutions. In the case of two solutions, one of them appears on the UR branch and the other  appears on the NR branch.
 
\end{itemize}
% The code is implemented in a way that 
% In summary, given values of $(\tilde{m}_{R},\ p_{F0})$, one obtains
% corresponding values of $(N,\ m_{\nu})$ via
% \begin{equation}
% (\tilde{m}_{R},\ p_{F0})\Rightarrow(\tilde{m}_{0},\ p_{F0})\Rightarrow(N,\ m_{\nu}).\label{eq:-172-1}
% \end{equation}
% Since it is difficult to determine parameters backwardly, in practice,
% we have to tune $(\tilde{m}_{R},\ p_{F0})$ to reach specific values
% of $(N,\ m_{\nu})$. 

\section{ Annihilation cross sections \label{sec:MM}}
In this appendix, we present the detailed calculations of the annihilation cross sections. 
First, we need to compute the squared amplitude. For $\nu+\overline{\nu}$
annihilation, it is computed as follows:
\begin{align}
|{\cal M}|^{2} & =y^{4}|\overline{v_{2}}\left[\frac{i}{\slashed{q}-m_{\nu}}+\frac{i}{\slashed{q}'-m_{\nu}}\right]u_{1}|^{2}\nonumber \\
 & =y^{4}{\rm tr}\left[\left(\slashed{p}_{2}-m_{\nu}\right)\left(\frac{1}{\slashed{q}-m_{\nu}}+\frac{1}{\slashed{q}'-m_{\nu}}\right)\left(\slashed{p}_{1}+m_{\nu}\right)\left(\frac{1}{\slashed{q}-m_{\nu}}+\frac{1}{\slashed{q}'-m_{\nu}}\right)\right]\nonumber \\
 & =2y^{4}\frac{-32m_{\nu}^{8}+16m_{\nu}^{6}(t+u)+m_{\nu}^{4}\left(t^{2}+30tu+u^{2}\right)-m_{\nu}^{2}(t+u)\left(t^{2}+14tu+u^{2}\right)+tu(t-u)^{2}}{\left(m_{\nu}^{2}-t\right)^{2}\left(m_{\nu}^{2}-u\right)^{2}},\label{eq:-79}
\end{align}
where $p_{1}$ and $p_{2}$ are the two initial momenta, $q$ and
$q'$ are the momenta of $t$- and $u$-channel neutrino propagators
($t=q\cdot q$ and $u=q'\cdot q'$). They are related to the two final
momenta $p_{3}$ and $p_{4}$ by $q=p_{1}-p_{3}$ and $q'=p_{1}-p_{4}$.
In the second row, we have used $v_{2}\overline{v}_{2}=\slashed{p}_{2}-m_{\nu}$,
$u_{1}\overline{u}_{1}=\slashed{p}_{1}+m_{\nu}$. 

In the non-relativistic limit and in the center-of-mass frame, we
have 
\begin{equation}
t=-m_{\nu}^{2}+2c_{\theta}\Delta\sqrt{\Delta^{2}+m_{\nu}^{2}}-2\Delta^{2},u=-m_{\nu}^{2}-2c_{\theta}\Delta\sqrt{\Delta^{2}+m_{\nu}^{2}}-2\Delta^{2},\label{eq:-80}
\end{equation}
where $\Delta=|\mathbf{p}_{1}|=|\mathbf{p}_{2}|$ and $c_{\theta}$
is the cosine of the angle between $\mathbf{p}_{1}$ and $\mathbf{p}_{3}$. 

Plugging Eq.~\eqref{eq:-80} to Eq.~\eqref{eq:-79} and expanding
it in terms of $\Delta$, we obtain
\begin{equation}
|{\cal M}|^{2}=8y^{4}\frac{4-3c_{\theta}^{2}}{m_{\nu}^{2}}\Delta^{2}+{\cal O}\left(\frac{\Delta^{4}}{m_{\nu}^{4}}\right).\label{eq:-184}
\end{equation}

For $\nu+\nu$ annihilation, the calculation is similar:
\begin{align}
|{\cal M}|^{2} & =y^{4}|\overline{u_{2}}\left[\frac{i}{\slashed{q}-m_{\nu}}+\frac{i}{\slashed{q}'-m_{\nu}}\right]u_{1}|^{2}\nonumber \\
 & =y^{4}{\rm tr}\left[\left(\slashed{p}_{2}+m_{\nu}\right)\left(\frac{1}{\slashed{q}-m_{\nu}}+\frac{1}{\slashed{q}'-m_{\nu}}\right)\left(\slashed{p}_{1}+m_{\nu}\right)\left(\frac{1}{\slashed{q}-m_{\nu}}+\frac{1}{\slashed{q}'-m_{\nu}}\right)\right]\nonumber \\
 & =2y^{4}\frac{8m_{\nu}^{4}-4m_{\nu}^{2}(t+u)+(t-u)^{2}}{\left(m_{\nu}^{2}-t\right)\left(m_{\nu}^{2}-u\right)}.\label{eq:-79-1}
\end{align}
The main difference is that $\overline{v_{2}}$ has been replaced
by $\overline{u_{2}}$.

Plugging Eq.~\eqref{eq:-80} to Eq.~\eqref{eq:-79-1} and expanding
it in terms of $\Delta$, we obtain 

\begin{equation}
|{\cal M}|^{2}=8y^{4}\left(1+\frac{2c_{\theta}^{2}-1}{m_{\nu}^{2}}\Delta^{2}\right)+{\cal O}\left(\frac{\Delta^{4}}{m_{\nu}^{4}}\right).\label{eq:-185}
\end{equation}
In summary, the squared amplitude in the non-relativistic regime reads:
\begin{equation}
|{\cal M}|^{2}\approx\begin{cases}
8y^{4}\left(4-3\cos^{2}\theta\right)\left(\frac{v}{2}\right)^{2} & \text{for }\nu+\overline{\nu}\rightarrow\phi+\phi\\
8y^{4} & \text{for }\nu+\nu\rightarrow\phi+\phi
\end{cases}.\label{eq:-96}
\end{equation}
With the above result of $|{\cal M}|^{2}$ , the total cross section
of annihilation can be computed by
\begin{equation}
\sigma=\frac{1}{64\pi\left[\left(p_{1}\cdot p_{2}\right)^{2}-m_{\nu}^{4}\right]}\int|{\cal M}|^{2}dq^{2}.\label{eq:-104}
\end{equation}
Here $q^{2}=(p_{3}-p_{1})^{2}\approx m_{\nu}^{2}(-1+vc_{\theta}$
varies in the range $m_{\nu}^{2}(-1-v)\lesssim q^{2}\lesssim m_{\nu}^{2}(-1+v)$.
Integrating $q^{2}$ in the kinetically allowed domain, we obtain
the result in Eqs.~\eqref{eq:-186} and \eqref{eq:-187}.

%%%%%%%%%%%%%%%%%%%%%%%%%%%%%%%%%%%%%%%%%%%%%%%%%%%%%%%%%%%%%%%%%%%%%%%%%%%
\section{Energy loss due to $\phi$ bremsstrahlung \label{sub:energy-loss}}
%%%%%%%%%%%%%%%%%%%%%%%%%%%%%%%%%%%%%%%%%%%%%%%%%%%%%%%%%%%%%%%%%%%%%

Due to  $\phi$ bremsstrahlung
and neutrino  annihilation, the energy loss can cause further contraction 
of the virialized halo. 
Let us first consider the bremsstrahlung process in two-neutrino collision:
$\nu+\nu\rightarrow\nu+\nu+\phi$ with $\phi$ exchange in $t$ channel. 
The differential cross-section can be estimated as
\begin{equation}
\frac{d\sigma}{dq^2}    
\sim \frac{y^{2}}{8\pi^{2}} \frac{y^{4}}{16\pi v|q^{2}|^{2}} \thinspace,
\label{eq:-126}
\end{equation}
where  $q$ is the momentum carried by the $t$-channel $\phi$ mediator and 
$v$ is the relative velocity of two colliding neutrinos.  The energy taken away
by $\phi$  equals  roughly $E_{\phi}\simeq |q^{2}| /2m_{\nu}$, 
%%Although the cross section at low $q^{2}$ is large, the energy emission
%%is small. So we are more interested in 
so that according to (\ref{eq:-126}) the energy loss is proportional to 
\begin{equation}
\langle\sigma E_{\phi}\rangle \simeq 
\frac{y^{2}}{8\pi^{2}}\int^{q^2_{\max}}_{q^2_{\min}} 
\frac{y^{4}}{16\pi v(|q^{2}| + m_\phi^2)^{2}} \frac{|q^{2}|}{2m_{\nu}}dq^{2} 
\simeq 
\frac{y^{6}}{256\pi^{3}vm_{\nu}} \log\left(q_{{\rm max}}^{2}/m_\phi^{2}\right).  
\label{eq:-138}
\end{equation}
Here  $q_{{\rm max}}^{2}$ and $q_{{\rm min}}^{2}$
are the maximal and minimal values of $|q^{2}|$. The former is determined
by kinematics, $q_{{\rm max}}^{2}\sim(m_{\nu}v)^{2}$, and for the latter  we took
$q_{{\rm min}}^{2} \approx 0$. 
% latter we take $q_{{\rm min}}^{2}\sim r_{{\rm mean}}^{-2}$,  where
% $r_{{\rm mean}}\sim n^{-1/3}$ is the mean distance between neutrinos.
Consequently, the rate of energy loss of a given neutrino
can be estimated as
\begin{equation}
\frac{dE_{{\rm loss}}}{dt}\sim\langle\sigma E_{\phi}\rangle nv\thinspace. 
\label{eq:-139}
\end{equation}

If $q^{2}\ll r_{{\rm mean}}^{-2}$,   a given neutrino interacts 
coherently with  neutrinos within the radius 
$$
%\begin{equation}
r_{{\rm coherent}}\sim 1/p \thinspace,
%%r_{{\rm coherent}}\sim 1/|q|\thinspace,
%\label{eq:-133}
%\end{equation}
$$
and the differential cross section can be enhanced by
a factor of $\tilde{N}^{2}$,  where 
$\tilde{N}\sim\frac{4}{3}\pi nr_{{\rm coherent}}^{3}$
is the number of neutrinos within the coherence volume.   On the
other hand, taking $\tilde{N}$ neutrinos as a target, the effective
number density of scatterers is reduced to $n/\tilde{N}$. As a result, the rate 
in Eq.~\eqref{eq:-139}  still increases by a factor of $\tilde{N}$  
%which means we can simply take $\tilde{N} \sim N$ to estimate the largest energy loss rate.
$$
%\begin{equation} 
\tilde{N} = {\cal O}(10).  
%\label{eq:tilde}
%\end{equation}
$$
The energy loss rate per neutrino reads
\begin{equation}
\frac{dE_{{\rm loss}}}{dt}\sim\frac{y^{6} \tilde{N}n}{256\pi^{3}m_{\nu}} 
\log\left(q_{\rm max}^2/ m_\phi^2 \right).
\label{eq:-140}
\end{equation}
The time it takes for the system to cool down and lose half of its kinetic energy can be estimated by
\begin{equation}
\tau_{{\rm cooling}}\sim
\frac{E_{K}^{({\rm vir})}}{2 \tilde{N} (dE_{{\rm loss}}/dt})\thinspace,
\label{eq:-141}
\end{equation}
where $E_{K}^{\rm vir}$  is determined in Eq.~\eqref{eq:-124}.
We take the  virialized radius 
$R_{{\rm vir}}\sim R_{\rm in}/2$ where $R_{\rm in}$ is given 
Eq.~\eqref{eq:rin1}. Then  using $n=N/(\frac{4}{3}\pi R_{{\rm vir}}^{3})$ and 
$v^{2}=E_{K}^{({\rm vir})}/(\frac{1}{2}m_{\nu}N)$,
we obtain 
$$
%\begin{equation}
\tau_{{\rm cooling}}\sim  10^{34} {\rm sec}  \frac{1}{\tilde{N}} 
\left(\frac{10^{-7}}{y}\right)^{3}\left(\frac{0.1\ 
{\rm eV}}{m_{\nu}}\right)\left(\frac{p_{F0}}{0.1m_{\nu}}\right)^{-5/2},
%\label{eq:20210526-6}
%\end{equation}
$$
which is much larger than the age of the Universe 
%%\begin{equation}
$\tau_{\text{universe}}=4.35\times10^{17}$\ sec.
 %, unless $y > 10^{-4}$ [[???]] 
%%\thinspace.\label{eq:20210526-7}
%%\end{equation}
% Actually  for thermal distribution $\tilde{N} \sim {\cal O}(1)$. 

%%%%%%%%%%%%%%%%%%%%%%%%%%%%%%%%%%%%%%%%%%%%%%%%%%%%%%%%%%%
\section{Annihilation\label{subsec:Rates}}
%%%%%%%%%%%%%%%%%%%%%%%%%%%%%%%%%%%%%%%%%%%%%%%%%%%%%%%%%%

\begin{figure}[h]
\includegraphics{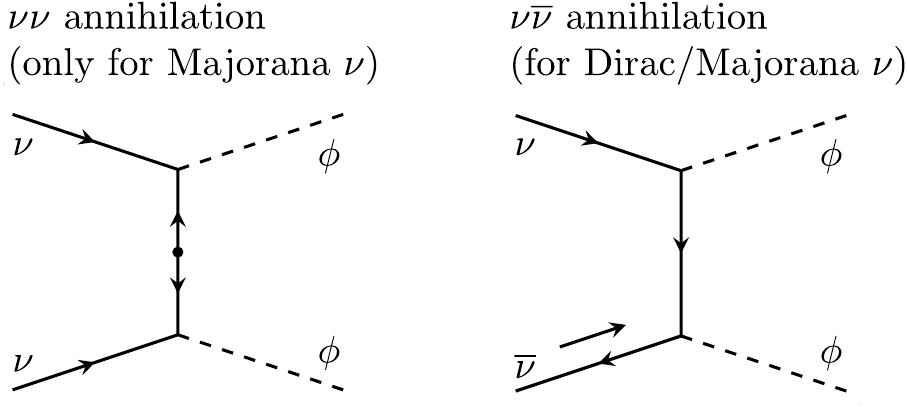}
\caption{Majorana and Dirac neutrino annihilation.
\label{fig:Majorana-and-Dirac}}
\end{figure}

Neutrino annihilation rate 
depends on whether neutrinos are Dirac or Majorana particles. For Dirac
neutrinos, the annihilation channel 
is $\nu+\overline{\nu}\rightarrow\phi+\phi$ 
(see the
second diagram in Fig.~\ref{fig:Majorana-and-Dirac}). 
%%provided sufficiently high number densities  of both $\nu$ and $\overline{\nu}$
%%in the $N$-body system. 
For the Majorana neutrinos, there is an additional 
%%to $\nu+\overline{\nu}\rightarrow\phi+\phi$,
process $\nu+\nu\rightarrow\phi+\phi$ with Majorana mass insertion 
(see the first diagram in Fig.~\ref{fig:Majorana-and-Dirac} ).
% [[interchange the diagrams]] 

%%%%%%%%%%%%%%%%%%ffff8 %%%%%%%%%%%%%%%%%%%%%%%%%%%%%%%%%%%%%%
%%%%%%%%%%%%%%%%%%%%%%%%%%%%%%%%%%%%%%%%%%%%%%%%%%%%%%%%%%%%%%%%%%%%%%%%%%%%%%%%%%

In the non-relativistic case the cross sections read (see Appendix~\ref{sec:MM}
for details of the calculation):
\begin{align}
\sigma^{\nu \bar{\nu}}  \equiv \sigma (\nu+\overline{\nu}\rightarrow\phi+\phi) & 
\approx\frac{3vy^{4}}{16\pi m_{\nu}^{2}}\thinspace,
\label{eq:-186}\\
\sigma^{\nu \nu}  \equiv \sigma (\nu+\nu\rightarrow\phi+\phi) & 
\approx\frac{y^{4}}{4\pi m_{\nu}^{2}v}\thinspace, 
\label{eq:-187}
\end{align}
where $v$ is the relative velocity of colliding neutrinos: 
$$
%\begin{equation}
v  =\sqrt{v_{1}^{2}+v_{2}^{2}-2v_{1}v_{2}\cos\theta_{12}}\thinspace,
%\label{eq:velo}
%\end{equation}
$$
and $v_{i}=|\mathbf{p}_{i}|/m_{\nu}$  ($i = 1,2$)  are the velocities of
the colliding neutrinos and $\theta_{12}$ the angle between $\mathbf{p}_{1}$
and $\mathbf{p}_{2}$.

The key difference is dependence on $v$: 
$\sigma^{\nu \bar{\nu}} \propto v$,  while 
$\sigma^{\nu \nu} \propto v^{-1}$. 
For the Majorana case (both $\nu+\overline{\nu}$ and $\nu+\nu$
channels are present) in the limit $v \ll 1$, 
the left diagram dominates. Hence for Dirac or 
Majorana neutrinos, one only needs to consider
$\nu+\overline{\nu}$ or $\nu+\nu$ annihilation, respectively. 

The change of number density due to annihilation can be computed as 
\begin{equation}
\frac{dn}{dt}= - \int f_{1}(\mathbf{p}_{1})f_{2}(\mathbf{p}_{2}) \sigma (v) v 
\frac{d^{3}\mathbf{p}_{1}}{(2\pi)^{3}}\frac{d^{3}\mathbf{p}_{2}}{(2\pi)^{3}}\thinspace,
\label{eq:-103}
\end{equation}
where $f_{1}$ and $f_{2}$ are the distribution functions
of the initial neutrinos.   
Introducing integration over velocities the  rate \eqref{eq:-103} becomes  
$$
%\begin{equation}
\frac{dn}{dt} =-\int f_{1}f_{2} \sigma (v) v
\frac{m_{\nu}^{3}4\pi v_{1}^{2}dv_{1}}{(2\pi)^{3}}
\frac{m_{\nu}^{3}2\pi v_{2}^{2}dv_{2}
d\cos\theta_{12}}{(2\pi)^{3}}\thinspace,
%\label{eq:-105}
%\end{equation}
$$
The result reads
\begin{align}
\frac{dn^{\nu \bar{\nu}}}{dt} & \approx-\frac{p_{F}^{8}y^{4}}{160\pi^{5}m_{\nu}^{4}}, ~~~~~
%%\ \ \text{for }\nu+\overline{\nu}\rightarrow\phi+\phi\thinspace,\\
\frac{dn^{\nu \nu}}{dt} \approx-\frac{p_{F}^{6}y^{4}}{144\pi^{5}m_{\nu}^{2}}.\ 
%%\ \text{for }\nu+\nu\rightarrow\phi+\phi\thinspace.
\label{eq:-127}
\end{align}

Using Eq.~\eqref{eq:-12}, we express the number density $n$
in terms of $p_{F}$, thus obtaining the differential equation for $p_{F}$: 
\begin{align}
\frac{dp_F^{\nu \bar{\nu}}}{dt} & =-\frac{p_{F}^{6}y^{4}}{80\pi^{3}m_{\nu}^{4}}\,, 
%%\ \text{for }\nu+\overline{\nu}\rightarrow\phi+\phi\thinspace,
\label{eq:-99}\\
\frac{dp_F^{\nu \nu}}{dt} & =-\frac{p_{F}^{4}y^{4}}{72\pi^{3}m_{\nu}^{2}}\,,
%%\ \text{for }\nu+\nu\rightarrow\phi+\phi\thinspace.
\end{align}
The solution is 
\begin{align}
p_{F}^{\nu \bar{\nu}}(t) & = p_{F}^0\left[1+\frac{p_{F0}^{5}y^{4}}{16\pi^{3}m_{\nu}^{4}}t\right]^{-1/5}\,,
%%\ \text{for }\nu+\overline{\nu}\rightarrow\phi+\phi\thinspace,
\label{eq:-128}\\
p_{F}^{\nu \nu}(t) & =p_{F}^0\left[1+\frac{p_{F0}^{3}y^{4}}{24\pi^{3}m_{\nu}^{2}}t\right]^{-1/3}\,,
%%\ \ \text{for }\nu+\nu\rightarrow\phi+\phi\thinspace.
\label{eq:-129}
\end{align}
where $p_{F}^0$  is the Fermi momentum in the beginning of the cooling phase. 
The annihilation lifetime $\tau_{{\rm anni}}$  
can be defined as the time during which the Fermi momentum reduces 
by a factor of 2: $p_{F}(\tau_{{\rm anni}})=p_{F}^0/2$,
which corresponds to the density being reduced by a factor of 8. 
Then from \eqref{eq:-128} and \eqref{eq:-129} 
we obtain 
\begin{align}
\tau^{\nu \bar{\nu}} & =\frac{496\pi^{3}m_{\nu}^{4}}{y^{4}p_{F0}^{5}}
%%\ \ \text{for }\nu+\overline{\nu}\rightarrow\phi+\phi\thinspace,
= 1.0\times10^{23}\text{seconds}\times\left(\frac{0.1\ {\rm eV}}{m_{\nu}}\right)
\left(\frac{0.1m_{\nu}}{p_{F0}}\right)^{5}\left(\frac{10^{-7}}{y}\right)^{4}, 
\label{eq:-136}\\
\tau^{\nu \nu} & =\frac{168\pi^{3}m_{\nu}^{2}}{y^{4}p_{F0}^{3}} 
%%\ \text{for }\nu+\nu\rightarrow\phi+\phi\thinspace,
= 3.4\times10^{20}\text{seconds}\times\left(\frac{0.1\ {\rm eV}}{m_{\nu}}\right)
\left(\frac{0.1m_{\nu}}{p_{F0}}\right)^{3}\left(\frac{10^{-7}}{y}\right)^{4}. 
\label{eq:-181}
\end{align}
Therefore for typical values of parameters presented in Eqs.~\eqref{eq:-136}
and \eqref{eq:-181},  one finds that 
that  $\tau_{{\rm anni}}\gg\tau_{\text{universe}}=4.35\times10^{17} {\rm seconds}$.

\newpage

\begin{acknowledgments}
    %%%%%%%%%%%%%%%%%%%%%%%%%%%%%%%%%%%%%%%%%%%%%%%%%%%%%%%%%%%%%%
    The authors would like to thank Julian Heeck and Arvind Rajaraman for useful discussions 
    on this topic in 2019, Jerome Vandecasteele and Peter Tinyakov for discussions on similar fermionic systems in neutron stars, and Laura Lopez Honorez for suggesting references. X.J.X is supported in part by the National Natural Science Foundation of China under grant No. 12141501.
    % This work is supported by the ``Probing dark matter with neutrinos'' 
    % ULB-ARC convention and by the F.R.S./FNRS under the Excellence of Science (EoS) 
    % project No.\ 30820817 - be.h ``The $H$ boson gateway to physics beyond the Standard Model''.
\end{acknowledgments}

\bibliographystyle{apsrev4-1}
\bibliography{ref}

\end{document}